\documentclass[12pt]{article}
\usepackage{epsfig}
\usepackage{amsmath}
\usepackage{hhline}
\usepackage{amssymb}
\usepackage{times}
\usepackage{cite}

\newlength{\dinwidth}
\newlength{\dinmargin}
\begin{document}
\newcommand{\ssss}{Schildknecht, Schuler and Surrow}
\newcommand{\mrtt}{Martin, Ryskin and Teubner}
\newcommand{\rcc}{Royen and Cudell}
\newcommand{\ikk}{Ivanov and Kirschner}
\newcommand{\mphi}{\mbox{$m_{\phi}$}}        
\newcommand{\Gphi}{\mbox{$\Gamma_{\phi}$}}   
\newcommand{\cosths}{\mbox{$\cos \theta$}}
\newcommand{\cosdelta}{\mbox{$\cos \delta$}}
\def\mbig#1{\mbox{\rule[-2. mm]{0 mm}{6 mm}#1}}

%
%
\newcommand{\s}{\mbox{$s$}}
\newcommand{\ttra}{\mbox{$t$}}
\newcommand{\modt}{\mbox{$|t|$}}
\newcommand{\eminpz}{\mbox{$E-p_z$}}
\newcommand{\eminpzs}{\mbox{$\Sigma(E-p_z)$}}
\newcommand{\rap}{\ensuremath{\eta^*} }
\newcommand{\W}{\mbox{$W$}}
\newcommand{\w}{\mbox{$W$}}
\newcommand{\Q}{\mbox{$Q$}}
\newcommand{\q}{\mbox{$Q$}}
\newcommand{\xB}{\mbox{$x$}}  
\newcommand{\xF}{\mbox{$x_F$}}  
\newcommand{\xg}{\mbox{$x_g$}}  
\newcommand{\xbj}{x}
\newcommand{\xpom}{x_{\PO}}
\newcommand{\y}{\mbox{$y~$}}
\newcommand{\Qsq}{\mbox{$Q^2$}}
\newcommand{\qsq}{\mbox{$Q^2$}}
\newcommand{\kjet}{\mbox{$k_{T\rm{jet}}$}}
\newcommand{\xjet}{\mbox{$x_{\rm{jet}}$}}
\newcommand{\Ejet}{\mbox{$E_{\rm{jet}}$}}
\newcommand{\thjet}{\mbox{$\theta_{\rm{jet}}$}}
\newcommand{\pjet}{\mbox{$p_{T\rm{jet}}$}}
\newcommand{\et}{\mbox{$E_T~$}}
\newcommand{\kt}{\mbox{$k_T~$}}
\newcommand{\ptrans}{\mbox{$p_T~$}}
\newcommand{\pth}{\mbox{$p_T^h~$}}
\newcommand{\pte}{\mbox{$p_T^e~$}}
\newcommand{\ptsq}{\mbox{$p_T^{\star 2}~$}}
\newcommand{\as}{\mbox{$\alpha_s~$}}
\newcommand{\ycut}{\mbox{$y_{\rm cut}~$}}
\newcommand{\gx}{\mbox{$g(x_g,Q^2)$~}}
\newcommand{\xpart}{\mbox{$x_{\rm part~}$}}
\newcommand{\mrsdm}{\mbox{${\rm MRSD}^-~$}}
\newcommand{\mrsdmp}{\mbox{${\rm MRSD}^{-'}~$}}
\newcommand{\mrsdn}{\mbox{${\rm MRSD}^0~$}}
\newcommand{\lambdams}{\mbox{$\Lambda_{\rm \bar{MS}}~$}}
%
%
\newcommand{\gp}{\ensuremath{\gamma}p }
\newcommand{\gammasp}{\ensuremath{\gamma}*p }
\newcommand{\gammap}{\ensuremath{\gamma}p }
\newcommand{\gsp}{\ensuremath{\gamma^*}p }
\newcommand{\dsiget}{\ensuremath{{\rm d}\sigma_{ep}/{\rm d}E_t^*} }
\newcommand{\dsigrap}{\ensuremath{{\rm d}\sigma_{ep}/{\rm d}\eta^*} }
\newcommand{\epem}{\mbox{$e^+e^-$}}
\newcommand{\ep}{\mbox{$ep~$}}
\newcommand{\epl}{\mbox{$e^{+}$}}
\newcommand{\emi}{\mbox{$e^{-}$}}
\newcommand{\epm}{\mbox{$e^{\pm}$}}
\newcommand{\se}{section efficace}
\newcommand{\ses}{sections efficaces}
%
%
\newcommand{\phib}{\mbox{$\varphi$}}
\newcommand{\rh}{\mbox{$\rho$}}
\newcommand{\rhz}{\mbox{$\rh^0$}}
\newcommand{\ph}{\mbox{$\phi$}}
\newcommand{\om}{\mbox{$\omega$}}
\newcommand{\jpsi}{\mbox{$J/\psi$}}
\newcommand{\pipi}{\mbox{$\pi^+\pi^-$}}
\newcommand{\pip}{\mbox{$\pi^+$}}
\newcommand{\pim}{\mbox{$\pi^-$}}
\newcommand{\kk}{\mbox{K^+K^-$}}
\newcommand{\bsl}{\mbox{$b$}}
\newcommand{\alp}{\mbox{$\alpha^\prime$}}
\newcommand{\alpom}{\mbox{$\alpha_{\PO}$}}
\newcommand{\alpomp}{\mbox{$\alpha_{\PO}^\prime$}}
\newcommand{\rzzzz}{\mbox{$r_{00}^{04}$}}
\newcommand{\rzqzz}{\mbox{$r_{00}^{04}$}}
\newcommand{\rzquz}{\mbox{$r_{10}^{04}$}}
\newcommand{\rzqumu}{\mbox{$r_{1-1}^{04}$}}
\newcommand{\ruuu}{\mbox{$r_{11}^{1}$}}
\newcommand{\ruzz}{\mbox{$r_{00}^{1}$}}
\newcommand{\ruuz}{\mbox{$r_{10}^{1}$}}
\newcommand{\ruumu}{\mbox{$r_{1-1}^{1}$}}
\newcommand{\rduz}{\mbox{$r_{10}^{2}$}}
\newcommand{\rdumu}{\mbox{$r_{1-1}^{2}$}}
\newcommand{\rcuu}{\mbox{$r_{11}^{5}$}}
\newcommand{\rczz}{\mbox{$r_{00}^{5}$}}
\newcommand{\rcuz}{\mbox{$r_{10}^{5}$}}
\newcommand{\rcumu}{\mbox{$r_{1-1}^{5}$}}
\newcommand{\rsuz}{\mbox{$r_{10}^{6}$}}
\newcommand{\rsumu}{\mbox{$r_{1-1}^{6}$}}
\newcommand{\rzqik}{\mbox{$r_{ik}^{04}$}}
\newcommand{\rhzik}{\mbox{$\rh_{ik}^{0}$}}
\newcommand{\rhqik}{\mbox{$\rh_{ik}^{4}$}}
\newcommand{\rhaik}{\mbox{$\rh_{ik}^{\alpha}$}}
\newcommand{\rhzzz}{\mbox{$\rh_{00}^{0}$}}
\newcommand{\rhqzz}{\mbox{$\rh_{00}^{4}$}}
\newcommand{\raik}{\mbox{$r_{ik}^{\alpha}$}}
\newcommand{\razz}{\mbox{$r_{00}^{\alpha}$}}
\newcommand{\rauz}{\mbox{$r_{10}^{\alpha}$}}
\newcommand{\raumu}{\mbox{$r_{1-1}^{\alpha}$}}

\newcommand{\R}{\mbox{$R$}}
\newcommand{\rzero}{\mbox{$r_{00}^{04}$}}
\newcommand{\rone}{\mbox{$r_{1-1}^{1}$}}
\newcommand{\costh}{\mbox{$\cos\theta$}}
\newcommand{\cosp}{\mbox{$\cos\psi$}}
\newcommand{\costop}{\mbox{$\cos(2\psi)$}}
\newcommand{\cosd}{\mbox{$\cos\delta$}}
\newcommand{\cossqp}{\mbox{$\cos^2\psi$}}
\newcommand{\cossqt}{\mbox{$\cos^2\theta^*$}}
\newcommand{\sint}{\mbox{$\sin\theta^*$}}
\newcommand{\sintot}{\mbox{$\sin(2\theta^*)$}}
\newcommand{\sinsqt}{\mbox{$\sin^2\theta^*$}}
\newcommand{\costhst}{\mbox{$\cos\theta^*$}}
\newcommand{\vep}{\mbox{$V p$}}
\newcommand{\mpipi}{\mbox{$m_{\pi^+\pi^-}$}}
\newcommand{\mkk}{\mbox{$m_{KK}$}}
\newcommand{\mkaka}{\mbox{$m_{K^+K^-}$}}
\newcommand{\mpp}{\mbox{$m_{\pi\pi}$}}       
\newcommand{\mppsq}{\mbox{$m_{\pi\pi}^2$}}   
\newcommand{\mpi}{\mbox{$m_{\pi}$}}          
\newcommand{\mrho}{\mbox{$m_{\rho}$}}        
\newcommand{\mrhosq}{\mbox{$m_{\rho}^2$}}    
\newcommand{\Gmpp}{\mbox{$\Gamma (\mpp)$}}   
\newcommand{\Gmppsq}{\mbox{$\Gamma^2(\mpp)$}}
\newcommand{\Grho}{\mbox{$\Gamma_{\rho}$}}   
\newcommand{\grho}{\mbox{$\Gamma_{\rho}$}}   
\newcommand{\Grhosq}{\mbox{$\Gamma_{\rho}^2$}}   
%
%
\newcommand{\cm}{\mbox{\rm cm}}
\newcommand{\GeV}{\mbox{\rm GeV}}
\newcommand{\gev}{\mbox{\rm GeV}}
\newcommand{\GeVx}{\rm GeV}
\newcommand{\gevx}{\rm GeV}
\newcommand{\GeVc}{\rm GeV/c}
\newcommand{\gevc}{\rm GeV/c}
\newcommand{\MeVc}{\rm MeV/c}
\newcommand{\mevc}{\rm MeV/c}
\newcommand{\MeV}{\mbox{\rm MeV}}
\newcommand{\mev}{\mbox{\rm MeV}}
\newcommand{\MeVx}{\mbox{\rm MeV}}
\newcommand{\mevx}{\mbox{\rm MeV}}
\newcommand{\GeVsq}{\mbox{${\rm GeV}^2$}}
\newcommand{\gevsq}{\mbox{${\rm GeV}^2$}}
\newcommand{\gevsqc}{\mbox{${\rm GeV^2/c^4}$}}
\newcommand{\gevcsq}{\mbox{${\rm GeV/c^2}$}}
\newcommand{\mevcsq}{\mbox{${\rm MeV/c^2}$}}
\newcommand{\GeVsqm}{\mbox{${\rm GeV}^{-2}$}}
\newcommand{\gevsqm}{\mbox{${\rm GeV}^{-2}$}}
\newcommand{\nb}{\mbox{${\rm nb}$}}
\newcommand{\nbinv}{\mbox{${\rm nb^{-1}}$}}
\newcommand{\pbinv}{\mbox{${\rm pb^{-1}}$}}
\newcommand{\mm}{\mbox{$\cdot 10^{-2}$}}
\newcommand{\mmm}{\mbox{$\cdot 10^{-3}$}}
\newcommand{\mmmm}{\mbox{$\cdot 10^{-4}$}}
\newcommand{\degr}{\mbox{$^{\circ}$}}
%
%
\newcommand{\F}{$ F_{2}(x,Q^2)\,$}  
\newcommand{\Fc}{$ F_{2}\,$}    
\newcommand{\XP}{x_{{I\!\!P}/{p}}}       
\newcommand{\TOSS}{x_{{i}/{\PO}}}        
\newcommand{\un}[1]{\mbox{\rm #1}} 
\newcommand{\LO}{Leading Order}
\newcommand{\NLO}{Next to Leading Order}
\newcommand{\ft}{$ F_{2}\,$}
%
%
\newcommand{\mc}{\multicolumn}
\newcommand{\bce}{\begin{center}}
\newcommand{\ece}{\end{center}}
\newcommand{\beq}{\begin{equation}}
\newcommand{\eeq}{\end{equation}}
\newcommand{\bea}{\begin{eqnarray}}
\newcommand{\eea}{\end{eqnarray}}
%
%
\def\lsim{\mathrel{\rlap{\lower4pt\hbox{\hskip1pt$\sim$}}
    \raise1pt\hbox{$<$}}}         
\def\gsim{\mathrel{\rlap{\lower4pt\hbox{\hskip1pt$\sim$}}
    \raise1pt\hbox{$>$}}}         
%
%
\newcommand{\pom}{{I\!\!P}}
\newcommand{\PO}{I\!\!P}
\newcommand{\slowpi}{\pi_{\mathit{slow}}}
\newcommand{\fiidiii}{F_2^{D(3)}}
\newcommand{\fiidiiiarg}{\fiidiii\,(\beta,\,Q^2,\,x)}
\newcommand{\n}{1.19\pm 0.06 (stat.) \pm0.07 (syst.)}
\newcommand{\nz}{1.30\pm 0.08 (stat.)^{+0.08}_{-0.14} (syst.)}
\newcommand{\fiidiiiful}{F_2^{D(4)}\,(\beta,\,Q^2,\,x,\,t)}
\newcommand{\fiipom}{\tilde F_2^D}
\newcommand{\ALPHA}{1.10\pm0.03 (stat.) \pm0.04 (syst.)}
\newcommand{\ALPHAZ}{1.15\pm0.04 (stat.)^{+0.04}_{-0.07} (syst.)}
\newcommand{\fiipomarg}{\fiipom\,(\beta,\,Q^2)}
\newcommand{\pomflux}{f_{\pom / p}}
\newcommand{\nxpom}{1.19\pm 0.06 (stat.) \pm0.07 (syst.)}
\newcommand {\gapprox}
   {\raisebox{-0.7ex}{$\stackrel {\textstyle>}{\sim}$}}
\newcommand {\lapprox}
   {\raisebox{-0.7ex}{$\stackrel {\textstyle<}{\sim}$}}
\newcommand{\pomfluxarg}{f_{\pom / p}\,(x_\pom)}
\newcommand{\dsf}{\mbox{$F_2^{D(3)}$}}
\newcommand{\dsfva}{\mbox{$F_2^{D(3)}(\beta,Q^2,x_{I\!\!P})$}}
\newcommand{\dsfvb}{\mbox{$F_2^{D(3)}(\beta,Q^2,x)$}}
\newcommand{\dsfpom}{$F_2^{I\!\!P}$}
\newcommand{\gap}{\stackrel{>}{\sim}}
\newcommand{\lap}{\stackrel{<}{\sim}}
\newcommand{\fem}{$F_2^{em}$}
\newcommand{\tsnmp}{$\tilde{\sigma}_{NC}(e^{\mp})$}
\newcommand{\tsnm}{$\tilde{\sigma}_{NC}(e^-)$}
\newcommand{\tsnp}{$\tilde{\sigma}_{NC}(e^+)$}
\newcommand{\st}{$\star$}
\newcommand{\sst}{$\star \star$}
\newcommand{\ssst}{$\star \star \star$}
\newcommand{\sssst}{$\star \star \star \star$}
\newcommand{\tw}{\theta_W}
\newcommand{\sw}{\sin{\theta_W}}
\newcommand{\cw}{\cos{\theta_W}}
\newcommand{\sww}{\sin^2{\theta_W}}
\newcommand{\cww}{\cos^2{\theta_W}}
\newcommand{\trm}{m_{\perp}}
\newcommand{\trp}{p_{\perp}}
\newcommand{\trmm}{m_{\perp}^2}
\newcommand{\trpp}{p_{\perp}^2}
\newcommand{\ev}{\'ev\'enement}
\newcommand{\evs}{\'ev\'enements}
\newcommand{\mdv}{mod\`ele \`a dominance m\'esovectorielle}
\newcommand{\mdmv}{mod\`ele \`a dominance m\'esovectorielle}
\newcommand{\mdm}{mod\`ele \`a dominance m\'esovectorielle}
\newcommand{\idiff}{interaction diffractive}
\newcommand{\idiffs}{interactions diffractives}
\newcommand{\pdmv}{production diffractive de m\'esons vecteurs}
\newcommand{\pdmr}{production diffractive de m\'esons \rh}
\newcommand{\pdmp}{production diffractive de m\'esons \ph}
\newcommand{\pdmo}{production diffractive de m\'esons \om}
\newcommand{\pdm}{production diffractive de m\'esons}
\newcommand{\pdiff}{production diffractive}
\newcommand{\diff}{diffractive}
\newcommand{\produ}{production}
\newcommand{\mv}{m\'eson vecteur}
\newcommand{\mvs}{m\'esons vecteurs}
\newcommand{\me}{m\'eson}
\newcommand{\mr}{m\'eson \rh}
\newcommand{\mph}{m\'eson \ph}
\newcommand{\mo}{m\'eson \om}
\newcommand{\mrs}{m\'esons \rh}
\newcommand{\mps}{m\'esons \ph}
\newcommand{\mos}{m\'esons \om}
\newcommand{\photo}{photoproduction}
\newcommand{\agq}{\`a grand \qsq}
\newcommand{\agqsq}{\`a grand \qsq}
\newcommand{\apq}{\`a petit \qsq}
\newcommand{\apqsq}{\`a petit \qsq}
\newcommand{\de}{d\'etecteur}
%
%
\newcommand{\sqrts}{$\sqrt{s}$}
\newcommand{\Oa}{$O(\alpha_s)$}
\newcommand{\Oaa}{$O(\alpha_s^2)$}
\newcommand{\PT}{p_{\perp}}
\newcommand{\sh}{\hat{s}}
\newcommand{\uh}{\hat{u}}
\newcommand{\ttbs}{\char'134}
\newcommand{\xpomlo}{3\times10^{-4}}
\newcommand{\xpomup}{0.05}
\newcommand{\llq}{$\alpha_s \ln{(\qsq / \Lambda_{QCD}^2)}$}
\newcommand{\llqx}{$\alpha_s \ln{(\qsq / \Lambda_{QCD}^2)} \ln{(1/x)}$}
\newcommand{\llx}{$\alpha_s \ln{(1/x)}$}
%
%
\newcommand{\Brodsky}{Brodsky {\it et al.}}
\newcommand{\FKS}{Frankfurt, Koepf and Strikman}
\newcommand{\Kop}{Kopeliovich {\it et al.}}
\newcommand{\Ginzburg}{Ginzburg {\it et al.}}
\newcommand{\Ryskin}{\mbox{Ryskin}}
\newcommand{\Kaidalov}{Kaidalov {\it et al.}}
%
%
\def\ar#1#2#3   {{\em Ann. Rev. Nucl. Part. Sci.} {\bf#1} (#2) #3}
\def\epj#1#2#3  {{\em Eur. Phys. J.} {\bf#1} (#2) #3}
\def\err#1#2#3  {{\it Erratum} {\bf#1} (#2) #3}
\def\ib#1#2#3   {{\it ibid.} {\bf#1} (#2) #3}
\def\ijmp#1#2#3 {{\em Int. J. Mod. Phys.} {\bf#1} (#2) #3}
\def\jetp#1#2#3 {{\em JETP Lett.} {\bf#1} (#2) #3}
\def\mpl#1#2#3  {{\em Mod. Phys. Lett.} {\bf#1} (#2) #3}
\def\nim#1#2#3  {{\em Nucl. Instr. Meth.} {\bf#1} (#2) #3}
\def\nc#1#2#3   {{\em Nuovo Cim.} {\bf#1} (#2) #3}
\def\np#1#2#3   {{\em Nucl. Phys.} {\bf#1} (#2) #3}
\def\pl#1#2#3   {{\em Phys. Lett.} {\bf#1} (#2) #3}
\def\prep#1#2#3 {{\em Phys. Rep.} {\bf#1} (#2) #3}
\def\prev#1#2#3 {{\em Phys. Rev.} {\bf#1} (#2) #3}
\def\prl#1#2#3  {{\em Phys. Rev. Lett.} {\bf#1} (#2) #3}
\def\ptp#1#2#3  {{\em Prog. Th. Phys.} {\bf#1} (#2) #3}
\def\rmp#1#2#3  {{\em Rev. Mod. Phys.} {\bf#1} (#2) #3}
\def\rpp#1#2#3  {{\em Rep. Prog. Phys.} {\bf#1} (#2) #3}
\def\sjnp#1#2#3 {{\em Sov. J. Nucl. Phys.} {\bf#1} (#2) #3}
\def\spj#1#2#3  {{\em Sov. Phys. JEPT} {\bf#1} (#2) #3}
\def\zp#1#2#3   {{\em Zeit. Phys.} {\bf#1} (#2) #3}
%
%
\newcommand{\clearemptydoublepage}{\newpage{\pagestyle{empty}\cleardoublepage}}
\newcommand{\scaption}[1]{\caption{\protect{\footnotesize  #1}}}
\newcommand{\proc}[2]{\mbox{$ #1 \rightarrow #2 $}}
\newcommand{\average}[1]{\mbox{$ \langle #1 \rangle $}}
\newcommand{\av}[1]{\mbox{$ \langle #1 \rangle $}}

\begin{titlepage}

\noindent
DESY 99-010  \hspace*{8.5cm} ISSN 0418-9833 \\
February 1999

\vspace{3.5cm}
 
\begin{center}
\begin{Large}
 
\boldmath
{\bf  Elastic Electroproduction of $\rho$ Mesons} \\
{\bf  at HERA}
\unboldmath
 
\vspace{2cm}
 
H1 Collaboration
 
\end{Large}
\end{center}
 
\vspace{2cm}
 
\begin{abstract}
\noindent
The elastic electroproduction of $\rho$ mesons is studied 
at HERA with the H1 detector for a photon virtuality in the range 
$1 < Q^2 < 60~{\rm GeV^2}$ and for a hadronic centre of mass energy in the 
range $30 < W < 140$~{\rm GeV}.
The shape of the ($\pi \pi$) mass distribution in the $\rho$ resonance region 
is measured as a function of $Q^2$. 
The full set of \rh\ spin density matrix elements is determined,
and evidence is found for a helicity flip amplitude at the level of
$8 \pm 3 \%$ of the non-flip amplitudes.
Measurements are presented of the dependence of the cross section on $Q^2$, 
$W$ and $t$ (the four-momentum transfer squared to the proton). 
They suggest that, especially at large \qsq, the $\gamma^*p$ 
cross section develops a stronger $W$ dependence than that expected from the
behaviour of elastic and total hadron$-$hadron cross sections.

\end{abstract}
 
\vspace{1.5cm}
 
\begin{center}
To be submitted to {\it Eur. Phys. J. C}. \\
\end{center}
 
\end{titlepage}

\noindent
\noindent
 C.~Adloff$^{34}$,                
 V.~Andreev$^{25}$,               
 B.~Andrieu$^{28}$,               
 V.~Arkadov$^{35}$,               
 A.~Astvatsatourov$^{35}$,        
 I.~Ayyaz$^{29}$,                 
 A.~Babaev$^{24}$,                
 J.~B\"ahr$^{35}$,                
 P.~Baranov$^{25}$,               
 E.~Barrelet$^{29}$
% ...

\noindent
% ...
$ ^{24}$ Institute for Theoretical and Experimental Physics,
          Moscow, Russia \\
 $ ^{25}$ Lebedev Physical Institute, Moscow, Russia$^{f,k}$ \\
% ...
$ ^{28}$ LPNHE, \'{E}cole Polytechnique, IN2P3-CNRS, Palaiseau, France \\
 $ ^{29}$ LPNHE, Universit\'{e}s Paris VI and VII, IN2P3-CNRS,
          Paris, France \\
% ...
$ ^{34}$ Fachbereich Physik, Bergische Universit\"at Gesamthochschule
          Wuppertal, Wuppertal, Germany$^a$ \\
 $ ^{35}$ DESY, Institut f\"ur Hochenergiephysik, Zeuthen, Germany$^a$ \\
% ...
$ ^a$ Supported by the Bundesministerium f\"ur Bildung, Wissenschaft,
        Forschung und Technologie, FRG,
        under contract numbers 7AC17P, 7AC47P, 7DO55P, 7HH17I, 7HH27P,
        7HD17P, 7HD27P, 7KI17I, 6MP17I and 7WT87P \\
% ...
$ ^f$ Supported by the Deutsche Forschungsgemeinschaft \\
% ...
$ ^k$ Supported by Russian Foundation for Basic Research 
       grant no. 96-02-00019 

 
\newpage

\section{Introduction}  \label{sect:introduction}

Measurements of the elastic electroproduction of vector mesons at HERA over a wide range of exchanged photon virtuality $Q^2$  
are of particular interest. For many years it has been known that at low $Q^2$, that is with no hard scale, vector meson electroproduction exhibits all the properties of a soft
diffractive process. Predictions of soft 
processes based on QCD calculations are however intractable. The presence of a hard scale, that is a significant $Q^2$, makes perturbative QCD calculations
possible. Measurements of the $Q^2$ dependences of
observables in vector meson electroproduction thereby provide insight into the transition and the
interplay between soft and hard processes in QCD.

This paper presents an analysis of elastic \rh\ meson electroproduction:
\begin{equation}
  e + p \rightarrow e + \rho + p \ , \ \ \ \
  \rho \rightarrow \pi^+ + \pi^- \ ,                        \label{eq:rh}
\end{equation}
in the \qsq\ range from 1 to 60~\gevsq\ 
($Q^2 = -q^2$, where $q$ is the four-momentum of the intermediate photon)
and the \w\ range from 
30 to 140~\gev\ (\w\ is the hadronic centre of mass energy).

The data were obtained with the H1 detector in two running periods of the HERA
collider, operated with 820~\gev\ protons and 27.5~\gev\ positrons.\footnote{
  In the rest of this paper, the word ``electron'' is generically used for 
  electrons and positrons.}
A low \qsq\ data set ($1 < \qsq < 5$~\gevsq) 
was obtained from a special run in 1995, with the $ep$ interaction 
vertex shifted by 70 cm in the outgoing proton beam direction; it
corresponds to an integrated luminosity of 125 \nbinv .
A larger sample with $2.5  < \qsq < 60$~\gevsq\ was obtained in 1996 
under normal running conditions; it corresponds to a luminosity 
of 3.87 \pbinv.

The present measurements provide detailed information in the region 
$1 \lsim  \qsq \lsim 8$~\gevsq\ and they increase the precision of the H1 
measurement of \rh\ electroproduction with $\qsq > 8$~\gevsq , which was 
first performed using data collected in 1994~\cite{H1_rho_94}.
They are compared to results of the ZEUS experiment~\cite{rho_zeus} at HERA
and of fixed target experiments~\cite{CHIO,NMC,E665}.

The H1 detector, the definition of the kinematic variables and the event 
selection  are introduced in section~\ref{sect:selection}.
Acceptances, efficiencies and background contributions are discussed
in section~\ref{sect:acceptances}.
The shape of the ($\pi \pi$) mass distribution and the evolution with \qsq\
of the skewing of this distribution are studied in section~\ref{sect:mass}.
Section~\ref{sect:helicity} is devoted to the study of the \rh\ meson
decay angular distributions and to the measurement of the 15 elements
of the spin density matrix, as a function of several kinematic variables.  
The \qsq\ dependence of the ratio $R$ of the longitudinal to transverse
$\gamma^*p$ cross sections is measured. 
The violation of $s$-channel helicity conservation, found to be small but
significant at lower energies~\cite{CHIO,joos}, is quantified.
Finally, section~\ref{sect:cross_section} presents the $t$ distribution and
the measurement of the $\gamma^*p \rightarrow \rh p$ cross section as a 
function of \qsq\ and \W. 
Predictions of several models are
compared to the measurements in sections~\ref{sect:helicity} and 
\ref{sect:cross_section}.

\section{H1 Detector, Kinematics and Event Selection}  \label{sect:selection}

Events corresponding to reaction (\ref{eq:rh}) are selected by requiring the
detection 
of the scattered electron and of a pair of 
oppositely charged particles originating from a common vertex.
The absence of additional activity in the detector is required, since the 
scattered proton generally escapes undetected into the beam pipe.

H1 uses a right-handed coordinate system with the $z$ axis taken along the 
beam direction, the $+z$
or ``forward" direction being that of the outgoing proton beam. The
$x$ axis points towards the centre of the HERA ring.

\subsection{The H1 Detector}  \label{sect:detector}

A detailed description of the H1 detector can be found in~\cite{H1_NIM}. 
Here only the detector components relevant for the present analysis are 
described.

The scattered electron is detected in the SPACAL~\cite{spacal},
a lead -- scintillating fibre calorimeter situated in the backward
region of the H1 detector, 152 cm from the nominal interaction point.
The calorimeter is divided into an electromagnetic and a hadronic part.
The electromagnetic section of the SPACAL, which covers the angular range
$ 153 ^{\rm \circ} < \theta < 177.5 ^{\rm \circ}$
(defined with respect to the nominal interaction point),
is segmented into cells of $4 \times 4$ {$\rm cm^2$} transverse size.\footnote{
  In this paper, ``transverse'' directions are relative to the beam direction.}
The hadronic section is used here to prevent hadrons from being misidentified as
the scattered electron.
In front of the SPACAL, a set of drift chambers, the BDC, allows the
reconstruction of electron track segments, providing a resolution in the
transverse direction of 0.5~mm.

The pion candidates are detected and their momentum is measured in the
central tracking detector.
The major components of this detector are two 2 m long coaxial cylindrical
drift chambers, the CJC chambers, with wires parallel to the beam
direction.
The inner and outer radii of the chambers are 203 and 451~mm, and 530 and 844~mm,
respectively.
In the forward region, the CJC chambers are supplemented by a set of drift chambers with
wires perpendicular to the beam direction.
The measurement of charged particle transverse momenta is performed
in a magnetic field of 1.15~T, uniform over the full tracker volume,
generated by a superconducting solenoidal magnet.
For charged particles emitted from the nominal vertex with polar angles
$ 20^{\rm \circ} < \theta < 160^{\rm \circ}$, the resolution on the
transverse momentum 
is $\Delta p_{t} / p_{t} \simeq\ 0.006 \ p_{t}$ (\gev).
Drift chambers with wires perpendicular to the beam direction, situated inside
the inner CJC and between the two CJC chambers, provide a measurement of
$z$ coordinates with a precision of 350~${\rm \mu m}$.

The ($x, y, z$) position of the interaction vertex is reconstructed for
each event by a global fit of all measured charged particle trajectories.
For each electron fill in the accelerator, a fit is performed of the
dependence on $z$ of the mean $x$ and $y$ positions of the vertices.
This provides a measurement of the corresponding beam direction, which varies
slightly from fill to fill.

The absence of activity in the H1 detector not associated with the scattered 
electron or the \rh\ decay is checked using several components of the detector.
The liquid argon (LAr) calorimeter, surrounding the tracking detector and
situated inside the solenoidal magnet, covers
the polar angular range $4^{\rm \circ} \leq \theta \leq  154^{\rm \circ}$
with full azimuthal acceptance.
The muon spectrometer (FMD), designed to identify and measure the momentum
of muons emitted in the forward direction, contains six active layers, each made 
of a pair of planes of drift cells, covering the polar angular region
$3^{\rm \circ} \leq \theta \leq 17^{\rm \circ}$.
The three layers situated between the main calorimeter and the toroidal
magnet of the FMD can be reached by secondary particles arising from the
interaction of small angle primary particles hitting the beam collimators or 
the beam pipe walls.
Secondary particles or the scattered proton at high \modt\
can reach a set of scintillators, the proton
remnant tagger (PRT), placed 24~m downstream of the interaction point and
covering the angles $0.06^{\rm \circ} \leq \theta \leq 0.17^{\rm \circ}$.

\subsection{Kinematic Variables}  \label{sect:kin_var}

The reconstruction method for the kinematic variables has been optimised for
the \rh\ measurement.

The \qsq\ variable is computed from $E_{\rm o}$, the incident electron beam
energy, and the polar angles $\theta_e$ and
$\theta_{\rho}$ of the electron and of the \rh\ meson candidates
\cite{Koijman_Workshop}:
\begin{equation}
  \qsq =  \frac{ 4 E_{\rm o}^2} {\tan (\theta_e / 2) \ 
                 (\tan (\theta_e / 2) + \tan (\theta_{\rho} / 2) \ ) } \ .
                                                             \label{eq:qsq}
\end{equation}
The electron emission angles are determined using the reconstructed vertex
position and the track segment in the BDC corresponding to the electron cluster
candidate. 
The momentum of the \rh\ meson is reconstructed as the sum of the momenta of 
the two pion candidates:
\begin{equation}
  \vec p_{\rho} = \vec {p}_{\pi^+} + \vec {p}_{\pi^-} \ .  \label{eq:prho}
\end{equation}

The inelasticity $y$ is defined as
\begin{equation}
  y = \frac {p \cdot q} {p \cdot k} \  ,  \label{eq:y_def}
\end{equation}
where $p$ and $k$ are the four-momenta of the incident
proton and of the incident electron, respectively.
For this analysis, $y$ is computed, with very good precision, 
using the energy, $E_{\rho}$, and the longitudinal momentum, $p_{z_{\rho}}$, of
the \rh\ meson candidate~\cite{Jacquet_Blondel}:
\begin{equation}
  y = \frac {E_{\rho} - p_{z_{\rho}}} {2 \ E_{\rm o}} \  .  \label{eq:y}
\end{equation}

The hadronic mass, \w, is computed using the relation
\begin{equation}
  W^2 =  y s  - \qsq  \ ,  \label{eq:w_def}
\end{equation}
where $s$ is the square of the $ep$ centre of mass energy.

The variable $t$ is the square of the four-momentum transfer to the proton.
At HERA energies, to very good precision, its absolute value is equal to the 
square of the transverse momentum of the outgoing proton.
The latter is computed,
under the assumption that the selected event corresponds
to reaction (\ref{eq:rh}), as the sum of
the transverse momenta $\vec p_{t_{\rho}}$ of the \rh\ meson 
candidate and $\vec p_{t_{e}}$ of the scattered
electron:
\begin{equation}
  t \simeq - |\vec p_{t_{\rho}} + \vec p_{t_{e}}|^2 \ .      \label{eq:t}
\end{equation}
The value of $t$ is thus distorted if the event is due to the production
of a hadron system of which the \rh\ is only part and of which the remaining 
particles were not detected.
For use in eq.~(\ref{eq:t}), $\vec{p}_{t_{e}}$ is determined from the \rh\ 
candidate measurement and the electron beam energy, such that 
%
\begin{equation}
  p_{t_{e}} = \frac {2  E_{\rm o} - E_{\rho} + p_{z_{\rho}}}
               {\tan (\theta_e / 2)} \ .
                                                      \label{eq:qsq1}
\end{equation}
This relation assumes reaction (\ref{eq:rh}) and the absence of QED radiation.

Finally, the total event \eminpz\ variable is computed as the sum of the differences
between the energies and the longitudinal momenta of the electron and pion
candidates, where the electron energy measured in the SPACAL calorimeter is
used.

\subsection{Trigger and Event Selection}  \label{sect:trigger}

The trigger and selection criteria for the events used in this analysis are
summarised in Table~\ref{Table:selection}.
Events are selected only from runs 
for which all relevant parts of the detector were functioning efficiently.

For the 1995 shifted vertex run, the trigger was based on the detection 
of a cluster in the electromagnetic section 
of the SPACAL calorimeter with energy greater than 12~\gev.
For the 1996 data, the energy threshold was increased to 15~\gev\ and,
in order to reduce the rate of
background events due to synchrotron radiation from the electron beam,
the centre of gravity of the cluster 
was required to lie outside the innermost part of the SPACAL, 
with $-16 < x < 8$~cm and $-8 < y < 16$~cm. 
Independent triggers were used to determine the efficiency of this trigger.

Off-line, electron candidates are defined as well identified electromagnetic
clusters in the SPACAL with energy larger than 17~GeV, correlated with a
track segment in the BDC.
The transverse position of the BDC track segment has to be more than 8.7 cm
from the beams for the 1995 data sample, and must correspond to the region of
the SPACAL included in the trigger for the 1996 data.
\begin{table}[h]
\begin{center}
\begin{tabular}{|l|ll|}
\hline
\hline
  Trigger   
            & \multicolumn{2}{|l|}{cluster in SPACAL with energy $>$ 12 (15) \gev\
            in 1995 (1996)} \\
            & \multicolumn{2}{|l|}{and with $-16 < x < 8$ cm and $-8 < y < 16$ cm (1996)}  \\
\hline
\hline
  Electron  
            & \multicolumn{2}{|l|}{cluster in electromagnetic SPACAL with energy $> 17$ \gev} \\
            & \multicolumn{2}{|l|}{distance between cluster c.o.g. and BDC track $<$ 3 cm}\\
            & \multicolumn{2}{|l|}{BDC segment $>$ 8.7 cm from the beams (1995)} \\
            & \multicolumn{2}{|l|}{transverse width of cluster $<$ 3.2 cm} \\
            & \multicolumn{2}{|l|}{energy in hadronic SPACAL $<$ 0.2 \gev} \\
\hline
  Pion candidates
            & \multicolumn{2}{|l|}{exactly two tracks with opposite signs} \\
            & \multicolumn{2}{|l|}{$20^{\rm \circ} < \theta < 160^{\rm \circ}$ (1996)} \\
            & \multicolumn{2}{|l|}{particle transverse momenta $p_t > 0.1$ \gev} \\
            & \multicolumn{2}{|l|}{vertex reconstructed within 30 cm of nominal position in $z$} \\
\hline
  Additional activity
            & \multicolumn{2}{|l|}{no cluster in LAr with energy $> 0.5$ \gev }\\
            & \multicolumn{2}{|l|}{at most 1 hit pair in FMD} \\
            & \multicolumn{2}{|l|}{no hit in PRT} \\
\hline
  Mass selection
            & \multicolumn{2}{|l|}{$0.6 < \mpp < 1.1$ \gev} \\
            & \multicolumn{2}{|l|}{$\mkk > 1.040$ \gev} \\
\hline
\hline
  Kinematic domain & &  \\
  1995 data 
            & $1.0 < \qsq < 5$ \gevsq ,   &  $40 < \w < 140$ \gev \\
  1996 data
            & $2.5 < \qsq < 4$ \gevsq ,   &  $30 < \w < 100$ \gev \\
            & $4.0 < \qsq < 6$ \gevsq ,   &  $30 < \w < 120$ \gev \\
            & $6.0 < \qsq < 14$ \gevsq ,  &  $40 < \w < 140$ \gev \\
            & $14.0 < \qsq < 60$ \gevsq , &  $50 < \w < 140$ \gev \\
\hline
  Other cuts
            & \multicolumn{2}{|l|}{$\modt < 0.5$ \gevsq} \\
            & \multicolumn{2}{|l|}{$\eminpz > 45$ \gev} \\
\hline
\hline
\end{tabular}
\caption{Summary of trigger conditions and event selection criteria (see text 
for details).}
\label{Table:selection}
\end{center}
\end{table}
 
Exactly two oppositely charged pion candidates are required, with polar 
angles of emission $20^{\rm \circ} < \theta < 160^{\rm \circ }$ (1996 data~\footnote{
For the 1995 data, no cut on the track polar angle is made.}),
and transverse momenta with respect to the beam direction $p_t > 0.1$~\gev , 
so that detection
and reconstruction in the central tracker are efficient. 
The reconstructed interaction vertex has to lie within 30 cm in $z$ of the 
nominal interaction point.

Rejection of \rh\ meson events with proton dissociation and of other
backgrounds is achieved using three selection criteria: 
there must be no cluster in the LAr calorimeter with energy greater 
than 0.5~GeV that is not associated with the pion candidates, there must be no 
more than one hit pair recorded in the FMD and there must be no signal in 
the PRT.
Given the limiting angle of $20^{\rm \circ}$ for pion candidates, this
corresponds to requiring
no activity for a range in pseudorapidity $1.75 < \eta \lsim 7.5$.\footnote{
 The pseudorapidity $\eta$ of an object detected with polar angle $\theta$
 is defined as $\eta = - \ln \ \tan (\theta / 2)$.}

The cuts $1 < \qsq < 60$~\gevsq\ and $30 < \w < 140$~\gev, which define the
kinematic domain under study, correspond to the region in which the electron 
and hadronic track acceptances are high.
A cut $|t| < 0.5$~\gevsq\ is also applied, the purpose of which is threefold.
Firstly, the acceptance for elastic events decreases at larger \modt~values,
because the probability becomes significant that the proton hits the beam pipe 
walls, thus producing a signal in the PRT.
Secondly, the \modt\ cut suppresses events from processes which are not elastic 
and have a flatter $t$ distribution, in particular \rh\ production with proton 
dissociation.
Thirdly, it suppresses the production of hadron systems of 
which the \rh\ is only part and in which the remaining particles were not 
detected, thereby distorting the measurement of $t$ (see eq.~\ref{eq:t}).
A further cut, $\eminpz > 45$~\gev, 
is designed to minimise the effects of initial state photon radiation from the 
electron.

The selected domain for \mpp, the invariant mass of the two pion candidates,
is restricted to $0.6 < \mpp < 1.1$~\gev, which covers the \rh\ meson mass peak
and avoids regions with large background contributions.
In order to minimise \ph\ meson contamination, the invariant mass of the
pion candidates is also computed  with the assumption that they are kaons, and
the cut $\mkk > 1.040$~\gev\ is applied on the corresponding \mkk\ mass.

After all selection cuts, the 1995 sample ($ 1 < Q^2 < 5$~\gevsq)
contains about 500 events, and the 1996 sample ($ 2.5 < Q^2 < 60$~\gevsq)
1800 events.

\section{Detector Effects and Background Contributions}  
                                                  \label{sect:acceptances}

\subsection{Acceptances and Efficiencies}
            \label{sect:DIFFVM}

\begin{figure}[p]
\begin{center}
\setlength{\unitlength}{1.0cm}
\begin{picture}(14.0,14.0)
\put(0.0,0.0){\epsfig{file=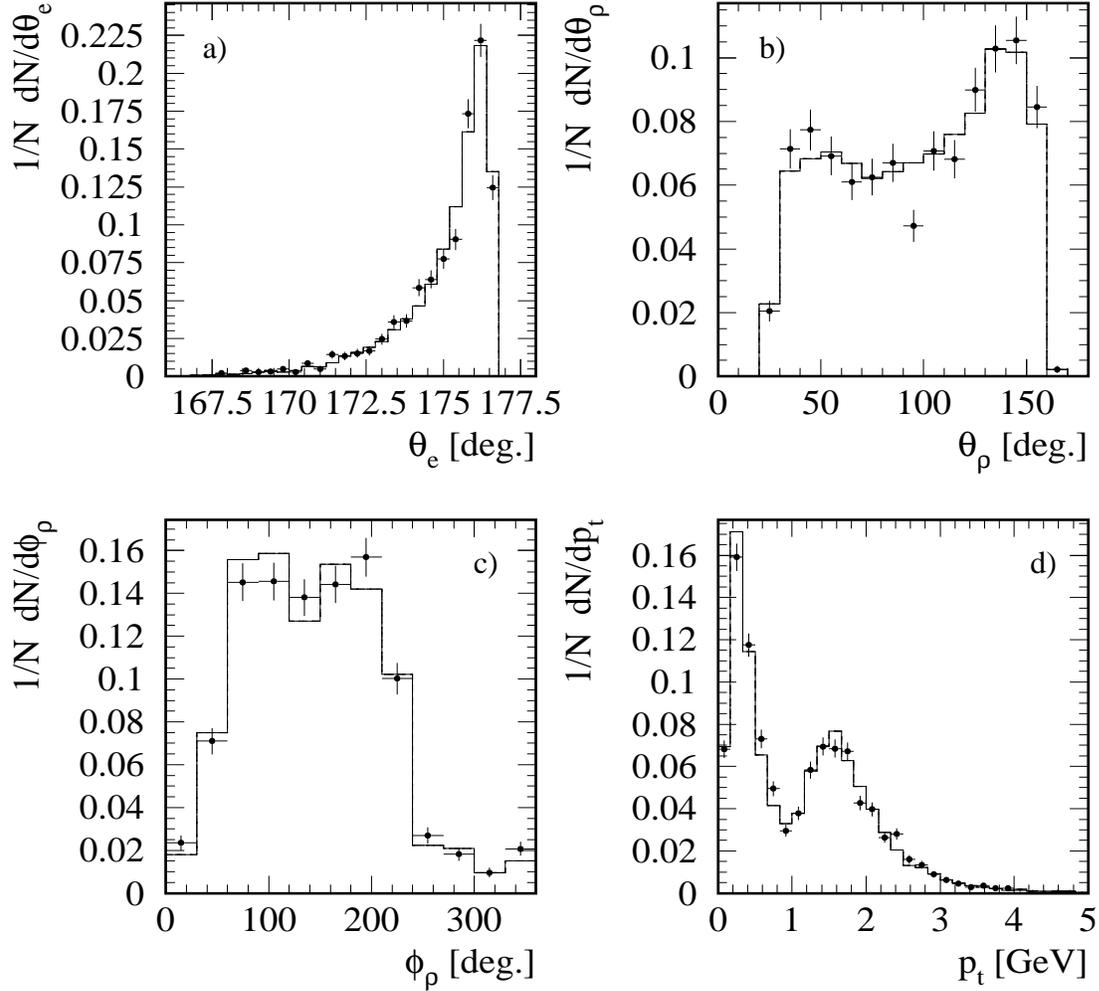,%
    bbllx=46pt,bblly=150pt,bburx=530pt,bbury=661pt,height=14.0cm,width=14.cm}}
\put(2.0,12.85){a)}
\put(9.4,12.85){b)}
\put(5.6,6.0){c)}
\put(13.0,6.0){d)}
\end{picture}
\caption{Uncorrected distributions of the polar angle of the scattered
electron, the polar angle of the \rh\ meson, the azimuthal angle of the \rh\ 
meson in the laboratory frame, and the transverse momenta of the two pion 
candidates,
for the 1996 data sample (points) and for the Monte Carlo simulation
(histograms), after all selection cuts.}
\label{fig:MC_comparison}
\end{center}
\end{figure}

Acceptances, efficiencies and detector resolution effects are determined using 
the DIFFVM Monte Carlo simulation~\cite{DIFFVM}, a program based on Regge
theory and the vector meson dominance model (VDM).
The simulation parameters are adjusted following the measurements presented 
below for the dependence of the cross section on \qsq, \w, $t$ and
for the \rh\ meson angular decay distributions.
The detector geometry and its response to generated particles are simulated in 
detail.
The same reconstruction procedures and event selection criteria as 
for real events are applied.
As an illustration of the good quality of the simulation,
Fig.~\ref{fig:MC_comparison} presents a comparison of the distributions of
several variables for the data and for the Monte Carlo simulation.
The distribution of the azimuthal angle of the \rh\ meson 
(Fig.~\ref{fig:MC_comparison}c) reflects the regions 
of the SPACAL that are active in the trigger.
The distribution of the transverse momenta of the pion candidates 
(Fig.~\ref{fig:MC_comparison}d) depends on the details of the \rh\ meson decay 
angular distribution.
It has been carefully checked that the Monte Carlo simulation reproduces well
the details of the tracker acceptance and efficiency, both for positively and 
for negatively charged pions.

In the kinematic domain defined in Table~\ref{Table:selection}, the acceptance 
depends most strongly on \qsq\ in a purely geometrical manner related to the
trigger conditions.
The cuts on the polar angles and on the minimum transverse momenta of the pion 
candidates induce \w -dependent acceptance corrections, which are sensitive 
to the angular decay distributions.
The \qsq\ and \W\ limits of the selected kinematic domain are such that 
the efficiency is almost constant over each bin.
The cut on \eminpz\ induces very small corrections.

For each of the measurements presented below, systematic errors are computed
by varying the reconstructed polar angle of the electron by $\pm 0.5$ mrad, 
which corresponds to the systematic uncertainty on this measurement,
and by varying 
in the Monte Carlo simulation the cross section dependence on \qsq, \w, $t$
and the \rh\ meson decay distributions
by the amount allowed by the present measurements 
(see~\cite{Barbara} for more details).
Small remaining uncertainties related to the simulation of the tracker
uniformity are neglected.
Further systematic uncertainties that affect only certain measurements are 
described where appropriate below.
The positive and negative variations are combined separately in the
form of quadratic sums, to compute the systematic errors.

In addition to the effects studied with the DIFFVM simulation, the trigger
efficiency is studied using several independent triggers. 
Regions of the SPACAL for which the trigger efficiency is below 94\% are
discarded from the measurement. 
Losses of elastic events due to noise in the LAr, FMD and PRT detectors are
computed 
from 
randomly triggered events in
the detector.
Radiative corrections are determined using the HERACLES program~\cite {HERACLES}.

\subsection{Background Contributions}  
            \label{sect:backgrounds}

The main background contributions to \rh\ meson elastic production are due to 
the elastic production of \om\ and \ph\ mesons and to diffractive \rh\ 
production with proton dissociation.

\boldmath
\subsubsection{Elastic Production of \om\ and \ph\ Mesons}  
            \label{sect:elastic_bg}
\unboldmath

The elastic production of \om\ mesons:
\begin{equation}
  e + p \rightarrow e + \omega + p                  \label{eq:omega}
\end{equation}
may produce background in the present data sample through the two \om\ 
decay modes~\cite{PDG} :
\begin{eqnarray}
  \omega \rightarrow \pi^+ + \pi^- + \pi^o \ \ \ \ \ \ \ \ \
                                  {\rm (BR = 88.8 \%)}         \\
  \omega \rightarrow \pi^+ + \pi^- \ \ \ \ \ \ \ \ \
                                  {\rm (BR = 2.2 \%)} \ .
                                              \label{eq:om_DK}
\end{eqnarray}
%
The contribution of the first decay mode is efficiently reduced by the mass 
selection cut, by requiring the absence in the LAr calorimeter of clusters 
with energy larger than 0.5~\gev\ which are not associated with a track, and by 
the cut on the variable $t$.
However, events from the second decay mode are selected within the present
sample. 
This background is subtracted statistically assuming the \om~:~\rh\ ratio of 
1~:~9 which is motivated by SU(3) flavour symmetry and is consistent with HERA 
photoproduction measurements~\cite{ZEUS_omega}.

The production rate of \ph\ mesons:
\begin{equation}
  e + p \rightarrow e + \phi + p                  \label{eq:phi}
\end{equation}
%
amounts to about 15\% of the \rh\ production rate for the present 
kinematic domain~\cite{ZEUS_phi,H1_phi,H1_rho_95}.
The following decay modes~\cite{PDG} may lead to the presence of background 
events in the selected sample:
\begin{eqnarray}
  \phi \rightarrow K^+ + K^-  \ \ \ \ \ \ \ \ \
                                  {\rm (BR = 49.1 \%)}         \\
  \phi \rightarrow \rh + \pi  \ \ \ \ \ \ \ \ \
                                  {\rm (BR = 12.9 \%)}         \\
  \phi \rightarrow \pi^+ + \pi^- + \pi^o \ \ \ \ \ \ \ \ \
                                  {\rm (BR = 2.7 \%)} \ .
                                              \label{eq:ph_DK}
\end{eqnarray}
The first contribution is mostly eliminated by the \mkk\ and the \mpp\
mass selection cuts, and the other two are significantly reduced 
by the cuts against additional particles
and by the $t$ and mass selection cuts.

Using the DIFFVM Monte Carlo simulation, the contribution of \om\ 
and \ph\ elastic production remaining in the selected sample is determined 
to be $3.3 \pm 2.0 \%$ in the invariant mass range 0.6 $<$ \mpp\ $<$ 1.1~GeV,
where 1.4\% and 1.9\% come from the \om\ and \ph\ contributions, respectively.
For the study of the shape of the mass distribution, the \mpp\ range 
used is 0.5 $<$ \mpp\ $<$ 1.1~GeV, where the contributions of \om\
and \ph\ elastic production are determined to be 4.7\% and 2.3\%,
respectively, and are subtracted statistically bin-by-bin from the mass 
distributions (see section~\ref{sect:mass}). 

\boldmath
\subsubsection{Diffractive Production of \rh\ Mesons with Proton Dissociation}  
            \label{sect:pdiss_bg}
\unboldmath

An important background to elastic \rh\ production is due to the diffractive
production of \rh\ mesons with proton dissociation
\begin{equation}
  e + p \rightarrow e + \rh + Y                   \label{eq:pdiss}
\end{equation}
when the baryonic system $Y$ is of relatively low mass $M_Y \lsim 1.6$~\gev\ 
and its decay products are thus not detected in the PRT, the FMD or the 
forward regions of the LAr calorimeter and the tracking detector.

The contamination from proton dissociation is determined using
the DIFFVM Monte Carlo.
The distribution of $M_Y$ is generated as
(see \cite{Goulianos}):
\begin{equation}
  \frac {{\rm d} \sigma } {{\rm d} M_Y^2} \propto
        \frac {1} {M_Y^{2} } \ .
                                                  \label{eq:M_Y}
\end{equation}
For $M_Y < 1.9$~\gev, the details of baryonic resonance production and decays are 
simulated following the Particle Data Group (PDG) tables~\cite{PDG}.
For larger masses, the system $Y$ is modelled as formed of a quark and a diquark,
which fragment according to the JETSET algorithm~\cite{jetset}.
The $t$ distribution of proton dissociation events is modelled by an
exponentially falling distribution with a slope parameter $b = 2.5$~\gevsqm\
(cf. the measurements in~\cite{H1_phi} and~\cite{pdiss_paper}).
The DIFFVM Monte Carlo is also used to compute the probability that the scattered 
proton in an  
elastic \rh\ event with $\modt < 0.5$~\gevsq\ gives a signal in the PRT.

The proton dissociation background in the selected sample of events is 
determined without making any hypothesis
for the relative production rates for elastic and inelastic events.
It is deduced using the total number of \rh\ events and the number of \rh\
events with no signal in the PRT or the FMD, given the probabilities  
of obtaining no signal in
these detectors for elastic interactions and for interactions with proton
dissociation.
These probabilities are determined using the Monte Carlo simulation.
The proton dissociation background in the present sample amounts to
$11 \pm 5 \%$.
The uncertainty on this number is estimated by varying by $\pm 0.3$ the
exponent of $M_Y$ in eq.~(\ref{eq:M_Y}), by varying the slope parameters of the
exponential \modt\ distributions of elastic and proton dissociation events within
the experimental limits (see section~\ref{sect:t}) and by computing
the correction using only the PRT or only the FMD~\cite{Barbara}.

\subsubsection{Other Background Contributions}  
            \label{sect:other_bg}

Other background contributions are negligibly small.
The background due to the $\pi^+ \pi^- \pi^{\rm \circ} \pi^{\rm \circ}$ decay 
mode of the $\rho^\prime (1450)$ meson is determined to be only $1 \pm 1 \% $,
due to the cuts against additional particles and the cut on the variable $t$.
The study of the mass distributions presented in section~\ref{sect:mass} also
indicates that events with photon dissociation into vector mesons other than
\rh, \om\ and \ph\ do not contribute more than 1\%.\footnote{
 In the analysis of the 1994 data~\cite{H1_rho_94}, events were accepted with
 a maximum energy of 1~GeV for clusters in the LAr calorimeter which are not 
 associated with tracks. 
 A contribution of $ 11 \pm 6 \% $ non-resonant background, concentrated mainly 
 at small \mpp\ masses, was thus subtracted from the cross section measurement. 
 For the present analysis, the limit on the cluster energy is 0.5~GeV, 
 leading to a small background contribution, but the losses of events due to 
 noise in the LAr calorimeter amount to $\simeq 10 \%$, as estimated using 
 random trigger data (section~\ref{sect:DIFFVM}).}
The background from photoproduction events with a hadron being misidentified
as the electron candidate in the SPACAL is extremely small, because of the high 
$E_e$ cut.
 

\section{Mass Distributions}  \label{sect:mass}

\begin{figure}[p]
  \begin{center}
        \epsfig{file=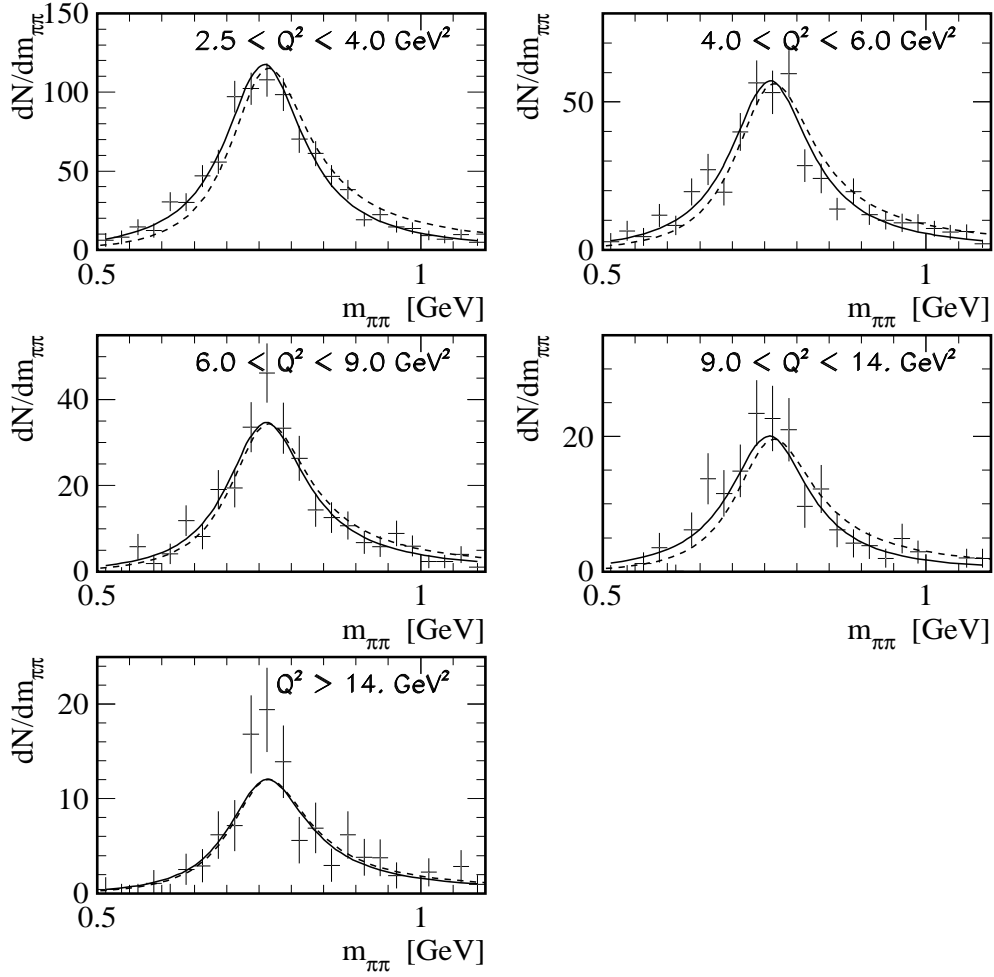,%
            bbllx=60pt,bblly=170pt,bburx=524pt,bbury=650pt,%
            height=13.0cm,width=13.cm}
  \end{center}
\vspace*{-0.5cm}
\caption{Acceptance corrected \mpp\ mass distributions for the 1996 data
sample, after statistical subtraction of the remaining \om\ and \ph\
background contributions, divided into five bins in \qsq.
The superimposed curves are the result of fits to skewed relativistic
Breit-Wigner distributions using the Ross-Stodolsky parameterisation of
eq.~(\ref{eq:ross}), with the \rh\ mass and width fixed at the
PDG values and assuming no non-resonant background.
The solid curves are the results of the fits, the dashed curves correspond to
the non-skewed Breit-Wigner contributions.
The errors on the data are statistical only.}
\label{fig:massRS}
\end{figure}

\begin{figure}[p]
  \begin{center}
        \epsfig{file=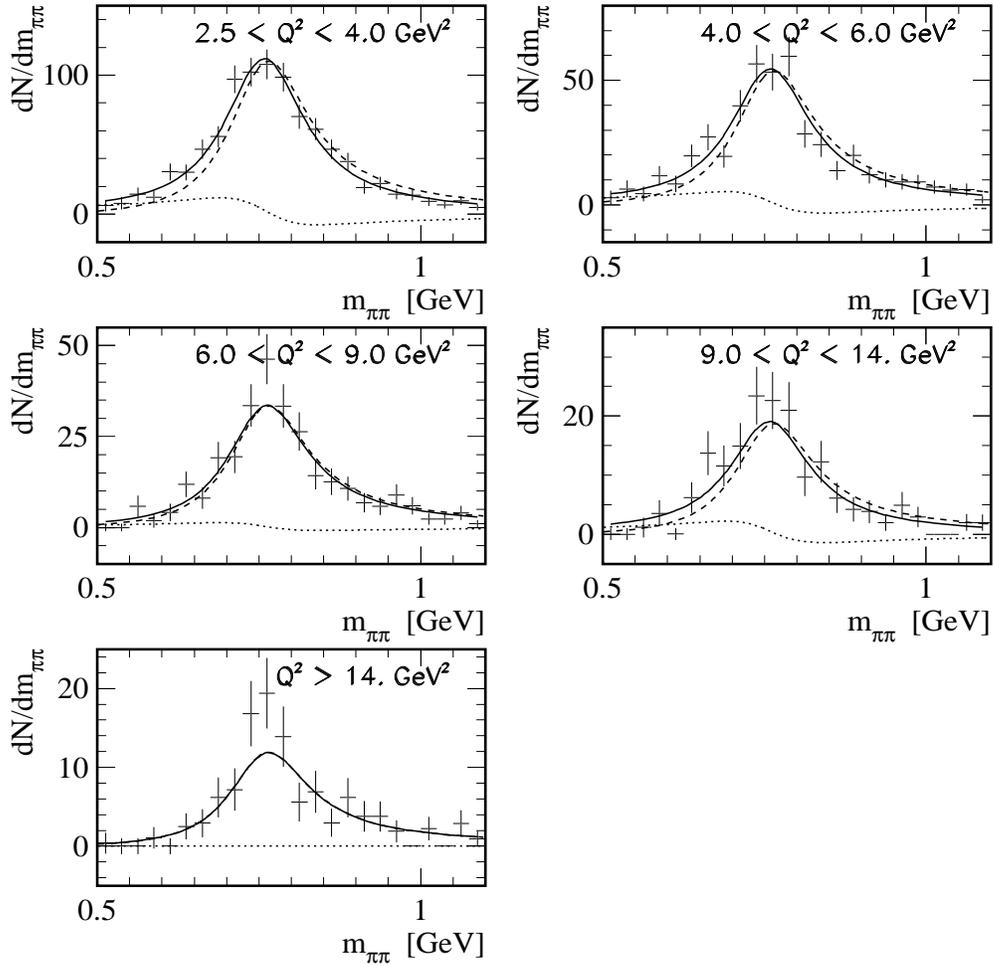,%
            bbllx=60pt,bblly=170pt,bburx=524pt,bbury=650pt,%
            height=13.0cm,width=13.cm}
  \end{center}
\vspace*{-0.5cm}
\caption{Same data as in Fig.~\ref{fig:massRS}, but compared 
to the S\"{o}ding parameterisation of eq.~(\ref{eq:sod}).
The solid curves are the results of the fits, the dashed curves correspond to
the non-skewed Breit-Wigner contributions, and the dotted curves correspond
to the interferences between the resonant and the non-resonant amplitudes.
The errors on the data are statistical only.}
\label{fig:massS}
\end{figure}

For the 1996 events passing the selection cuts of Table~\ref{Table:selection},
with $\langle Q^2 \rangle = 4.8$~\gevsq\ and $\langle W \rangle = 75$~\gev,
the distribution of \mpp, the invariant $\pi^+\pi^-$ mass, is presented in
Figs.~\ref{fig:massRS} and \ref{fig:massS} for five domains in \qsq.
The \om\ and \ph\ background contributions (see section~\ref{sect:elastic_bg}) 
are subtracted according to their mass distribution obtained from the DIFFVM
Monte Carlo simulation.

The mass distributions are skewed compared to a relativistic Breit-Wigner profile:
enhancement is observed in the low mass region and suppression in the
high mass side.
This effect has been attributed to an interference between the resonant and
the non-resonant production of two pions~\cite{drell}.
In order to extract the contribution of the resonant part of the cross
section, two different procedures are used. 


Following the phenomenological parameterisation
of Ross and Stodolsky~\cite{ross_sto}, the \mpp\ distribution is described
as:
\begin{equation}
\frac {{\rm d}N(\mpp)} {{\rm d}\mpp} = f_{\rho} \ BW_{\rho}(m_{\pi \pi})
      \ (\frac {m_{\rho}} {\mpp})^n + f_{bg} \ ,
                                                      \label{eq:ross}
\end{equation}
where $f_{\rh}$ is a normalisation constant and
\begin{equation}
    BW_{\rho}(\mpp) = \frac {\mpp \ \mrho \ \Gmpp}
               {(\mrhosq - \mppsq)^2 + \mrhosq \ \Gmppsq}
                                                      \label{eq:b_w}
\end{equation}
is a relativistic Breit-Wigner function with momentum dependent width
\cite{jackson}
\begin{equation}\Gmpp = \Gamma_{\rho} \ (\frac {q^*} {q_0^*})^3 \
        \frac {2} {1 + (q^* / q^*_0)^2} \ .
                                                     \label{eq:GJ}
\end{equation}
Here, \Grho\ is the \rh\ resonance width,
$q^*$ is the pion momentum in the ($\pi^+\pi^-$) rest frame and $q^*_0$ is
this momentum when $\mpp = \mrho$.
The factor $(\mrho /\mpp)^n$ in eq.~(\ref{eq:ross}) accounts for the
skewing of the shape of the \rh\ signal.
The background term $f_{bg}$ is parameterised using a distribution in phase 
space which includes the effect of the dipion threshold and an exponential 
fall off:
\begin{equation}
 f_{bg}= \alpha_1 \ (\mpp - 2 \mpi)^{\alpha_2 } \ e^{-\alpha_3 m_{\pi \pi}} \ ,
                                                   \label{eq:ps_bg}
\end{equation}
where \mpi\ is the pion mass and $\alpha_1 $, $\alpha_2 $ and $\alpha_3 $
are constants.

With eq.~(\ref{eq:ross}), the mass distribution for all selected events 
with 2.5 $<$ \qsq $<$ 60~\gevsq\ is fitted, after subtraction of  
the \om\ and \ph\ background contributions, over the range
$0.5 < \mpp < 1.1$~\gev, with the parameters $f_{\rho}$, $\mrho$, $\Grho$,
{\it n}, $\alpha_1$, $\alpha_2$ and $\alpha_3$ left free.
The resonance mass is found to be 0.766 $\pm$ 0.004~\gev\ and the width
$0.155 \pm 0.006$~\gev, in agreement with the PDG
values of 0.770 and 0.151~\gev~\cite{PDG}.
The fit value of the skewing parameter is $n$ = 1.4 $\pm$ 0.2 and the
background contribution corresponds to 1 $\pm$ 1\% of the number of events
in the peak.
The fit is of good quality: $\chi^2 / {\rm ndf} = 20.3/17 $.

For the five \qsq\ domains presented in Fig.~\ref{fig:massRS},
fits to the form of eq.~(\ref{eq:ross}) are thus performed with 
the mass and the width of the \rh\ meson fixed to the PDG values
and assuming the absence of
non-resonant background ($f_{bg}=0$).
This leaves two free parameters: the overall normalisation $f_{\rho}$ and the 
skewing parameter $n$.
The results of the fits are presented in Fig.~\ref{fig:massRS},
the $\chi^2 / {\rm ndf}$ values being good in all \qsq\ bins.

The data are also analysed using the parameterisation proposed by
S\"{o}ding~\cite{soding}, in which the skewing of the mass 
spectrum is explained by the interference of a resonant 
$ \rh \rightarrow \pi^+\pi^- $ amplitude
and a {\it p}-wave $\pi \pi$ Drell-type background term:
\begin{eqnarray}
\frac {{\rm d}N(\mpp)} {{\rm d}\mpp} = f_{\rho} \ BW_{\rho}(m_{\pi \pi}) +
   f_I \ I(m_{\pi \pi}) + f_{bg} \ ,                     \label{eq:sod} \\
   I(m_{\pi \pi})=  \frac { \mrhosq - \mppsq }
             {(\mrhosq - \mppsq)^2 + \mrhosq \ \Gmppsq } \ ,
                                                     \label{eq:int}
\end{eqnarray}
where $f_I$ is a constant fixing the relative normalisation of the
interference contribution.
In view of the uncertainty in the phase between the resonant and the
non-resonant amplitudes, no constraint is imposed on the relative contributions
of the background and interference terms.

The S\"{o}ding parameterisation also describes well the \qsq\ integrated data 
in the range $0.5 < \mpp < 1.1~\gev$, with values for the resonance mass
and width in agreement with the PDG values and non-resonant background
compatible with zero.
For the five selected \qsq\ bins, the width and the mass of the \rh\ meson are
thus fixed and $f_{bg}$ is taken to be zero.
Fits to the normalisation and the skewing parameter $f_I/f_{\rho}$ are
again of good quality, and the results are presented in Fig.~\ref{fig:massS}.

\begin{figure}[p]
\begin{center}
\setlength{\unitlength}{1.0cm}
\begin{picture}(13.0,13.0)
\put(0.0,0.0){\epsfig{file=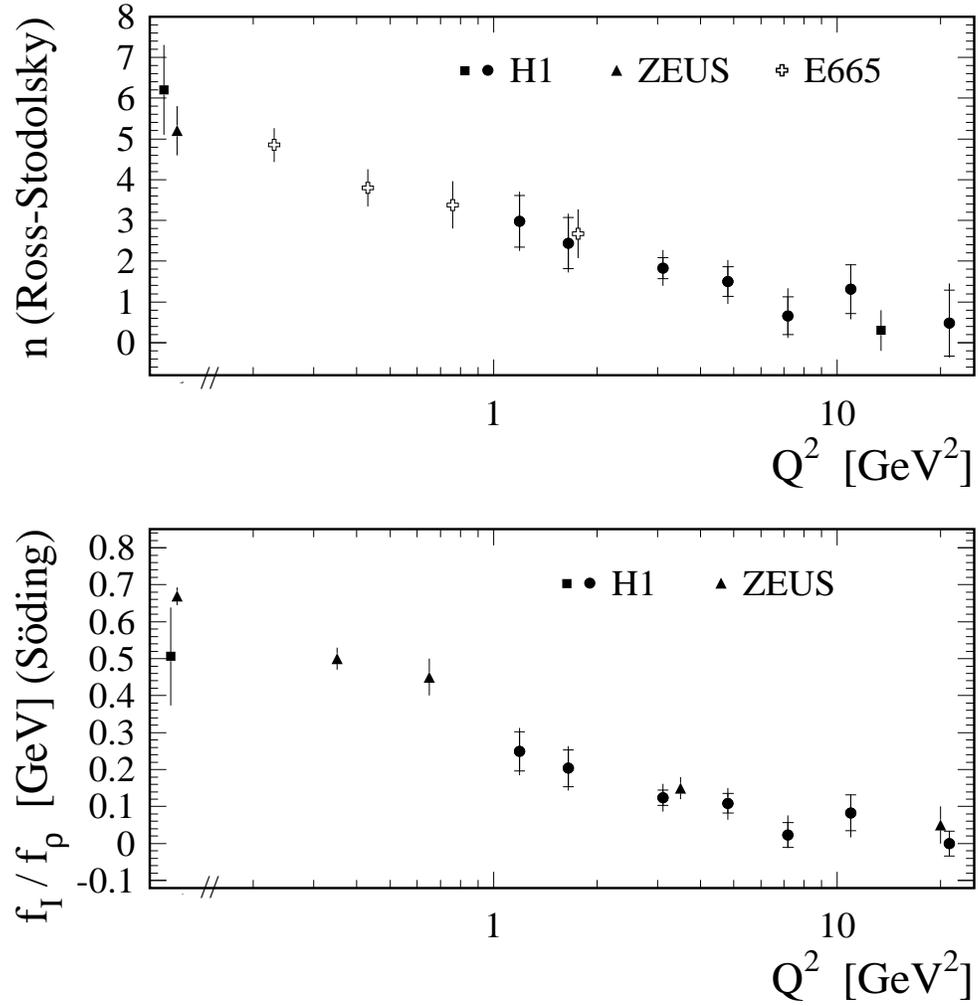,%
            bbllx=20pt,bblly=154pt,bburx=523pt,bbury=667pt,%
            height=13.0cm,width=13.cm}}
\put(2.7,1.37){//}
\put(1.2,0.22){\epsfig{file=whitebox.eps,height=1.2cm,width=1.5cm}}
\put(2.7,8.15){//}
\put(1.2,7.02){\epsfig{file=whitebox.eps,height=1.2cm,width=1.5cm}}
\end{picture}
\caption{\qsq\ dependence of the skewing parameters for elastic \rh\
production:
{\it n}, for the Ross-Stodolsky parameterisation of eq.~(\ref{eq:ross}),
and $f_I/f_{\rh}$, for the S\"{o}ding parameterisation of eq.~(\ref{eq:sod}).
For the present measurements (full circles), the inner error bars are statistical, 
and the full error bars include the systematic errors added in quadrature.
The other measurements are from H1~\protect\cite{h1_phot} and 
ZEUS~\protect\cite{zeus_phot,zeus_phottt} in photoproduction, 
and from H1~\protect\cite{H1_rho_94}, ZEUS~\protect\cite{rho_zeus} and
E665~\protect\cite{E665} in electroproduction.}
\label{fig:skewq}
\end{center}
\end{figure}

Fig.~\ref{fig:skewq} shows the fit values of the skewing parameters
as a function of \qsq, together with the results of other measurements in 
photoproduction~\cite{h1_phot,zeus_phot,zeus_phottt} and in 
electroproduction~\cite{H1_rho_94,rho_zeus,E665}.
The systematic errors are computed as described in section~\ref{sect:DIFFVM}, 
and include in addition the effect of the variation by 50\% 
of the \om\ and \ph\ background contributions.
The skewing of the mass distribution is observed to decrease with \qsq.
No significant \w\ or $t$ dependence of the skewing is observed within
the data.

\newpage

\section{Helicity Study}  \label{sect:helicity}

\subsection{Angular Decay Distributions}  
            \label{sect:angles}

\begin{figure}[t]
    \begin{center}
    \epsfig{file=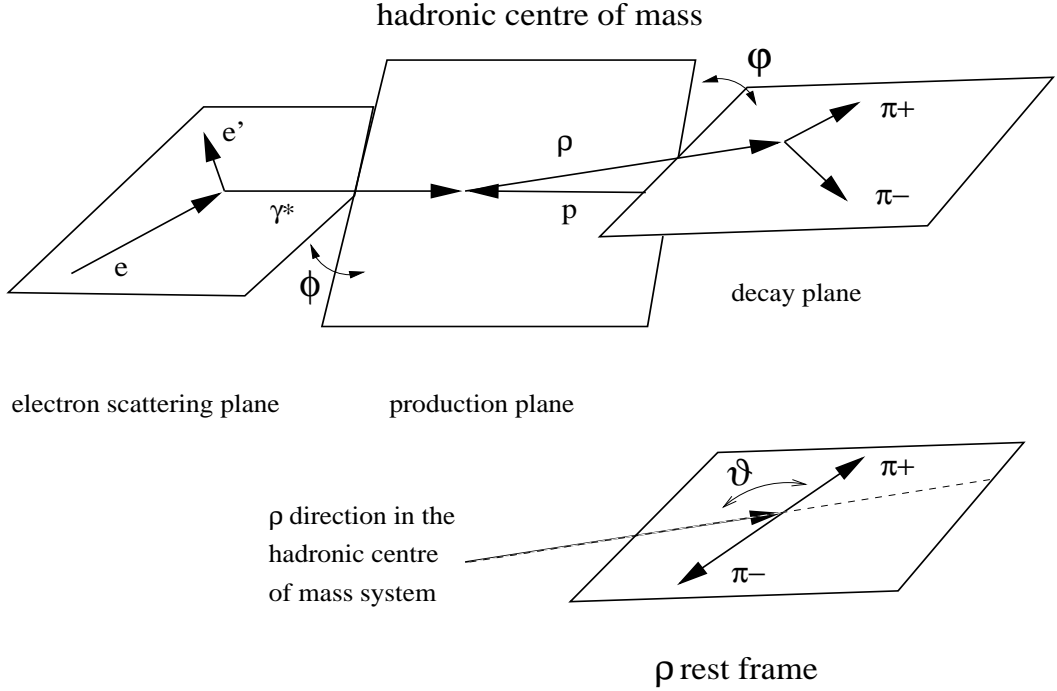,%
     height=9.0cm,width=14.cm}
\caption{Angle definition for the helicity system in elastic \rh\ meson
 production.}
    \label{fig:dec_ang}
    \end{center}
\end{figure}

The study of the angular distributions of the production and decay of the 
\rh~meson gives information on the photon and \rh\ polarisation states. 
The decay angles can be defined in several reference frames~\cite{bauer}.
In the helicity system, used for the present
measurement, three angles are defined as follows (Fig.~\ref{fig:dec_ang}). 
The angle $\phi$, defined in the hadronic centre of mass system (cms), is the 
azimuthal angle between the electron scattering plane and the plane containing 
the \rh\ and the scattered proton.
The \rh\ meson decay is described by the polar angle $\theta$ and the azimuthal 
angle \phib\ of the positive pion in the $\pi^+\pi^-$ rest frame, with the 
quantisation axis taken as the direction opposite to that of the outgoing 
proton in the hadronic cms. 

The normalised angular decay distribution $W$(\cosths, \phib, $\phi$) is 
expressed following the formalism used in~\cite{shilling-wolff} as a function of 
15 spin density matrix elements in the form 
%
%
%
\begin{eqnarray}
&& W(\costh, \phib, \phi) = \frac{3}{4\pi}\ \; \left\{ \ \
  \frac{1}{2} (1 - \rzqzz) + \frac{1}{2} (3 \ \rzqzz -1) \
  \cos^2\theta \right. \nonumber \\
&&  - \sqrt{2}\ {\rm Re} \ \rzquz\ \sin 2\theta  \cos\phib
    - \rzqumu\ \sin^2\theta \cos 2\phib \nonumber \\
&&  - \varepsilon\ \cos2\phi  \left(
      \ruuu\ \sin^2\theta + \ruzz\ \cos^2\theta
      - \sqrt{2}\ {\rm Re} \ \ruuz\ \sin2\theta \cos\phib \right.
\nonumber \\
&&    \ \ \ \ \ \ \ \ \ \ \ \ \ \ \ \ \ \
      - \left. \ruumu\ \sin^2\theta \cos2\phib \frac{}{} \right) \nonumber \\
&&    - \varepsilon\ \sin2\phi  \left( \sqrt{2}\ {\rm Im}\ \rduz\
\sin2\theta
\sin\phib
      +  {\rm Im}\ \rdumu\ \sin^2\theta \sin2\phib \right) \nonumber
\\
&&  + \sqrt{2\varepsilon\ (1+\varepsilon)}\ \cos\phi\ \left(  \frac{ }{ } \rcuu\
\sin^2\theta
      + \rczz\ \cos^2\theta \right. \nonumber \\
&&    \ \ \ \ \ \ \ \ \ \ \ \ \ \ \ \ \ \ - \sqrt{2}\ {\rm Re} \ \rcuz\
\sin 2\theta \cos\phib
       - \left. \rcumu\ \sin^2\theta \cos2\phib \frac{ }{ } \right) \nonumber \\
&&  + \sqrt{2\varepsilon\ (1+\varepsilon)}\ \sin\phi\ \left( \sqrt{2}\
{\rm Im}
\
          \rsuz\ \sin2\theta \sin\phib  \right. \nonumber \\
&&    \ \ \ \ \ \ \ \ \ \ \ \ \ \ \ \ \ \ \left.
        \left. + {\rm Im}\ \rsumu\ \sin^2\theta \sin2\phib \frac{ }{ } \right)\
\right\} \ ,
                                        \label{eq:W}
\end{eqnarray}
where $\varepsilon$~is the polarisation parameter of the virtual photon:
\begin{equation}
  \varepsilon \simeq \frac {1 - y} {1 - y + y^2/2} \ ,   \label{eq:epsilon}
\end{equation}
with $\av {\varepsilon} \approx 0.99$ in the present data.\footnote{
 In general, there are further contributions to the angular decay distribution, 
 which vanish for unpolarised leptons and for $\varepsilon = 1$ 
 (see~\cite{shilling-wolff}).}

The spin density matrix elements correspond to different bilinear combinations 
of the helicity amplitudes 
$T_{\lambda_{\rho} \lambda_{N'}, \lambda_{\gamma} \lambda_{N}}$ for \rh\ meson 
production, where
$\lambda_{\rho}$ and $\lambda_{\gamma}$ are the helicities of the \rh\ 
and of the photon, respectively, and $\lambda_{N}$ and $\lambda_{N'}$  
the helicities of the incoming and outgoing proton.
The upper indices 1 and 2 of the matrix elements refer to the production of \rh\
mesons by transverse photons, the index 04 corresponds to a combination
of transverse and longitudinal photons, 
and the indices 5 and 6 correspond to the interference between \rh\ production by
transverse photons and by longitudinal photons.
The lower indices of the matrix elements refer to the values of the \rh\
meson helicity $\lambda_{\rho}$ entering the combination of amplitudes.

Specific relations between the amplitudes, leading to predictions for the values
of several matrix elements, follow from additional hypotheses.
\begin{itemize}
\boldmath
\item {\bf $s$--channel helicity conservation} \\
\unboldmath
For the case of $s$-channel helicity conservation (SCHC), the helicity of
the virtual photon is retained by the \rh~meson and the helicity of the
proton is unchanged:
\begin{equation}
  T_{\lambda_{\rho} \lambda_{N'},\lambda_{\gamma} \lambda_{N}} =
  T_{\lambda_{\rho} \lambda_{N'},\lambda_{\gamma} \lambda_{N}} \
  \delta_{\lambda_{\rho} \lambda_{\gamma}} \ \delta_{\lambda_{N'} \lambda_{N}} \ .
                         \label{eq:schc}
\end{equation}
Single and double helicity flip amplitudes then vanish so that (omitting 
the nucleon helicities):
\begin{eqnarray}
  T_{\lambda_{\rho} \lambda_{\gamma}} =
  T_{01} = T_{10} = T_{0-1} = T_{-10} =0, \\
  T_{-11} = T_{1-1} = 0,               
\end{eqnarray}
%
and all matrix elements become zero, except five:
\begin{equation}
\rzqzz,   \ \ \ \ruumu , \ \ \ {\rm Im}\ \rdumu, 
\ \ \ {\rm Re}\ \rcuz , \ \ \ {\rm Im}\ \rsuz \ . 
                          \label{eq:nonnullschc}
\end{equation}
Furthermore, the following relationships occur between these elements:
\begin{equation}
          \ \ \ \ruumu \ = \ - \ {\rm Im} \ \rdumu, 
\ \ \ {\rm Re} \ \rcuz \ = \ - \ {\rm Im} \ \rsuz \ . 
                          \label{eq:pairing}
\end{equation}

\item {\bf Natural parity exchange} \\
Natural parity exchange (NPE) is defined by the following relations between the
amplitudes:\footnote{
  For unnatural parity exchange, an additional factor $(-1)$ appears in the
  right hand side of eq. (\ref{eq:NPE})~\cite{shilling-wolff}.}
%
\begin{equation}
  T_{-\lambda_{\rho} \lambda_{N'}, -\lambda_{\gamma} \lambda_{N}} =
  (-1)^{\lambda_{\rho}-\lambda_{\gamma}} \
  T_{\lambda_{\rho} \lambda_{N'}, \lambda_{\gamma} \lambda_{N}} \ .
                        \label{eq:NPE}
\end{equation}
  
\end{itemize}

In Table~\ref{Table:amplitudes}, the expressions of the matrix elements are 
given in terms of the helicity amplitudes for two specific sets of assumptions.
In column~2, the double helicity flip amplitudes $T_{1-1}$ and $T_{-11}$ and
the single flip amplitudes $T_{10}$ and $T_{-10}$ for the production of 
transversely polarised \rh\ mesons by longitudinal photons are neglected,
and the NPE relations $T_{0-1} = - T_{01}$ and $T_{-1-1} = T_{11}$ are assumed 
(see the discussion in section~\ref{sect:flip} and the presentation of the 
QCD model~\cite{ivanov}, in particular eq. (\ref{eq:hierarchy}), 
in section~\ref{sect:modelik}).
In column 3, the matrix elements are given for the case of SCHC 
(i.e. neglecting all helicity flip amplitudes) and assuming the NPE 
relation $T_{-1-1} = T_{11}$.
The nucleon helicities $\lambda_{N}$ and 
$\lambda_{N'}$ are omitted from the amplitudes, $T$, for brevity.

\begin{center}
\begin{figure}[htbp]
\setlength{\unitlength}{1.0cm}
\begin{picture}(13.0,18.0)
\put(0.0,0.0){\epsfig{file=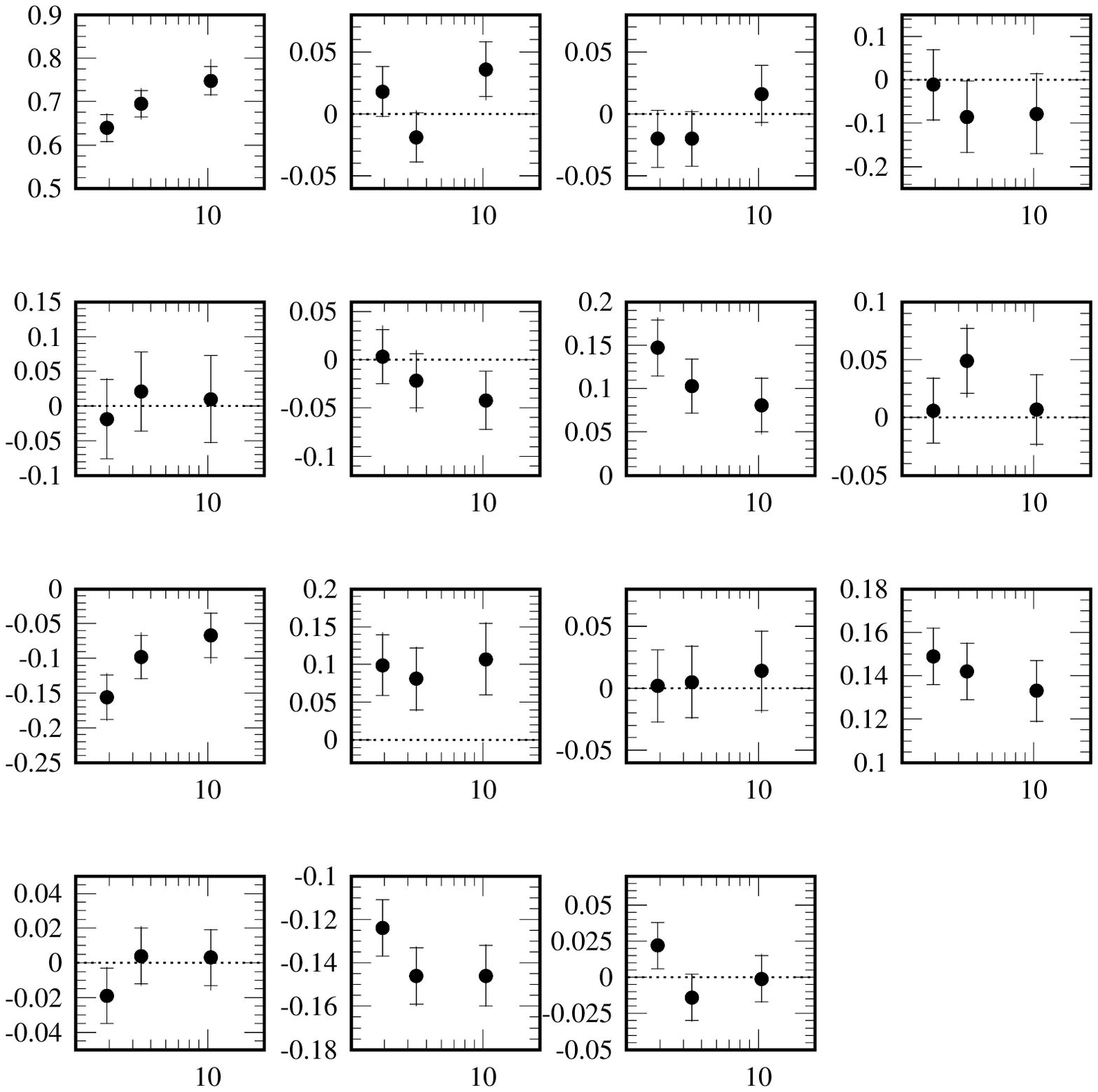,%
            bbllx=55pt,bblly=189pt,bburx=450pt,bbury=693pt,%
            height=17.0cm,width=13.cm}}
\put(6.7,17.0){\Large H1 data}
\put(2.4,15.7){\large \rzqzz}
\put(5.9,15.7){\large {\rm Re} \rzquz}
\put(10.0,15.7){\large \rzqumu}
\put(13.8,15.7){\large \ruzz}
\put(2.4,11.65){\large \ruuu}
\put(5.9,11.65){\large {\rm Re} \ruuz}
\put(10.0,11.65){\large \ruumu}
\put(13.5,11.65){\large {\rm Im} \rduz}
\put(2.1,7.6){\large {\rm Im} \rdumu}
\put(6.2,7.6){\large \rczz}
\put(10.0,7.6){\large \rcuu}
\put(13.5,7.6){\large {\rm Re} \rcuz}
\put(2.4,3.6){\large \rcumu}
\put(5.9,3.6){\large {\rm Im} \rsuz}
\put(9.6,3.6){\large {\rm Im} \rsumu}
\put(9.5,-0.5){\large \qsq~[\gevsq] }
\put(13.15,4.33){3}
\put(13.15,8.37){3}
\put(13.15,12.45){3}
\put(9.40,0.31){3}
\put(9.40,4.33){3}
\put(9.40,8.37){3}
\put(9.40,12.45){3}
\put(5.58,0.31){3}
\put(5.58,4.33){3}
\put(5.58,8.37){3}
\put(5.58,12.45){3}
\put(1.79,0.31){3}
\put(1.79,4.33){3}
\put(1.79,8.37){3}
\put(1.79,12.45){3}
\end{picture}
\vspace*{1cm}
\caption{Spin density matrix elements for elastic electroproduction 
  of \rh\ mesons, measured for three values of \qsq\ with the 1996 data sample. 
  The inner error bars are statistical and the full error bars include 
  the systematic errors added in quadrature.  
  The dashed lines indicate the expected null values in the case of SCHC.}
\label{fig:matqsq}
\end{figure}
\end{center}

\begin{center}
\begin{figure}[htbp]
\setlength{\unitlength}{1.0cm}
\begin{picture}(13.0,18.0)
\put(0.0,0.0){\epsfig{file=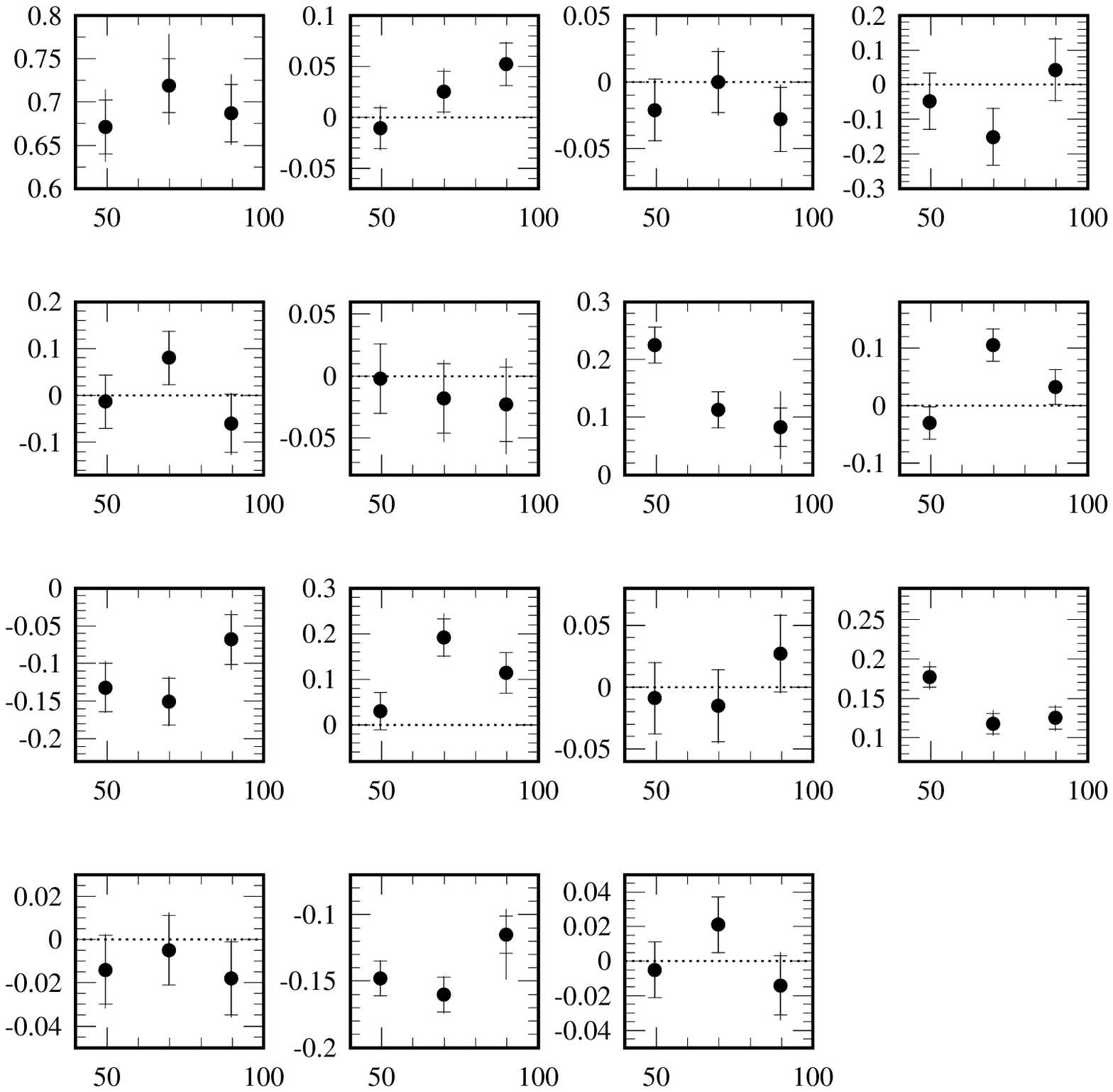,%
            bbllx=55pt,bblly=189pt,bburx=450pt,bbury=693pt,%
            height=17.0cm,width=13.cm}}
\put(6.7,17.0){\Large H1 data}
\put(2.4,15.7){\large \rzqzz}
\put(5.9,15.7){\large {\rm Re} \rzquz}
\put(10.0,15.7){\large \rzqumu}
\put(13.8,15.7){\large \ruzz}
\put(2.4,11.65){\large \ruuu}
\put(5.9,11.65){\large {\rm Re} \ruuz}
\put(10.0,11.65){\large \ruumu}
\put(13.5,11.65){\large {\rm Im} \rduz}
\put(2.1,7.6){\large {\rm Im} \rdumu}
\put(6.2,7.6){\large \rczz}
\put(10.0,7.6){\large \rcuu}
\put(13.5,7.6){\large {\rm Re} \rcuz}
\put(2.4,3.6){\large \rcumu}
\put(5.9,3.6){\large {\rm Im} \rsuz}
\put(9.6,3.6){\large {\rm Im} \rsumu}
\put(9.5,-0.5){\large \W~[\gev] }
\end{picture}
\vspace*{1cm}
\caption{Spin density matrix elements for elastic electroproduction 
  of \rh\ mesons, measured for three values of \w\ with the 1996 data sample. 
  The inner error bars are statistical and the full error bars include 
  the systematic errors added in quadrature.  
  The dashed lines indicate the expected null values in the case of SCHC.}
\label{fig:matrelw}
\end{figure}
\end{center}

\begin{center}
\begin{figure}[htbp]
\setlength{\unitlength}{1.0cm}
\begin{picture}(13.0,18.0)
\put(0.0,0.0){\epsfig{file=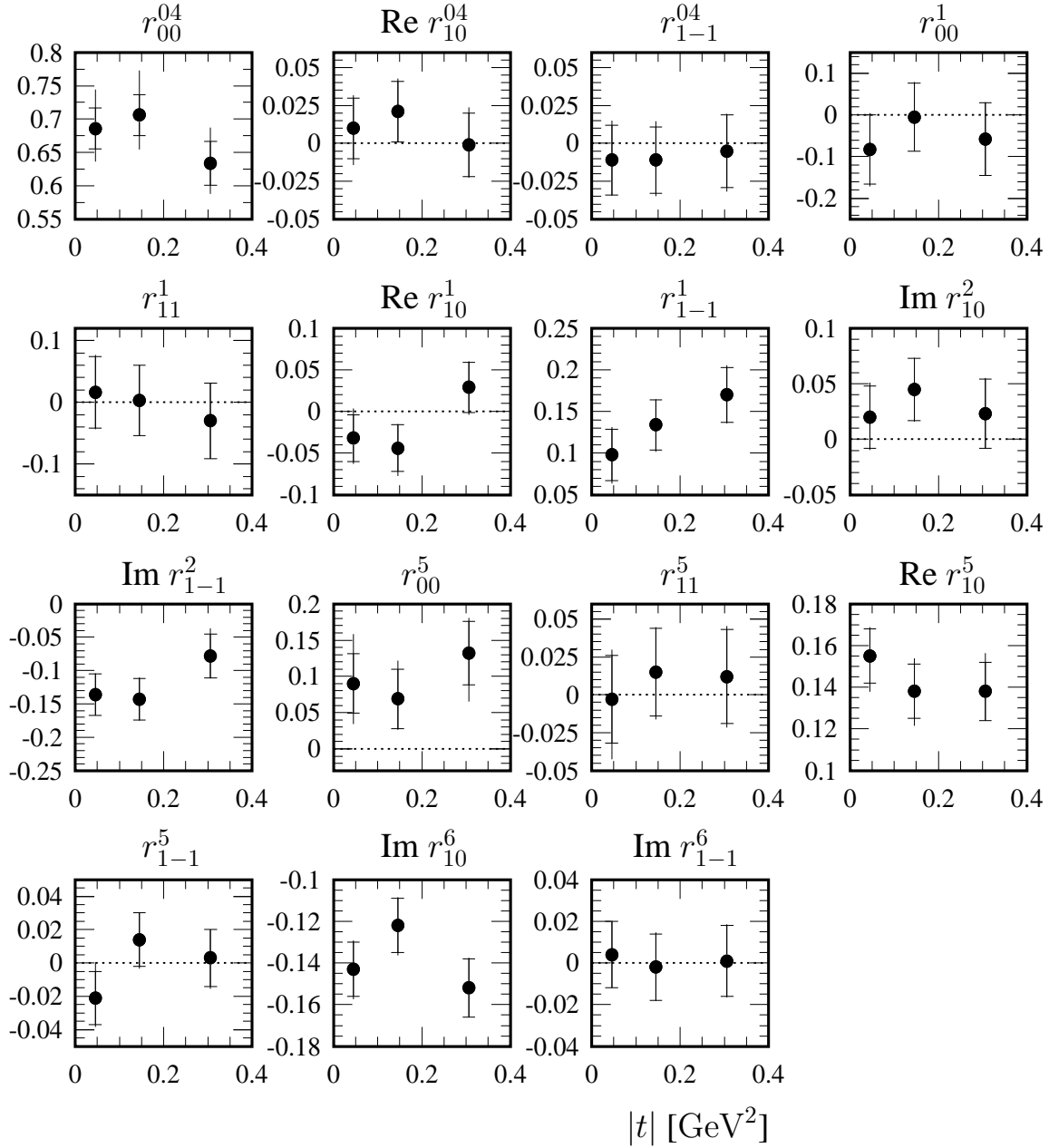,%
            bbllx=55pt,bblly=189pt,bburx=450pt,bbury=693pt,%
            height=17.0cm,width=13.cm}}
\put(6.7,17.0){\Large H1 data}
\put(2.4,15.7){\large \rzqzz}
\put(5.9,15.7){\large {\rm Re} \rzquz}
\put(10.0,15.7){\large \rzqumu}
\put(13.8,15.7){\large \ruzz}
\put(2.4,11.65){\large \ruuu}
\put(5.9,11.65){\large {\rm Re} \ruuz}
\put(10.0,11.65){\large \ruumu}
\put(13.5,11.65){\large {\rm Im} \rduz}
\put(2.1,7.6){\large {\rm Im} \rdumu}
\put(6.2,7.6){\large \rczz}
\put(10.0,7.6){\large \rcuu}
\put(13.5,7.6){\large {\rm Re} \rcuz}
\put(2.4,3.6){\large \rcumu}
\put(5.9,3.6){\large {\rm Im} \rsuz}
\put(9.6,3.6){\large {\rm Im} \rsumu}
\put(9.5,-0.5){\large $|t|$~[\gevsq] }
\end{picture}
\vspace*{1cm}
\caption{Spin density matrix elements for elastic electroproduction 
  of \rh\ mesons, measured for three values of \ttra\ with the 1996 data sample. 
  The inner error bars are statistical and the full error bars include 
  the systematic errors added in quadrature.  
  The dashed lines indicate the expected null values in the case of SCHC.}
\label{fig:matrelt}
\end{figure}
\end{center}

The matrix elements can be measured as the projections of the decay angular 
distribution (eq.~\ref{eq:W}) onto orthogonal trigonometric functions of the angles
$\theta$, \phib\ and $\phi$, which are listed in Appendix C of 
ref.~\cite{shilling-wolff}.
The average values of these functions, for the 1996 data and for the kinematic 
domain defined in Table~\ref{Table:selection}, provide the measurements 
presented in Table~\ref{Table:moments}.
The results are also presented in Figs.~\ref{fig:matqsq}$-$\ref{fig:matrelt} 
(and in Tables~\ref{table:matqsq}$-$\ref{table:matrelt}) as a function of \qsq, 
$W$ and $t$.
Statistical and systematic errors are given separately, 
the systematic errors being computed here, and in the rest of 
section~\ref{sect:helicity}, as described in section~\ref{sect:DIFFVM}.
The data sample is not corrected for the small backgrounds due to
proton dissociation,\footnote{
 The measurements in~\cite{H1_phi} and~\cite{pdiss_paper} indicate that,
 within errors, elastic and proton dissociation events have the same \rh\
 meson decay angular distributions.} 
\om\ and \ph\ production and radiative effects. 

Within the measurement precision, the matrix elements presented 
in Table~\ref{Table:moments} and in Figs.~\ref{fig:matqsq}$-$\ref{fig:matrelt} 
generally follow the SCHC predictions (with the NPE relation 
$T_{-1-1} = T_{11}$).
This is not the case, however, for the \rczz\ element, which is 
significantly different from zero (see also the discussion of the 
distribution of the angle $\phi$ in section~\ref{sect:flip}).
It has been checked that this effect is not an artifact 
of the Monte Carlo simulation used to correct the data for detector acceptance 
effects~\cite{Barbara}.

As will be discussed in section~\ref{sect:flip}, the violation of 
SCHC is small.
Information on the photon polarisation can thus be obtained from the
measurements of the spin density matrix elements using SCHC as a first
order approximation. 
This analysis is performed in section~\ref{sect:photon}.
The violation of SCHC is then studied in more detail in section~\ref{sect:flip}.
For these analyses, the good description of the data provided by the
function $W$(\cosths, \phib, $\phi$) is verified through various
angular distributions.
Finally, section~\ref{sect:model} presents comparisons of the results with model
predictions.

\begin{table}[htbp]
\begin{center}
\begin{tabular}{|c|c|c|}
\hline
\hline
 Element  & NPE and $T_{1-1}$ = $T_{10}$ = 0   & NPE and  SCHC    \\
\hline
\hline
 & & \\

$ r^{04}_{00}$                &  $\frac{1}{1+\epsilon R} \
  \left( \frac{|T_{01}|^2}{|T_{11}|^2+|T_{01}|^2} +\epsilon R \right)$ 
  & $\frac{\epsilon R}{1+\epsilon R}$  \\ & & \\    
$ {\rm Re} \ r^{04}_{10}$    & $\frac{1}{2} \ \frac{1}{1+\epsilon R} \
  \frac{1}{|T_{11}|^2+|T_{01}|^2} \ {\rm Re} \ (T_{11}T^*_{01}$) & 0 \\ & & \\
$ r^{04}_{1 -1} $            &  0 & 0 \\ & & \\
$ r^{1}_{00}$                &  $\frac{-1}{1+\epsilon R} \
   \frac{|T_{01}|^2}{|T_{11}|^2+|T_{01}|^2} $ & 0 \\ & & \\
$ r^{1}_{11}$                & 0 & 0 \\ & & \\
$ {\rm Re} \ r^{1}_{10}$     & $- {\rm Re} \ \rzquz$ & 0 \\ & & \\
$ r^{1}_{1 -1}$              &  $\frac{1}{2} \ \frac{1}{1+\epsilon R} \
   \frac{|T_{11}|^2}{|T_{11}|^2+|T_{01}|^2}$
  & $\frac{1}{2}\frac{1}{1+\epsilon R}$  \\ & & \\
$ {\rm Im} \ r^{2}_{10}$     & ${\rm Re} \ \rzquz$  & 0 \\ & & \\     
$ {\rm Im} \ r^{2}_{1-1}$    & $- \ruumu$  & $-r^{1}_{1-1}$  \\ & & \\
$ r^{5}_{00}$                &  $ \frac{\sqrt{2 R}}{1+\epsilon R} \
 \frac{1}{|T_{00}| \sqrt{|T_{11}|^2+|T_{01}|^2}}  \ {\rm Re} \ (T_{00}T^*_{01})$  
 & 0  \\ & & \\
$ r^{5}_{11}$                & 0  & 0 \\ & & \\
$ {\rm Re} \ r^{5}_{10}$     & $ \frac{1}{2 \sqrt{2}} \ \frac{\sqrt{R}}{1+\epsilon R} \
  \frac{1}{|T_{00}| \sqrt{|T_{11}|^2+|T_{01}|^2}} \ {\rm Re} \ (T_{11}T^*_{00})$
  & $ \frac{1}{2 \sqrt{2}} \ \frac {\sqrt {R}}{1+\epsilon R} \
  \frac {1}{|T_{11}| |T_{00}|} {\rm Re} \ (T_{11} T^*_{00})$  \\ & & \\
$ r^{5}_{1-1}$               & 0  & 0  \\ & & \\
$ {\rm Im} \ r^{6}_{10}$     & $- {\rm Re} \ \rcuz$ & $-{\rm Re} \ r^{5}_{10}$  \\ & 
  & \\
$ {\rm Im} \ r^{6}_{1-1}$    & 0  & 0 \\ & & \\
\hline
\hline
\end{tabular}
\end{center}
 \caption{Spin density matrix elements for the elastic 
  electroproduction of \rh\ mesons, expressed as a function of the 
  helicity amplitudes $T_{\lambda_{\rho} \ \lambda_{\gamma}}$:
  second column: 
  the single-flip $T_{10}$ and double-flip $T_{1-1}$ amplitudes are 
  neglected and the NPE relations~(\ref{eq:NPE}) are assumed for the
  other amplitudes;
  third column:  the SCHC conditions and the NPE relation 
  $T_{-1-1} = T_{11}$ are assumed (i.e. 
  the $T_{01}$ helicity flip amplitude is also neglected).
  $R$ is the ratio of cross sections for \rh\ production by longitudinal and 
  transverse photons.
  The nucleon helicities are omitted for brevity.}
\label{Table:amplitudes}
\end{table}

\begin{table}[htbp]
\begin{center}
\begin{tabular}{|c|ccc|}
\hline
\hline
 Element         
   & \multicolumn{3}{|c|}{Measurement} \\
\hline
\hline
$  r^{04}_{00}  $     
   & 0.674 & $\pm$ 0.018 & \mbig{$^{+0.051}_{-0.036}$} \\     
$ {\rm Re} \ r^{04}_{10}$
   & 0.011 & $\pm$ 0.012 & \mbig{$^{+0.007}_{-0.001}$} \\         
$ r^{04}_{1 -1} $      
   &-0.010 & $\pm$ 0.013 & \mbig{$^{+0.004}_{-0.003}$} \\        
$ r^{1}_{00}$       
   &-0.058 & $\pm$ 0.048 & \mbig{$^{+0.013}_{-0.011}$} \\       
$ r^{1}_{11}$
   & 0.002 & $\pm$ 0.034 & \mbig{$^{+0.006}_{-0.006}$} \\       
$ {\rm Re} \ r^{1}_{10}$
   &-0.018 & $\pm$ 0.016 & \mbig{$^{+0.010}_{-0.014}$} \\  
$ r^{1}_{1 -1}$       
   & 0.122 & $\pm$ 0.018 & \mbig{$^{+0.004}_{-0.005}$} \\           
$ {\rm Im} \ r^{2}_{10}$
   & 0.023 & $\pm$ 0.016 & \mbig{$^{+0.010}_{-0.009}$} \\        
$ {\rm Im} \ r^{2}_{1-1}$
   &-0.119 & $\pm$ 0.018 & \mbig{$^{+0.010}_{-0.005}$} \\
$ r^{5}_{00}$
   & 0.093 & $\pm$ 0.024 & \mbig{$^{+0.019}_{-0.010}$} \\       
$ r^{5}_{11}$
   & 0.008 & $\pm$ 0.017 & \mbig{$^{+0.008}_{-0.012}$} \\   
$ {\rm Re} \ r^{5}_{10}$
   & 0.146 & $\pm$ 0.008 & \mbig{$^{+0.006}_{-0.006}$} \\         
$ r^{5}_{1-1}$
   &-0.004 & $\pm$ 0.009 & \mbig{$^{+0.001}_{-0.003}$} \\        
$ {\rm Im} \ r^{6}_{10}$
   &-0.140 & $\pm$ 0.008 & \mbig{$^{+0.002}_{-0.004}$} \\   
$ {\rm Im} \ r^{6}_{1-1}$
   & 0.002 & $\pm$ 0.009 & \mbig{$^{+0.003}_{-0.000}$} \\                       
\hline
\hline
\end{tabular}
\end{center}
\caption{Spin density matrix elements for the elastic electroproduction 
  of \rh\ mesons, measured for the 1996 data sample as the average values of 
  the corresponding orthogonal functions of the \rh\ meson production and decay 
  angles (see Appendix C of ref.~\protect\cite{shilling-wolff}).
  The first errors are statistical, the second systematic.}
\label{Table:moments}
\end{table}


\newpage

\subsection{Helicity Conserving Amplitudes}  
            \label{sect:photon}

\boldmath
\subsubsection{Ratio of the Longitudinal and Transverse Cross Sections}  
            \label{sect:r0400}
\unboldmath

\begin{figure}[b]
  \begin{center}
        \epsfig{file=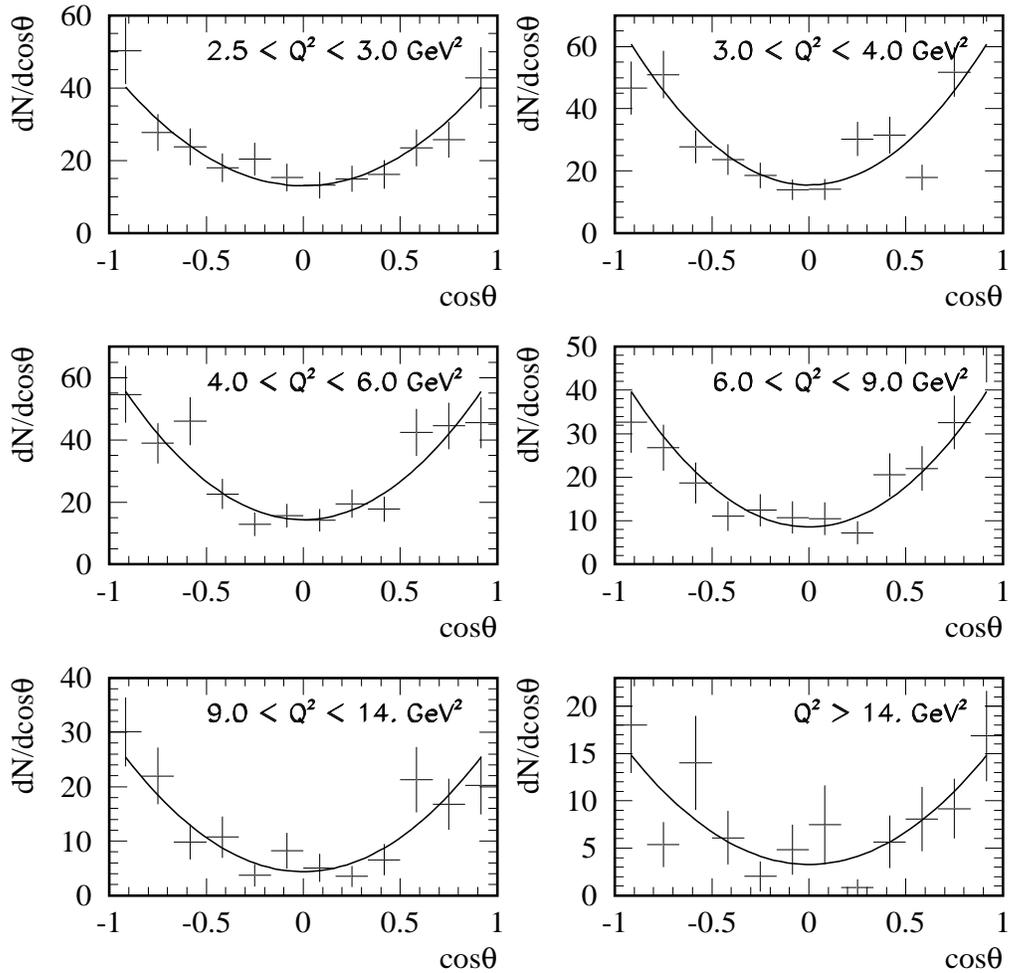,%
         bbllx=60pt,bblly=170pt,bburx=524pt,bbury=650pt,%
         height=13.0cm,width=13.cm}
  \end{center}
\caption{Acceptance corrected \cosths\ distributions for the 1996 data sample 
  in six bins in \qsq.
  The curves are the result of fits to the form of eq.~(\ref{eq:cosths}).
  The errors on the data points are statistical only.}
\label{fig:costh}
\end{figure}

After integration over the angles \phib\ and $\phi$, the angular distribution 
(eq.~\ref{eq:W}) takes the form
\begin{equation}
W(\costh) \propto  1 - \rzzzz + (3 \ \rzzzz -1) \ \cos^2\theta \ .
                        \label{eq:cosths}
\end{equation}
In Fig.~\ref{fig:costh}, the \cosths~distributions for the 1996 data are 
presented for six bins in \qsq, and the results of fits to 
eq.~(\ref{eq:cosths}) are superimposed. 
As can be observed from the figures, the quality of the fits is good.
The resulting measurements of $r^{04}_{00}$ are in good agreement with those 
presented in Figs.~\ref{fig:matqsq}$-$\ref{fig:matrelt}
and in Tables~\ref{table:matqsq}$-$\ref{table:matrelt}.

In the case of SCHC, the matrix element
\rzzzz\ provides a direct measurement of $R$, the ratio of cross sections for 
\rh\ production by longitudinal and transverse virtual photons 
(see Table~\ref{Table:amplitudes}, column~3):
\begin{equation}
\R = \frac{\sigma_L}{\sigma_T} = \frac{1}{\varepsilon} \ 
          \frac{\rzzzz}{1-\rzzzz} \ .
                                              \label{eq:R}
\end{equation}
As the SCHC violating amplitudes are small compared to the helicity conserving 
ones (see section~\ref{sect:flip}), eq.~(\ref{eq:R}) can be used assuming 
SCHC to estimate $R$.\footnote{
 The $T_{01}$ amplitude, which appears to be the dominant helicity-flip 
 amplitude, 
 corresponds to $8 \pm 3 \%$ of the non-flip amplitudes 
 $\sqrt{|T_{00}|^2 + |T_{11}|^2}$ (see section~\ref{sect:flip}).  
 A comparison of the forms of \rzzzz\ in columns 2 and 3 of 
 Table~\ref{Table:amplitudes} indicates that
 the effect of SCHC violation on the measurement of $R$ is  
 $2.5 \pm 1.5 \%$. This is neglected.}

\begin{figure}[t]
  \begin{center}
        \epsfig{file=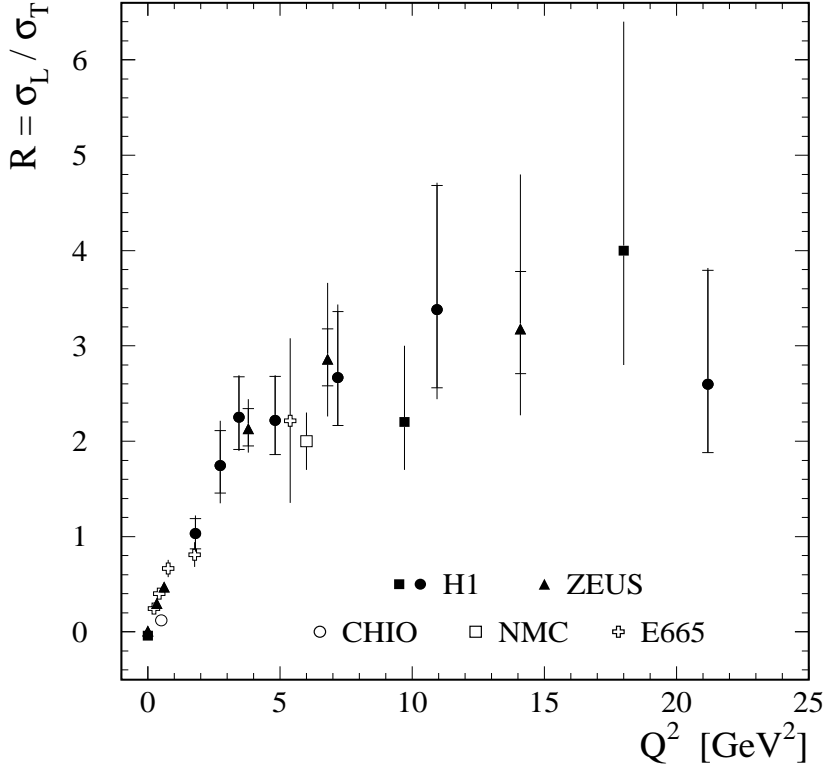,%
         bbllx=60pt,bblly=170pt,bburx=524pt,bbury=650pt,%
         height=10.0cm,width=10.cm}
  \end{center}
\caption{The ratio $R$ of cross sections for elastic \rh\ meson electroproduction 
  by longitudinal and transverse photons, measured in the SCHC approximation 
  and presented as a function of \qsq.
  For the present measurements (full circles), the inner error bars are statistical 
  and 
  the full error bars include the systematic errors added in quadrature.
  The other measurements are from H1~\protect\cite{h1_phot} and 
  ZEUS~\protect\cite{zeus_phottt} in photoproduction, 
  and from CHIO~\protect\cite{CHIO}, NMC~\protect\cite{NMC},
  E665~\protect\cite{E665}, H1~\protect\cite{H1_rho_94} and
  ZEUS~\protect\cite{rho_zeus} in electroproduction.}
\label{fig:R}
\end{figure}

The values of $R$ deduced from eq. (\ref{eq:R}) using the results of the fits 
of the \cosths~distributions to eq.~(\ref{eq:cosths}) 
are presented in Fig.~\ref{fig:R} (and in Table~\ref{table:R})
as a function of \qsq, together with other measurements performed assuming
SCHC~\cite{H1_rho_94,rho_zeus,CHIO,NMC,E665,h1_phot,zeus_phottt}.
It is observed that $R$ rises steeply at small \qsq, and that the longitudinal
cross section dominates over the transverse cross section for 
$Q^2 \gsim$ 2~\gevsq. 
However, the rise is non-linear, with a weakening dependence at large $Q^2$ 
values.
No significant $W$ dependence of the behaviour of $R$ as a function of \qsq\ is 
suggested by the comparison of the fixed target and HERA results.

\subsubsection{Longitudinal$-$Transverse Interference}  
            \label{sect:interference}

In the case of NPE\footnote{
 The asymmetry $P_\sigma$ between natural ($\sigma^N$) and unnatural 
 ($\sigma^U$) parity 
 exchange can be determined, for transverse photons, from the measured matrix
 elements as: 
 $$ P_\sigma = \frac {\sigma^N - \sigma^U} {\sigma^N + \sigma^U}
            = (1 + \varepsilon R) \ (2 r^1_{1-1} - r^1_{00})\ , $$ 
 and is found to be compatible with 1, as at lower energy~\cite{CHIO,joos}.
 This implies that NPE holds in the data at least for transverse photons.
 The measurement of the corresponding asymmetry for longitudinal photons 
 would require two different values of $\varepsilon$, i.e. two beam energies
 (see eq. (103) in ref.~\cite{shilling-wolff}).} 
and SCHC, the decay angular 
distribution $W$(\cosths, \phib, $\phi$) reduces to a function of two variables, 
\cosths\ and $\psi$, where $\psi = \phi - \phib$ is the angle between the 
electron scattering plane and the \rh\ meson decay plane:
\begin{eqnarray}
&& W(\costh, \psi) = \frac{3}{8\pi}\ \frac{1}{1+\varepsilon\ R}\ 
  \left\{ \ \sin^2\theta \ (1+ \varepsilon\ \cos2\psi)  \frac{ } { }  
  \right. \nonumber \\
&&  + \left. 2\ \varepsilon\ R\ \cos^2\theta
  - \sqrt{2\varepsilon\ (1+\varepsilon)\ R}\ \cos\delta \sin2\theta
 \cos\psi \right\} \ .
                         \label{eq:Wcosdelta}
\end{eqnarray}
Here $\delta$ is the phase between the transverse $T_{11}$ and the 
longitudinal $T_{00}$ amplitudes:
\begin{equation}
T_{00} \ T_{11}^* = |T_{00}| \ |T_{11}| \ e^{-i\delta}
\end{equation}
and
\begin{equation}
\cos\delta = \frac{1+\varepsilon\ R}{ \sqrt{R/2} } \left( {\rm Re}\ \rcuz -
 {\rm Im}\ \rsuz \right) \ .
                                                  \label{eq:cosd}
\end{equation}

\begin{figure}[htbp]
  \begin{center}
        \epsfig{file=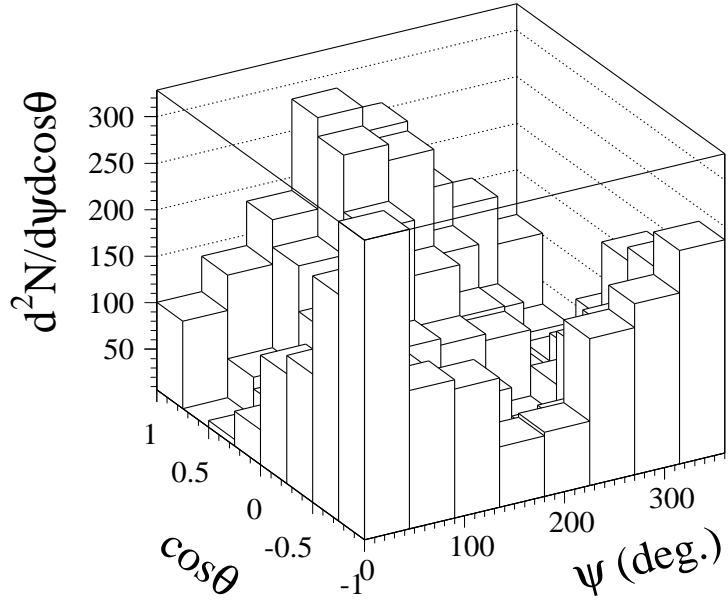,%
         bbllx=23pt,bblly=160pt,bburx=540pt,bbury=688pt,%
         height=10.0cm,width=10.cm}
  \end{center}
\caption{Acceptance corrected plot of the event distribution in \cosths\
  and $\psi$ for the 1996 data sample.}
\label{fig:cthpsi}
\end{figure}

A two-dimensional plot of the \cosths\ and $\psi$ variables is presented 
in Fig.~\ref{fig:cthpsi} for the 1996 data. 
A fit of eq.~(\ref{eq:Wcosdelta}) to these data gives:
\begin{equation}
\cos \delta = 0.925 \pm 0.022 \ ^{+0.011}_{-0.022}\ .
                  \label{eq:vcosd}
\end{equation}
This number is in agreement within errors with the value of $\cos\delta$
computed from eqs.~(\ref{eq:R}) and (\ref{eq:cosd}) using the measurements of 
\rzzzz, {\rm Re}\ \rcuz\ and {\rm Im}\ \rsuz\ given in 
Table~\ref{Table:moments}.

\begin{center}
\begin{figure}[htbp]
\setlength{\unitlength}{1.0cm}
\begin{picture}(13.0,5.0)
\put(1.0,0.0){\epsfig{file=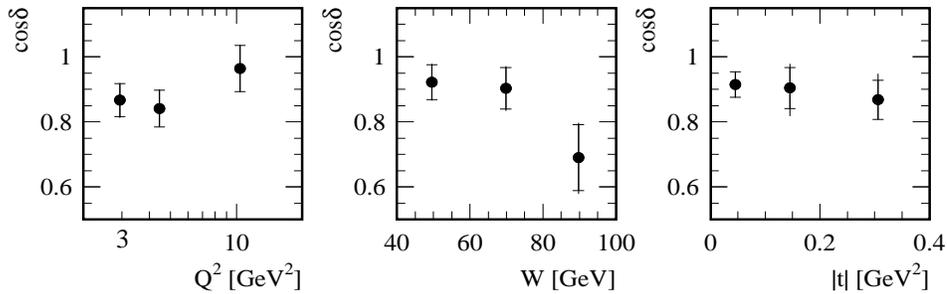,%
         bbllx=43pt,bblly=485pt,bburx=533pt,bbury=685pt,%
         height=5.0cm,width=13.cm}}
\put(2.9,0.83){\footnotesize 3}
\end{picture}
\caption{Measurements of the \cosdelta\ parameter as a function of \qsq,
  \W\ and $t$, obtained assuming SCHC and NPE from fits to the 
  (\cosths, $\psi$) distributions.
   The inner error bars are statistical and the full
   error bars include the systematic errors added in quadrature.}
\label{fig:cdel}
\end{figure}
\end{center}

Fig.~\ref{fig:cdel} (and Table~\ref{table:cdel}) presents the measurements 
of $\cos\delta$ as a function of \qsq, \W\ and~\ttra.
No significant evidence is found for a variation in the phase between the
transverse and longitudinal amplitudes with these variables.
That these amplitudes are nearly in phase was already observed at 
lower energy~\cite{CHIO,joos,delpapa}.

%
\boldmath
\subsubsection{The $\psi$ Distribution}  
            \label{sect:r11m1}
\unboldmath

\begin{figure}[p]
 \begin{center}
        \epsfig{file=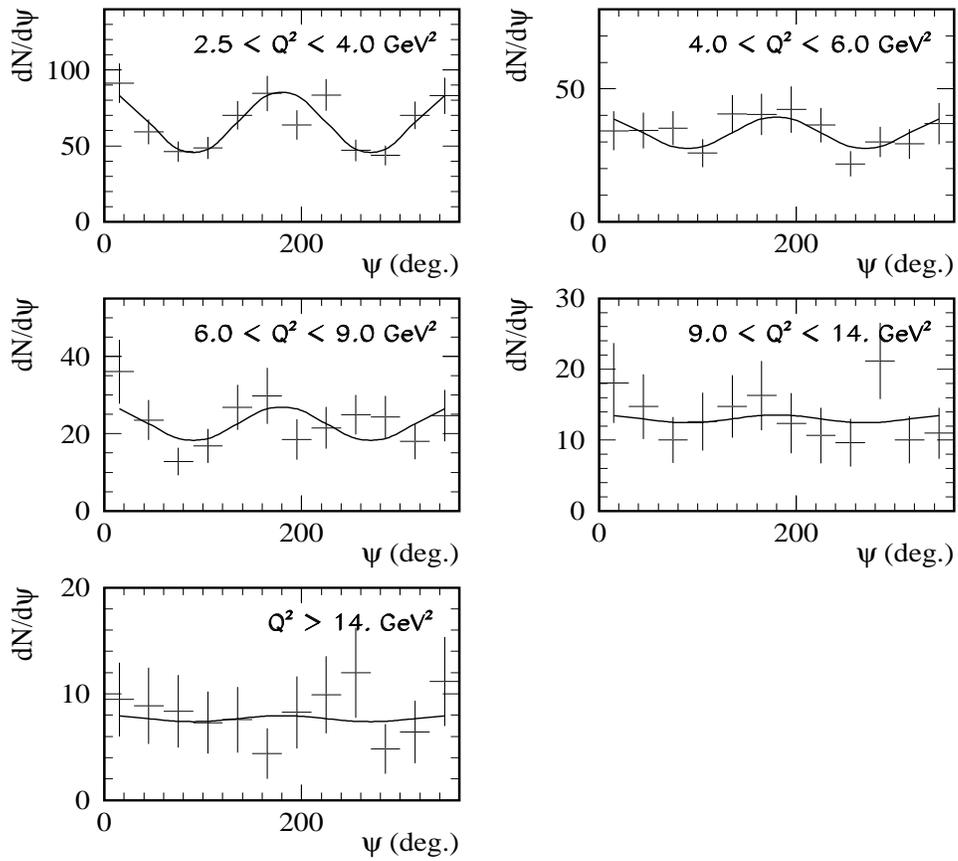,%
         bbllx=45pt,bblly=160pt,bburx=533pt,bbury=694pt,%
         height=13.0cm,width=13.cm}
 \end{center}
 \vspace*{-0.2cm}
\caption{Acceptance corrected distributions of the $\psi$
  angle for the 1996 data sample in five bins in \qsq.
  The curves are the result of fits to the form of eq.~(\ref{eq:psi}).
  The errors on the data points are statistical only.}
\label{fig:psi}
\end{figure}

Fig.~\ref{fig:psi} shows the distributions of the angle $\psi$ for five bins 
in \qsq.
They are well described by the function 
\begin{equation}
 W(\psi) = \frac{1}{2\pi} \ ( 1  +  2\ \varepsilon\ \rone\ \cos2\psi) \ ,                                             
                                                         \label{eq:psi}
\end{equation}
obtained from the integration over \costh\ of the function 
$W$(\costh, \phib, $\phi$) (eq. \ref{eq:W}), assuming SCHC.
Measurements of the $r^1_{1-1}$ matrix element extracted from fits to 
eq.~(\ref{eq:psi}) as a function of \qsq, \W\ and \ttra\ are in good agreement 
with the measurements presented in Figs.~\ref{fig:matqsq}$-$\ref{fig:matrelt}
and in Tables~\ref{table:matqsq}$-$\ref{table:matrelt}, which supports the fact 
that SCHC is a good approximation for the present data.

\subsection{Helicity Flip Amplitudes}  
            \label{sect:flip}

\begin{figure}[p]
 \begin{center}
        \epsfig{file=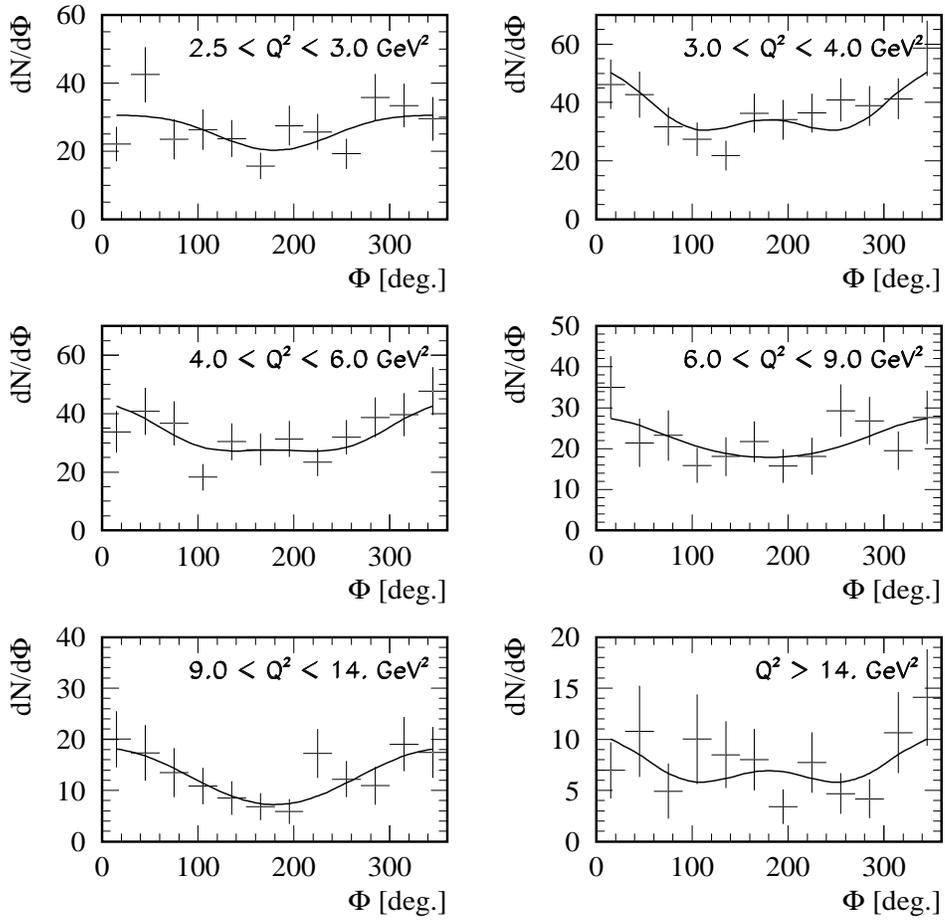,%
         bbllx=46pt,bblly=156pt,bburx=540pt,bbury=667pt,%
         height=13.0cm,width=13.cm}
 \end{center}
\caption{Acceptance corrected distributions of the $\phi$
  angle for the 1996 data sample in six bins in \qsq.
  The curves are the result of fits to the form of eq.~(\ref{eq:WPhi}).
  The errors on the data points are statistical only.}
\label{fig:Phiqsq}
\end{figure}

Distributions of the angle $\phi$ are presented in Fig.~\ref{fig:Phiqsq} 
for six bins in \qsq.
These distributions, as well as the corresponding distributions for bins in 
\w\ and \ttra, exhibit significant variation in $\cos \phi$.
Variation in $\cos2\phi$ is compatible with zero.
They are well described by the function
\begin{equation}
W(\phi) \propto  1 - \varepsilon\ \cos2\phi\ (2 \ \ruuu + \ruzz) + 
   \sqrt{2\varepsilon (1+\varepsilon)} \cos \phi\ (2 \ \rcuu + \rczz) \ ,
                                  \label{eq:WPhi}
\end{equation}
obtained from the integration of the decay angular distribution 
(eq.~\ref{eq:W}) over \cosths\ and $\phib$.

\begin{center}
\begin{figure}[htbp]
\setlength{\unitlength}{1.0cm}
\begin{picture}(13.0,9.0)
\put(1.5,-1.0){\epsfig{file=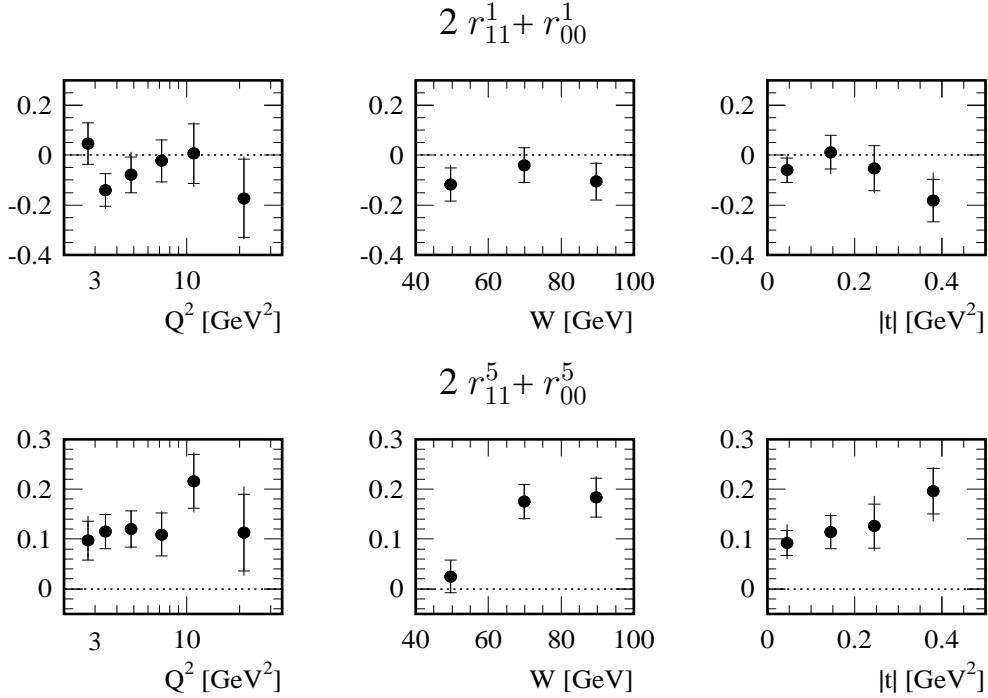,%
            bbllx=70pt,bblly=308pt,bburx=520pt,bbury=672pt,%
            height=10.0cm,width=13.cm}}
\put(7.3,8.9){\large 2 \ruuu + \ruzz}
\put(7.3,4.1){\large 2 \rcuu + \rczz}
\put(2.64,0.7){\footnotesize 3}
\put(2.64,5.5){\footnotesize 3}
\end{picture}
\caption{Measurements of the combinations of matrix elements
  $2 \ruuu + \ruzz$ and $2 \rcuu + \rczz$,
  as a function of \qsq, \W\ and \ttra, obtained from fits to the $\phi$ 
  distributions.
  The inner error bars are statistical and the full error bars include
  the systematic errors added in quadrature.
  The dashed lines indicate the null values which are expected in the case of
  SCHC.}
\label{fig:rfive}
\end{figure}
\end{center}

The combinations of the matrix elements $(2 \ \ruuu +\ruzz)$ and 
$(2 \ \rcuu +\rczz)$, extracted as a function of \qsq,
\W\ and \ttra, are presented in Fig.~\ref{fig:rfive} (and in 
Table~\ref{table:rfive}).
There is no indication for a significant deviation from zero of the combination
$( 2 \ \ruuu +\ruzz)$, which is consistent with the measurements presented in 
Figs.~\ref{fig:matqsq}$-$\ref{fig:matrelt}
and in Tables~\ref{table:matqsq}$-$\ref{table:matrelt}.
In contrast, the combination $( 2 \ \rcuu +\rczz)$ is significantly
different from zero.
As discussed in section~\ref{sect:angles}, this effect is attributed to a 
violation of SCHC for the matrix element \rczz.

As can be deduced from the second column of Table~\ref{Table:amplitudes},
the \rczz\ matrix element is approximately proportional to the amplitude $T_{01}$
for a transverse photon to produce a longitudinal \rh\ meson:
\begin{equation}
 r^{5}_{00} \simeq \frac {\sqrt{2 R}} {1 + \varepsilon R}
                   \frac {|T_{01}|} {|T_{11}|} \ ,
                         \label{eq:rczz}
\end{equation}                   
where the term $|T_{01}|^2$ has been neglected with respect to $|T_{11}|^2$ 
in the denominator and the amplitudes $T_{00}$
and $T_{01}$ are assumed to be in phase and purely imaginary~\cite{ivanov}.

With these approximations and with $\varepsilon \simeq\ 1$, the measurement 
of \rczz\ allows the determination of the ratio of the $T_{01}$ amplitude to 
the non-flip amplitudes $\sqrt{|T_{00}|^2 + |T_{11}|^2}$ for the present
\qsq\ domain:
\begin{eqnarray}
\frac {|T_{01}|}{\sqrt{|T_{00}|^2 + |T_{11}|^2}} 
   \simeq \frac{|T_{01}|}{|T_{11}| \sqrt{1+R}}  & 
   \simeq&  \ r^{5}_{00} \ \sqrt { \frac{1+R}{2 R} }  \\
 & \simeq & 8 \pm 3 \%\ ,
\end{eqnarray}
using the results in Table \ref{Table:moments} and eq. (\ref{eq:R}).
This value is of the order of magnitude, or slightly lower than those found,
with large errors, at lower energy and for 
$\langle Q^2 \rangle \simeq 0.5$ \gevsq\ ($15 - 20\%$ for 
$W \simeq$ 2.5~GeV~\cite{joos} 
and $11 - 14\%$ for $10 < W < 16$~GeV~\cite{CHIO}).

The other helicity flip amplitudes are consistent with zero within the present
measurement precision, as can be deduced from the fact
that among the matrix elements which vanish under SCHC only the \rczz\ 
element is measured to be non-zero.
This is confirmed by the study of the \phib\ distribution.
After integration over \cosths\ and $\phi$, the decay distribution~(\ref{eq:W}) 
reduces to
\begin{equation}
W(\phib) \propto  1 - 2\ \rzqumu \cos2\phib \ .
               \label{eq:Wphi}
\end{equation}


This distribution is compatible with being constant for all bins in
\qsq, $W$ and $t$, supporting the observation that the matrix element \rzqumu\
is consistent with zero.
The expression for this matrix element contains a term proportional to 
$T_{11}T^*_{1-1}$, the interference between the helicity conserving transverse
amplitude and the double-flip amplitude, and a term proportional to the square of 
the single flip contribution $T^2_{10}$ (NPE is assumed). 
The constant \phib\ distributions thus indicate that the helicity amplitudes 
$T_{1-1}$ and $T_{10}$ are compatible with zero.

Another way to study the amplitude $T_{10}$ is to compare the measured values of 
the \ruumu\ and \rzzzz\ matrix elements, which are related by 
\begin{equation}
  \ruumu~= \frac{1}{2} ~(1 - \rzzzz) \ 
               \label{eq:fff}
\end{equation}
if, and only if, $T_{10} = 0$ (NPE is assumed).
Relation~(\ref{eq:fff}) is satisfied within errors for the measurements 
presented in Figs.~\ref{fig:matqsq}$-$\ref{fig:matrelt}
and in Tables~\ref{table:matqsq}$-$\ref{table:matrelt}.

\subsection{Comparison with Models}  
            \label{sect:model}

Numerous models for the electroproduction of vector mesons based on VDM 
or QCD have been proposed.
Most of them predict, for the present \qsq\ domain,
a linear increase with \qsq\ of the ratio $R$ of the
longitudinal to transverse cross sections, in disagreement with the results
presented in Fig.~\ref{fig:R}.

However, several recent models predict a slower increase of $R$
at high \qsq~\cite{ivanov,sss,mrt,cudell}, which corresponds better to the 
trend in the data.
One of them offers in addition full predictions for the spin density matrix 
elements~\cite {ivanov}.
In the rest of this section, we concentrate on the comparison of 
these model predictions with the present measurements.

\subsubsection{Generalised Vector Dominance}
                          \label{sect:modelsss}

A calculation based on the Generalised Vector Dominance Model (GVDM)
has been performed by \ssss~\cite{sss}.
It takes into account a continuous mass spectrum of vector meson states,
with destructive interferences between neighbouring states. This leads
to a non-linear \qsq\ dependence for the ratio $R$, in contrast with
the conventional VDM predictions. 
The ratio $R$ tends asymptotically to a 
constant value, defined by effective transverse and longitudinal masses 
which must be obtained from a fit to experimental data. 
The domain of applicability of the model extends in \qsq\ down to 
photoproduction.

\begin{figure}[t]
  \begin{center}
        \epsfig{file=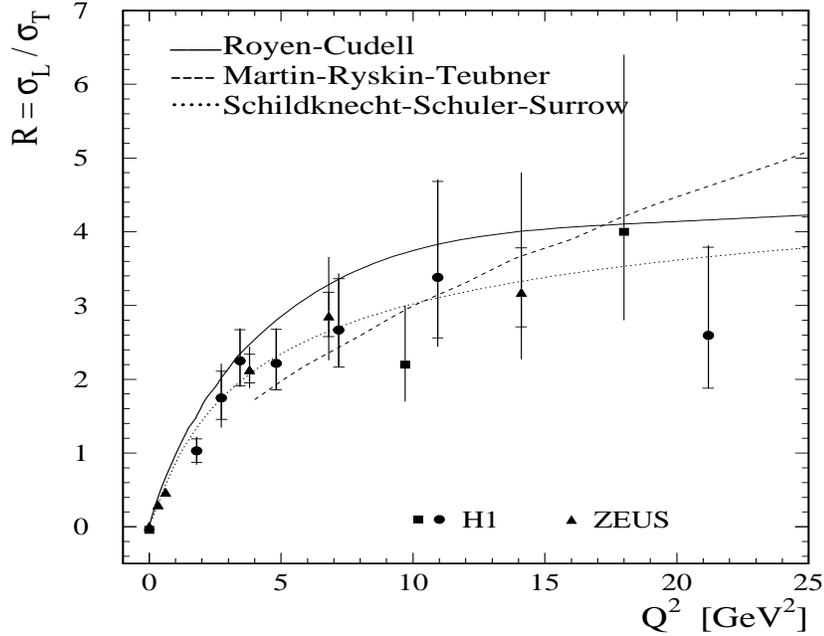,%
         height=9.5cm,width=12.cm}
  \end{center}
\vspace*{-0.2cm}
\caption{The ratio $R$ of the longitudinal to transverse photon cross 
sections for elastic \rh\ meson electroproduction as a function of \qsq. 
The data are the HERA measurements as in Fig.~\ref{fig:R}.
The curves are the predictions of the models of Royen and 
Cudell~\protect\cite{cudell} (solid), of Martin, Ryskin and Teubner 
(dashed)~\protect\cite{mrt} and of \ssss~\protect\cite{sss} (dotted), 
for the HERA energy range.}
\label{fig:rcud}
\end{figure}

In Fig.~\ref{fig:rcud}, the prediction of this model is compared to 
the measurement of $R$ as a function of \qsq, using the best set of parameters 
(``2-par. fit'' in~\cite{sss}).
The data are the HERA measurements presented in Fig.~\ref{fig:R}.

\subsubsection{Parton--Hadron Duality}
                          \label{sect:modelmrt}

Martin, Ryskin and Teubner have 
observed that QCD calculations of the \rh\ cross section that 
convolute the scattering amplitude with the \rh\ wave function give 
transverse cross sections which fall off too quickly with increasing \qsq\
and thus lead to values of $R$ which are too large at high \qsq~\cite{mrt}.
They have proposed an alternative approach, in which open $q \bar q$ production 
is considered in a broad mass interval containing the \rh\ meson.
Hadronisation proceeds predominantly into two pion states, following
phase space considerations. 
The hard interaction is 
modelled through two gluon exchange (or a gluon ladder), which induces a
dependence on the parameterisation of the gluon density in the proton. 
The main uncertainties of the model come from the higher order 
corrections and from the choice of the mass interval embracing 
the \rh\ meson. However, the prediction for the ratio $R$ of the cross 
sections has little sensitivity to these uncertainties.

Fig.~\ref{fig:rcud} presents the prediction of the model of \mrtt, computed
with the MRS(R4) parameterisation~\cite{pdflib,mrs} for the gluon content
of the proton.

\subsubsection{Quark Off-Shellness Model}
                          \label{sect:modelrc}

Another model based on lowest-order perturbative QCD calculations has been
proposed by \rcc~\cite{cudell}.
The \rh\ meson production is computed from the $q\overline{q}$ Fock state of 
the photon, convoluted with the amplitude for hard scattering modelled 
as two-gluon exchange.
A proton form factor and a meson vertex wave function, including Fermi
motion, are part of the calculation.
The specific feature of the model
is that the constituent quarks are allowed to go off-shell. 
The $W$ dependence of the cross section is not predicted, but the
\qsq\ and $t$ dependences are.
The uncertainties of the model come from the choice of the constituent 
quark mass $m_q$ and the Fermi momentum $p_F$.

The prediction of the model of \rcc\ is shown in Fig.~\ref{fig:rcud} for 
$m_q$ = 0.3~GeV and $p_F$ = 0.3~GeV.
When $m_q$ and $p_F$ are varied by $\pm$ 50 MeV, the $R$ value changes by
about 15\% and 30\%, respectively, for \qsq\ = 10~\gevsq.

\subsubsection{Predictions of Polarisation}
                          \label{sect:modelik}

\ikk\ have provided predictions for the full set of 15 elements of the spin 
density matrix, based on perturbative QCD~\cite{ivanov}.
This model predicts a violation of SCHC at high \qsq, the largest 
helicity-flip amplitude being $T_{01}$, with:
\begin{equation}
|T_{00}| > |T_{11}| > |T_{01}| > |T_{10}| > |T_{1-1}|
                                 \label{eq:hierarchy}
\end{equation}
for the HERA kinematical domain. 
The ratios $|T_{11}| / |T_{00}|$, $|T_{01}| / |T_{00}|$ and $|T_{10}| / |T_{00}|$
depend on $t$, \qsq, $M$ and $\gamma$, where $M$ is the invariant mass of the 
$q\overline{q}$ pair and $\gamma$ is the anomalous dimension of the gluon density
($xg(x,\qsq) \propto Q^{2\gamma}$). 
The ratio $|T_{1-1}| / |T_{00}|$ depends also on the gluon density at the scale 
$Q^2/4$.

\begin{center}
\begin{figure}[bp]
\setlength{\unitlength}{1.0cm}
\begin{picture}(13.0,18.0)
\put(0.0,0.0){\epsfig{file=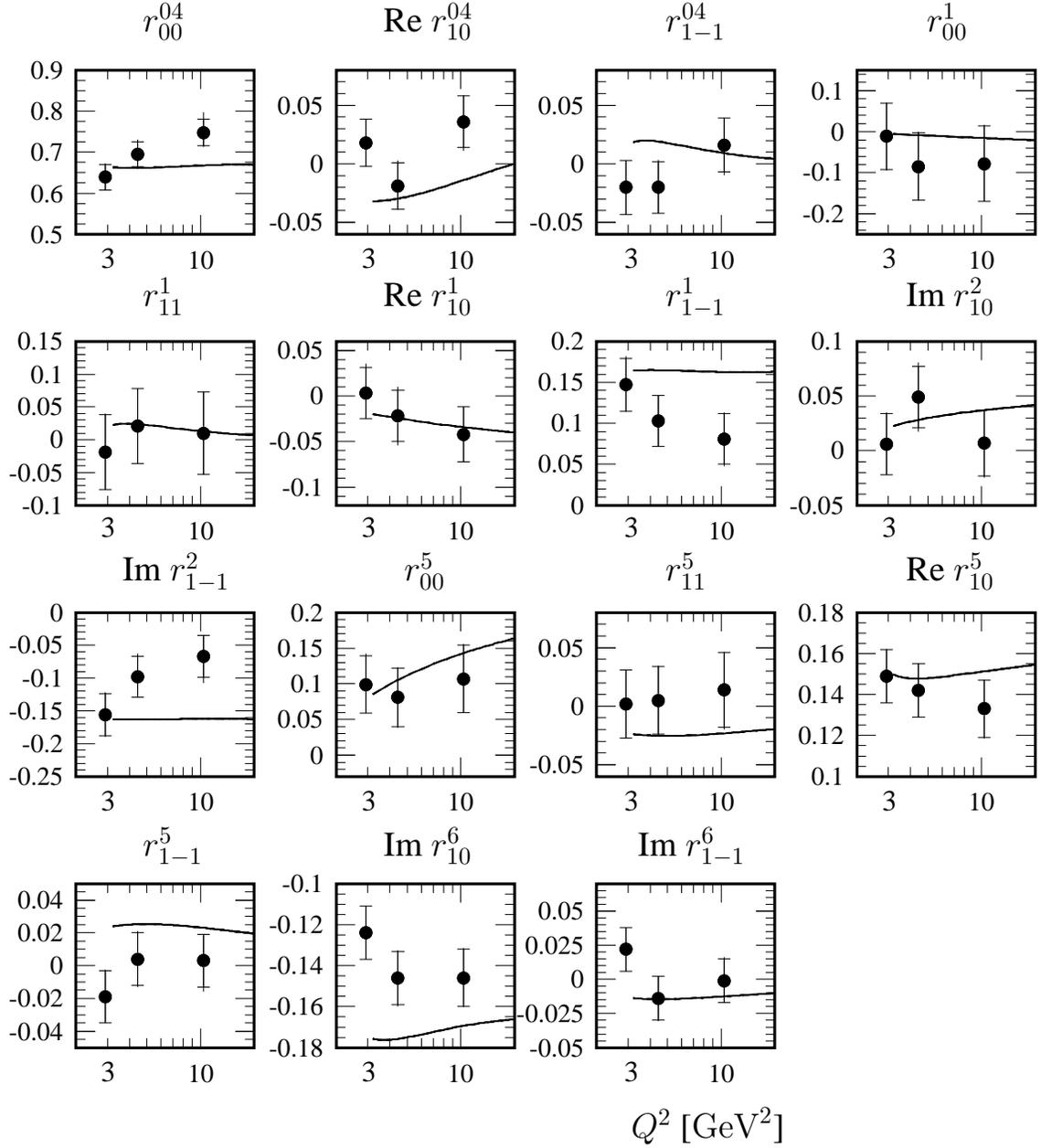,%
            bbllx=55pt,bblly=189pt,bburx=450pt,bbury=693pt
            height=17.0cm,width=13.cm}}
\put(6.7,17.0){\Large H1 data}
\put(2.4,15.6){\large \rzqzz}
\put(5.9,15.6){\large {\rm Re} \rzquz}
\put(10.0,15.6){\large \rzqumu}
\put(13.8,15.6){\large \ruzz}
\put(2.4,11.6){\large \ruuu}
\put(5.9,11.6){\large {\rm Re} \ruuz}
\put(10.0,11.6){\large \ruumu}
\put(13.5,11.6){\large {\rm Im} \rduz}
\put(2.1,7.6){\large {\rm Im} \rdumu}
\put(6.2,7.6){\large \rczz}
\put(10.0,7.6){\large \rcuu}
\put(13.5,7.6){\large {\rm Re} \rcuz}
\put(2.4,3.6){\large \rcumu}
\put(5.9,3.6){\large {\rm Im} \rsuz}
\put(9.6,3.6){\large {\rm Im} \rsumu}
\put(9.5,-0.5){\large \qsq~[\gevsq] }
\put(13.15,4.26){3}
\put(13.15,8.24){3}
\put(13.15,12.15){3}
\put(9.40,0.31){3}
\put(9.40,4.26){3}
\put(9.40,8.24){3}
\put(9.40,12.15){3}
\put(5.58,0.31){3}
\put(5.58,4.26){3}
\put(5.58,8.24){3}
\put(5.58,12.15){3}
\put(1.79,0.31){3}
\put(1.79,4.26){3}
\put(1.79,8.24){3}
\put(1.79,12.15){3}
\end{picture}
\vspace*{1cm}
\caption{Spin density matrix elements for elastic electroproduction 
  of \rh\ mesons, for three values of \qsq. 
  The data are the same as in Fig.~\ref{fig:matqsq}. 
  The curves are the predictions of the model of \ikk~\protect\cite{ivanov} 
  for the GRV 94HO parameterisation of the gluon density in the proton.}
\label{fig:mativa}
\end{figure}
\end{center}

Fig.~\ref{fig:mativa} shows the predicted values of the matrix elements
obtained with the parameterisation GRV 94HO of the gluon density in the proton
\cite{pdflib,grv}, compared to the measurements presented in Fig.~\ref{fig:matqsq}.
This density is assumed to be valid throughout the range of \qsq\ of the data.
For higher \qsq\ values, other parameterisations 
give predictions differing by much less than the measurement uncertainties.
Reasonable agreement of the model predictions with the data is observed,
with a correct prediction of the hierarchy between the amplitudes which are 
measured to be non-zero, and of the magnitude of the matrix element \rczz.

\section{Cross Sections}  \label{sect:cross_section}

\boldmath
\subsection{$t$ Dependence of the $ep$ Cross Section}  
            \label{sect:t}
\unboldmath

\begin{figure}[p]
 \begin{center}
        \epsfig{file=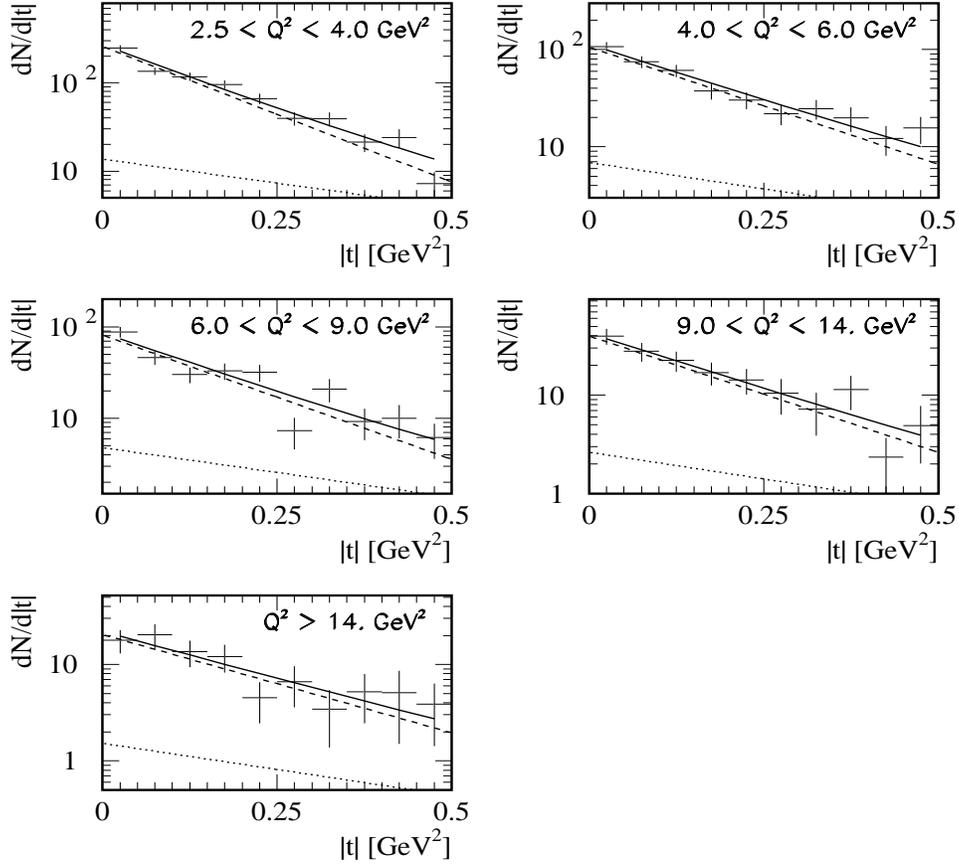,%
         bbllx=37pt,bblly=157pt,bburx=533pt,bbury=694pt,%
         height=13.0cm,width=13.cm}
 \end{center}
\vspace*{-0.2cm}
\caption{Acceptance corrected $t$ distributions for the 1996
data sample, for five bins in \qsq.
The full curves correspond to a fit of the distributions as the sum of
three exponentials, corresponding to the elastic signal (dashed curves), 11\% 
background of proton dissociation events with slope $b_{pd}$ = 2.5~\gevsqm\ 
(dotted), and 1\% non-resonant background with slope $b_{nr}$ = 0.3~GeV$^{-2}$
(not visible on the plots).
The errors on the data points are statistical only.}
\label{fig:t}
\end{figure}

The acceptance corrected $t$ distributions of the selected events with 
$\modt < 0.5~\gevsq$
are presented in Fig.~\ref{fig:t} for five bins in \qsq.
To study the $t$ dependence of elastic \rh\ production, these
distributions are fitted as the sum of three exponentials corresponding
to the elastic component, the diffractive component with proton dissociation
and the non-resonant two-pion background.
The elastic component is fitted with a free slope parameter $b$, whereas the
contribution of diffractive \rh\ events with proton dissociation, which
amounts to $11 \pm 5 \%$ of the elastic signal, has a fixed slope parameter
$b_{pd} = 2.5 \pm 1.0~\gevsqm$ (see section~\ref{sect:pdiss_bg}).\footnote{
  It should be noted that the slope parameter 
  for low mass excited proton states could be larger than in the high mass
  region, from which the parameter $b_{pd}$ is extracted. 
  The corresponding uncertainty is covered by the systematic errors quoted below.}
The non-resonant background, amounting to $1 \pm 1 \%$ of the signal,
also has a fixed slope parameter, $b_{nr}$ = 0.3 $\pm$ 0.1~\gevsqm , extracted 
from the present data at large \modt\ values.

\begin{figure}[p]
 \begin{center}
        \epsfig{file=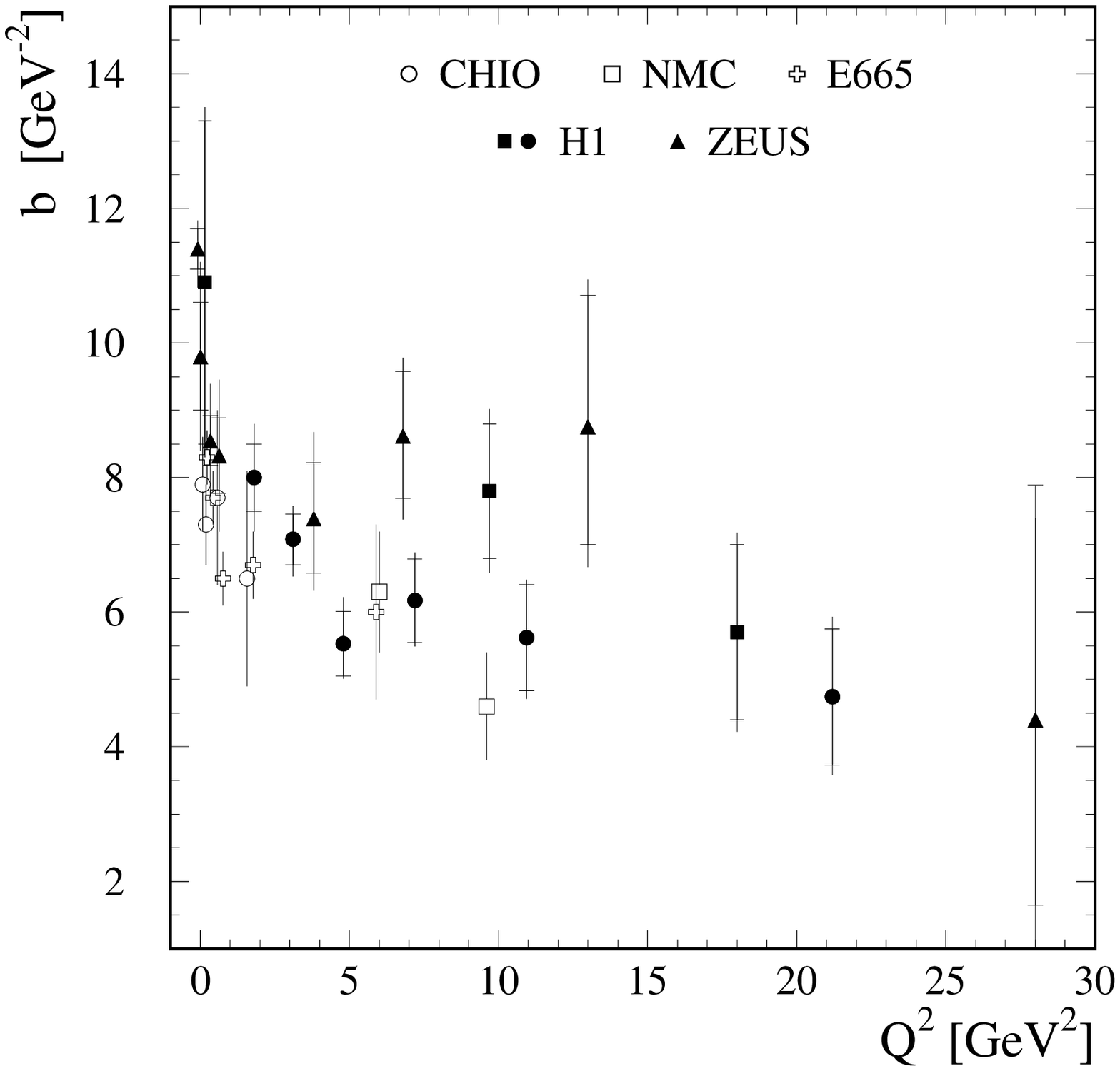,%
         bbllx=21pt,bblly=146pt,bburx=533pt,bbury=661pt,%
         height=10.0cm,width=12.cm}
 \end{center}
\vspace*{-0.6cm}
\caption{Measurement of the slope parameter $b$ of the exponential $t$ dependence
for elastic \rh\ production.
For the present measurements (full circles), the inner error bars are statistical 
and 
the full error bars include the systematic errors added in quadrature.
The other measurements are from H1~\protect\cite{h1_phot} and 
  ZEUS~\protect\cite{zeus_phottt,zeus_phott} in photoproduction, 
  and from CHIO~\protect\cite{CHIO}, NMC~\protect\cite{NMC},
  E665~\protect\cite{E665}, H1~\protect\cite{H1_rho_94} and
  ZEUS~\protect\cite{rho_zeus} in electroproduction.
It should be noted that the definition of the parameter $b$ is not
unique (see text).}
\label{fig:b}
\end{figure}

The fitted exponential slope parameters, $b$, for elastic events are presented as 
a function of \qsq\ in Fig.~\ref{fig:b} (and in Table~\ref{tab:b}), 
together with H1~\cite{H1_rho_94,h1_phot},
ZEUS~\cite{rho_zeus,zeus_phottt,zeus_phott} and fixed target~\cite{CHIO,NMC,E665}
measurements.\footnote{
  For the ZEUS measurements, the definitions of the slope differ somewhat: 
  in the photoproduction case~\cite{zeus_phottt}, the exponent of the $t$ 
  distribution was parameterised in a parabolic form, and only the linear term 
  is plotted here;
  the fit in~\cite{zeus_phott} was restricted to 
  \modt\ $<$ 0.4~\gevsq\ and that in ~\cite{rho_zeus} was performed for 
  \modt\ $<$ 0.3~\gevsq.} 
The systematic errors are computed by varying the parameters of the 
Monte Carlo simulation used for the acceptance corrections 
(see section~\ref{sect:DIFFVM}), by varying the amounts of background
contributions and 
their slopes within the quoted errors, and by varying the binning and the 
limits of the fits.

The present measurements confirm the decrease of $b$ when \qsq\ increases from
photoproduction to the deep-inelastic domain,
presumably reflecting the decrease of the transverse size of the virtual photon.
It is also observed in Fig.~\ref{fig:b} that at low \qsq\ 
(\qsq\ $\lsim$ 2~\gevsq) measurements at HERA lie systematically 
above the low energy fixed target results.
This may indicate shrinkage of the diffractive peak as $W$ increases.
At higher \qsq, given the experimental errors, no significant 
information on a possible shrinkage of the $t$ distribution can be extracted
within the \w\ range of the present experiment.

\begin{figure}[p]
 \begin{center}
        \epsfig{file=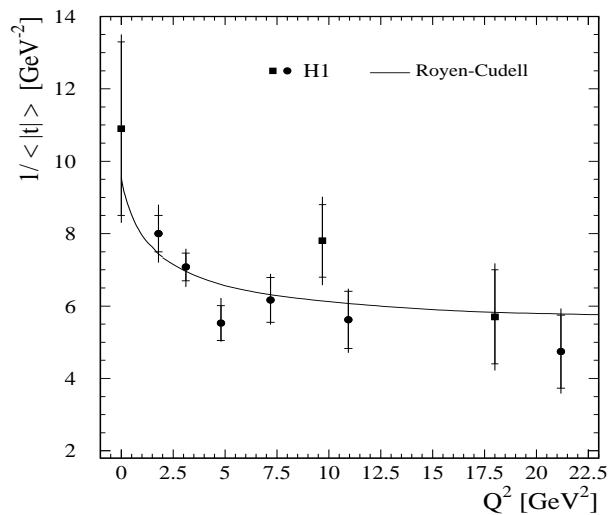,%
         bbllx=21pt,bblly=146pt,bburx=533pt,bbury=661pt,%
         height=7.cm,width=8.cm}
 \end{center}
\vspace*{-0.6cm}
\caption{\qsq\ dependence of the slope parameter $b$ for \rh\ elastic 
production by H1 (these and previous 
measurements~\protect\cite{H1_rho_94,h1_phot}), 
compared to the predictions of the
model of Royen and Cudell~\protect\cite{cudell} for the HERA energy range, 
presented in the form of the variable $1/\langle \modt \rangle$.}
\label{fig:b_cud}
\end{figure}

The \qsq\ evolution of the $t$ distribution in the model of Royen and 
Cudell~\cite{cudell} is compared in 
Fig.~\ref{fig:b_cud} to the present measurements, in the form of the variable 
$1/\langle \modt \rangle$, which coincides with $b$ for an exponential 
distribution.\footnote{
  In the present kinematic domain, the integration limits of \modt, 
  $|t|_{min}$ and $|t|_{max}$, are such that 
  $|t|_{min} \approx 0 \ll\langle \modt \rangle$ and $|t|_{max}$ $\gg$ $\langle \modt \rangle$.}
The trend of the data is reproduced.

\boldmath
\subsection{\qsq\ Dependence of the $\gamma^*p$ Cross Section}  
            \label{sect:qsq}
\unboldmath

The $\gamma^*p$ cross section for \rh\ elastic production is extracted from the
$ep$ cross section using the relation:
\begin{eqnarray}
   \frac { {\rm d}^2 \sigma (e p \rightarrow e \rho p)}
         { {\rm d} y \  {\rm d} \qsq} =
   \Gamma\ \sigma (\gamma^* p \rightarrow \rho p) =
   \Gamma\ \sigma_T (\gamma^* p \rightarrow \rho p) \ (1 + \varepsilon R) \ ,
                                                      \label{eq:gamma*p}
\end{eqnarray}
where $\Gamma$ is the flux of virtual photons~\cite{hand}, given by:
\begin{equation}
   \Gamma = \frac {\alpha \ (1 - y + y^2 / 2) }
          {\pi \ y \ \qsq} \ ,                        \label{eq:gamma}
\end{equation}
$\alpha$ being the fine structure constant.
The flux is integrated over each kinematic domain using the measured 
\qsq\ and \W\ dependences of the $\gamma^*p$ cross section.

\begin{figure}[p]
 \begin{center}
        \epsfig{file=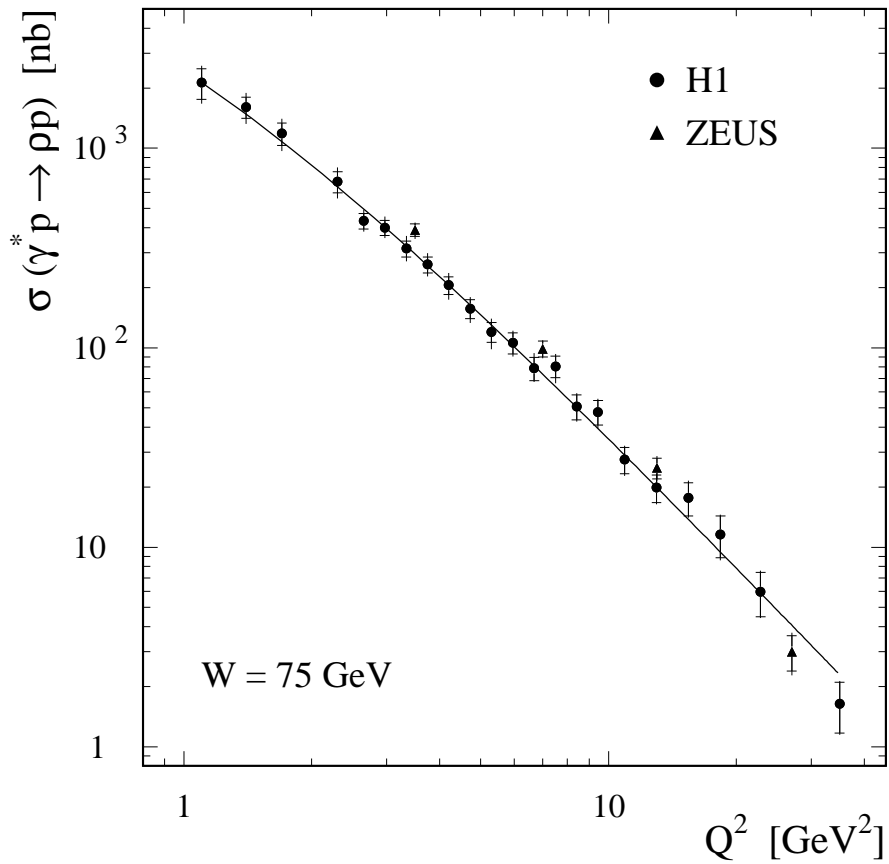,%
         height=13.0cm,width=13.cm}
\caption{Cross section measurements for the process 
$\gamma^* p \rightarrow \rho p$
plotted as a function of \qsq\ for $W = 75~\gev$ (the ZEUS 
measurements~\protect\cite{rho_zeus}
have been scaled to $W = 75~\gev$.
The inner error bars are statistical and the full error bars include
the systematic errors added in quadrature.
The curve corresponds to a fit to the present data of the form of 
eq.~(\ref{eq:qsqdistr}), with $n = 2.24$.}
\label{fig:s_qsq}
 \end{center}
\end{figure}

The $\gamma^*p$ cross section is presented in Fig.~\ref{fig:s_qsq} 
(and in Table~\ref{tab:s_qsq}) as a 
function of \qsq, for a common value $W = 75~\gev$.
It is obtained from the fits described in section~\ref {sect:mass}, which
take into account the \qsq\ dependent skewing of the \mpp\ mass distribution.
The cross section is quoted for a relativistic Breit-Wigner distribution of 
the \rh\ mass, described by eqs. (\ref{eq:b_w}) and (\ref{eq:GJ}), for the 
mass interval
\begin{equation}
   2 \ m_{\pi} \leq \mpp \leq \mrho \ + 5 \ \grho \ .  \label{eq:window}
\end{equation}

The use of two alternative forms to eq. (\ref{eq:GJ}) in parameterising the 
width \Gmpp~\cite{jackson} would cause an increase of the cross section 
by 5\%, which is included in the systematic errors.
The background contributions of \rh\ diffractive production with proton 
dissociation, of \om\ and \ph\ elastic production, and the non-resonant 
background are subtracted assuming the same distribution in \qsq\ as for
the signal.
The uncertainties in these backgrounds are included in the systematic errors.
The \qsq\ dependent losses induced by the $\modt < 0.5~\gevsq$ cut are
corrected for on a bin-by-bin basis, according to the measured $b$ slope
parameters (see section~\ref{sect:t}).
The data are corrected for the losses of events due to noise in the detectors
FMD and PRT (5 $\pm$ 3\%) and LAr (10 $\pm$ 3\%).
Acceptance and efficiency effects and their errors are determined as described 
in section~\ref{sect:DIFFVM}.
The errors on the extrapolations of the cross sections to the common
value $W = 75~\gev$ and to the quoted \qsq\ values are estimated 
by varying the assumed $W$ and \qsq\ dependences of the cross section according 
to the limits of the present measurements. 
The radiative corrections are very small for the chosen value of the
\eminpz\ cut and for the procedure used to compute the kinematic variables 
(see section~\ref{sect:kin_var}); 
an error of 4\% accounts for the relevant uncertainties in the \qsq\ 
and \w\ dependences of the cross section, for higher order processes,
and for detector effects not simulated in detail.
The systematic errors on the cross section measurements also include an 
uncertainty of 2\% in the luminosity, and the uncertainties due to limited
Monte Carlo statistics.

A parameterisation of the \qsq\ dependence of the cross section in the form
\begin{equation}
 \sigma (\gsp)  \propto \frac {1} {(\qsq + m^2_{\rh})^n}
                                                 \label{eq:qsqdistr}
\end{equation}
is shown superimposed on Fig.~\ref{fig:s_qsq}.
It is obtained by a fit to the present data with the result 
\begin{equation}
   n = 2.24 \pm 0.09 \ .
                                                 \label{eq:nqsqdistr}
\end{equation}
The uncertainty on this value is determined using the statistical and the 
non-correlated systematic errors only.
The nominal normalisations are used for the 1995 and 1996 data sets, which 
agree within one standard deviation.
The quality of the fit for the full \qsq\ range
$1 \leq \qsq \leq 35~\gevsq$ is good: $\chi^2 / {\rm ndf}$ = 13.3 / 20.

Fig.~\ref{fig:s_qsq} presents in addition the measurements of the ZEUS 
collaboration~\cite{rho_zeus}, scaled to the value W = 75~GeV.
Agreement is observed between the results of the two experiments.

\begin{figure}[t]
 \begin{center}
        \epsfig{file=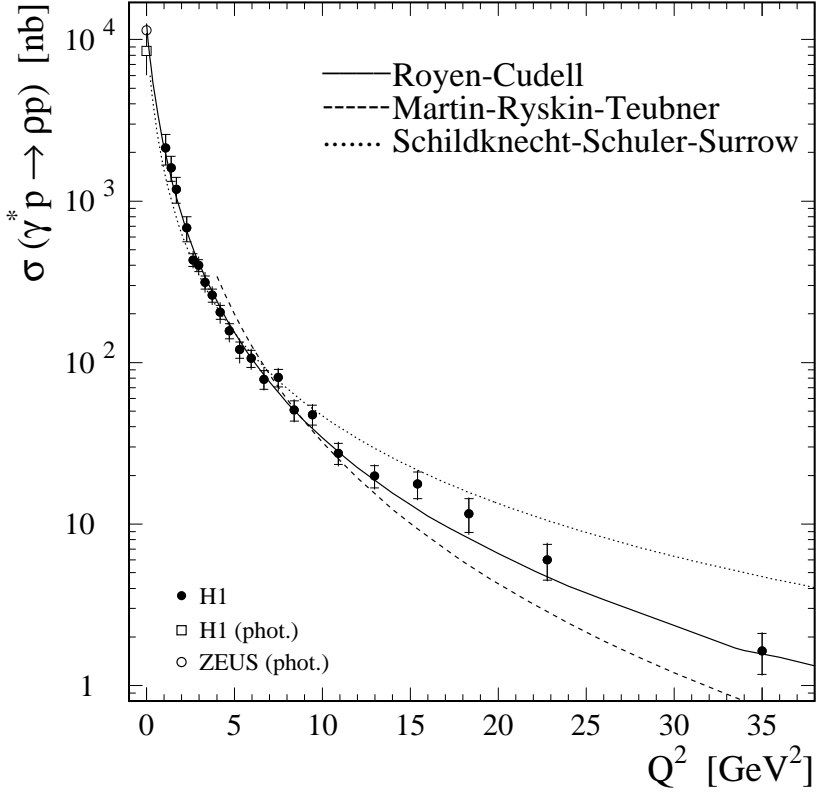,%
         height=12.0cm,width=12.cm}
\caption{Cross section for the process 
$\gamma^* p \rightarrow \rho p$,
plotted as a function of \qsq\ for $W = 75~\gev$ (the data are the same
as in Fig.~\ref{fig:s_qsq}). 
Photoproduction measurements by H1~\protect\cite{h1_phot} and 
ZEUS~\protect\cite{zeus_phottt} are also shown.
The curves are the predictions
of the models of Royen and Cudell~\protect\cite{cudell} (solid), of 
Martin, Ryskin and Teubner~\protect\cite{mrt} (dashed)
and of \ssss~\protect\cite{sss} (dotted).}
\label{fig:qsqcud}
 \end{center}
\end{figure}

In Fig.~\ref{fig:qsqcud}, the \qsq\ dependence of the  $\gamma^*p$ 
cross section, including photoproduction 
measurements~\cite{h1_phot,zeus_phottt}, is compared with the predictions of 
the models of \ssss~\cite{sss}, of Martin, Ryskin and Teubner~\cite{mrt} and 
of Royen and Cudell~\cite{cudell}. 
The latter model describes the data well down to the photoproduction region.

\newpage

\boldmath
\subsection{\w\ Dependence of the $\gamma^*p$ Cross Section}  
            \label{sect:w}
\unboldmath

\begin{figure}[htbp]
 \begin{center}
        \epsfig{file=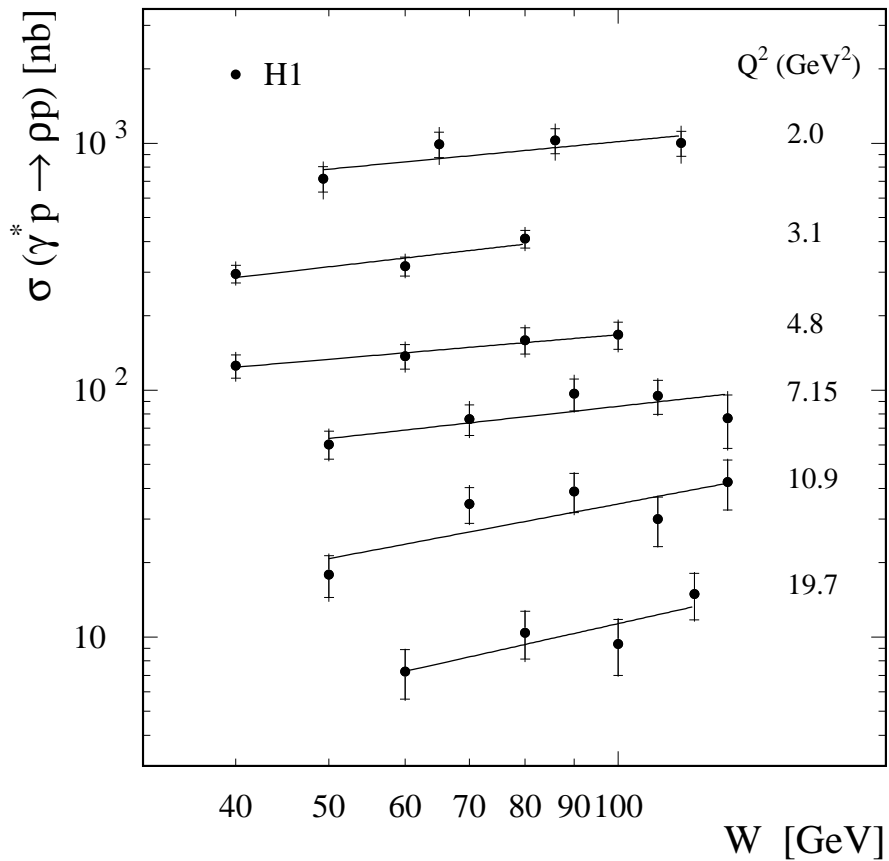,%
         height=13.0cm,width=13.cm}
\caption{Cross section for
the process $\gamma^* p \rightarrow \rho p$ as a function of $W$ for several 
values of \qsq.
The inner error bars are statistical and the full error bars include
the systematic errors added in quadrature.
The lines correspond to a fit of the form of eq.~(\ref{eq:wdistrdelta}).}
\label{fig:w}
 \end{center}
\end{figure}

The $\gamma^*p$ cross section for \rh\ elastic production is presented as a 
function of \w\ for six values of \qsq\ in Fig.~\ref{fig:w} (and in 
Table~\ref{tab:w}).
The extrapolations of the measured cross sections to the chosen  
\qsq\  values are performed using the \qsq\ dependence given by
eqs.~(\ref{eq:qsqdistr}) and (\ref{eq:nqsqdistr}).
Corrections and systematic errors are determined as described in 
section~\ref{sect:qsq}.

To quantify the $W$ dependence of the cross section,
a fit is performed for each \qsq\ bin to a power law:
\begin{equation}
 \sigma (\gsp)  \propto W^{\delta} \ ,
                                                 \label{eq:wdistrdelta}
\end{equation}
as shown in Fig.~\ref{fig:w}.
Only the statistical and the non-correlated systematic errors are used in the 
fits, and the values of $\chi^2 / {\rm ndf}$ are reasonable for all \qsq\ bins.


In a Regge context, the parameter $\delta$ can be related to the exchange
trajectory:\footnote{
   Strictly speaking, this applies if the \W\ dependence of the integrated
   cross section $\int {\rm d}\sigma / {\rm d}t \ {\rm d}t$ is the same, over the 
   relevant \W\ domain, as the \W\ dependence of the differential cross section 
   ${\rm d}\sigma / {\rm d}t$ for $t = \langle t \rangle$.}
\begin{equation}
 \delta \simeq 4 \ [ \alpha ( \langle t \rangle )  - 1] \ .
                                                 \label{eq:wdistr}
\end{equation}
The trajectory is assumed to take a linear form:
\begin{eqnarray}
   \alpha(t) &=& \alpha(0) + \alpha^\prime \ t \ .
                                                 \label{eq:alphaprim}
\end{eqnarray}

\begin{figure}[htbp]
\begin{center}
\setlength{\unitlength}{1.0cm}
\begin{picture}(12.0,7.0)
\put(0.0,0.0){\epsfig{file=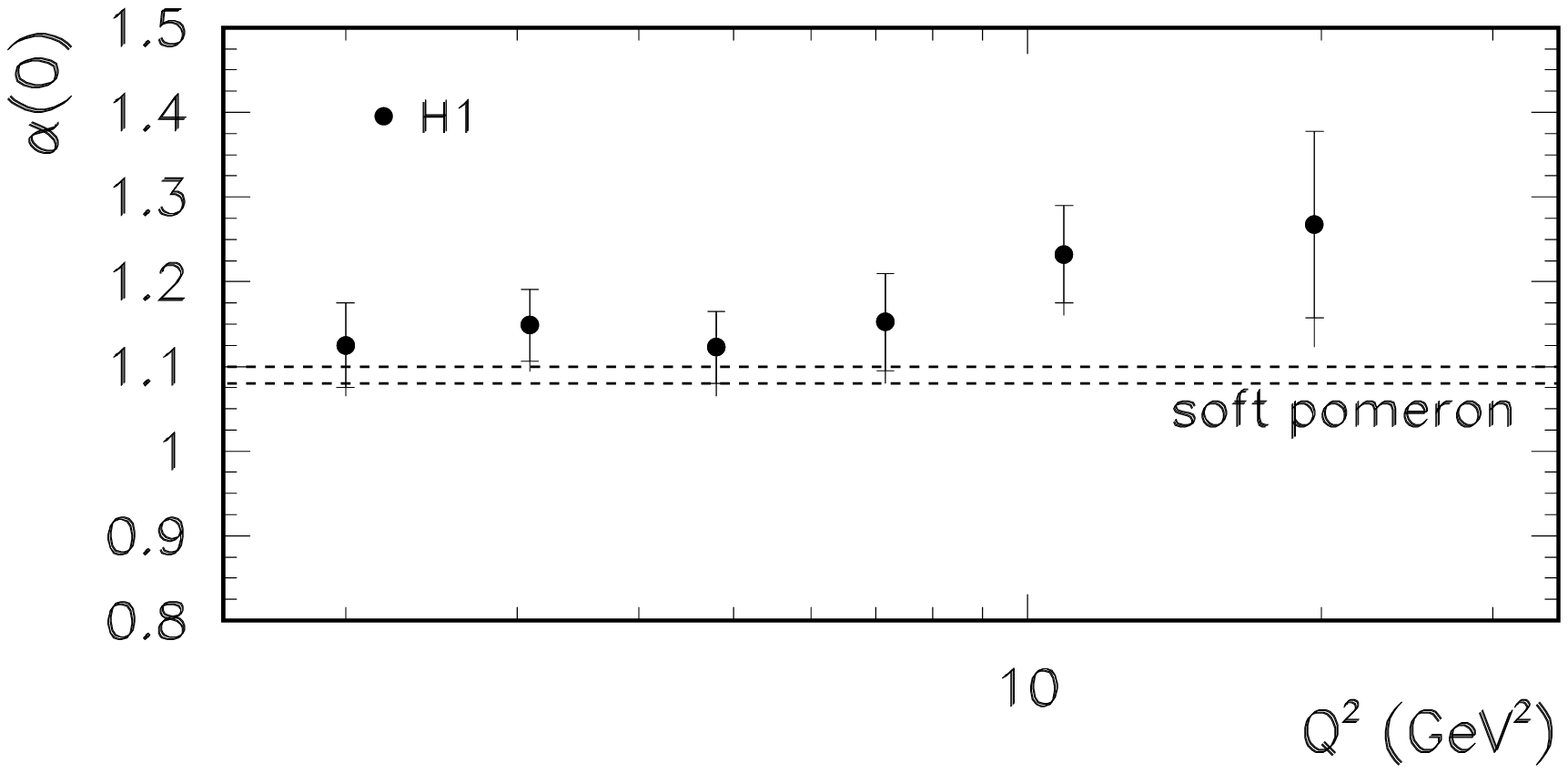,%
         bbllx=38pt,bblly=393pt,bburx=533pt,bbury=661pt,%
         height=7.0cm,width=12.cm}}
\put(2.35,1.0){\bf 2}
\end{picture}
\vspace*{-0.5cm}
\caption{\qsq\ dependence of the intercept $\alpha(0)$ (see 
eqs.~\ref{eq:wdistrdelta} $-$ \ref{eq:alphaprim}).
The inner error bars represent the statistical and non-correlated systematic 
uncertainties on the cross section measurements, the outer error bars include the 
variation of the intercept $\alpha(0)$ when assuming $\alpha^\prime = 0$, added in 
quadrature.
The dashed lines represent the range of values obtained for the ``soft pomeron" 
intercept, as derived from fits to total and elastic hadron$-$hadron cross 
section measurements~\protect\cite{dola,cudellfit}.}
\label{fig:epsilon}
\end{center}
\end{figure}

To extract the effective trajectory intercept $\alpha(0)$, 
$\langle \modt \rangle$ = 1/$b$  is taken from the measured values (see 
section~\ref{sect:t}).
In the absence of a measurement of the \qsq\ dependence of the shrinkage 
of the $t$ distribution with increasing $W$, 
the value $\alpha^\prime = 0.25~\gevsqm$ is assumed, as measured in 
hadron$-$hadron interactions~\cite{dola}.
The values obtained for the intercept $\alpha(0)$ as a function of 
\qsq\ are shown in Fig.~\ref{fig:epsilon} (and in Table~\ref{tab:epsilon}). 
The inner error bars come from the statistical and non-correlated 
systematic uncertainties on the cross section measurements. 
The sensitivity to the choice of $\alpha^\prime$ is shown by the outer bars,
which contain the variation due to the assumption $\alpha^\prime = 0$
(i.e. no shrinkage) added in quadrature.
The measurements are compared to the values $1.08 - 1.10$ obtained from fits to
the total and elastic hadron--hadron cross sections~\cite{dola,cudellfit}.
They suggest that the intercept of the effective trajectory 
governing high \qsq\ \rh\ electroproduction is larger than that describing 
elastic and total hadronic cross sections.


It should be noted that several studies (see e.g.~\cite{fks}) indicate that  
QCD based predictions for the $W$ dependence of the $\gamma^* p$ cross section
are affected by large uncertainties.
These are related particularly to the assumptions made concerning the 
appropriate factorisation scale and the \rh\ wave function, and also to the 
choice of the parameterisation of the gluon distribution in the proton.

\newpage

\section{Summary and Conclusions} \label{sect:summary}

The elastic electroproduction of \rh\ mesons has been studied at HERA
with the H1 detector, for $1~< Q^2 < 60$~GeV$^2$ and 30 $<$ \W\ $<$ 
140~GeV. 

The shape of the ($\pi\pi$) mass distribution has been studied as a function of
\qsq.
It indicates significant skewing at low \qsq, which gets smaller with increasing 
\qsq.

The full set of 15 elements of the \rh\ spin density matrix has been
measured as a function of \qsq, \W\ and \ttra, using the decay angular 
distributions defined in the helicity frame. 
Except for a small but significant deviation from zero of the
\rczz\ element, $s$-channel helicity conservation is found to be a good 
approximation. 
For \qsq\ $\gsim$ 2~\gevsq, the longitudinal $\gamma^*p$
cross section becomes larger than the transverse cross section, and the ratio
$R$ reaches the value $R$ $\simeq$ 3 for \qsq\ $\simeq$ 20~\gevsq.
The phase $\delta$ between the longitudinal and transverse amplitudes is measured 
to be $\cos\delta$ = 0.93 $\pm$ 0.03, assuming natural parity exchange and 
$s$-channel helicity conservation. 
The dominant helicity flip amplitude 
$T_{ \lambda_{\rho} = 0, \lambda_{\gamma} = 1}$
is found to be $8 \pm 3 \%$ of the non-flip amplitudes.
A model based on GVDM~\cite{sss} and models based on
perturbative QCD~\cite{mrt,cudell} reproduce the flattening of the ratio 
$R$ observed at high \qsq. 
A QCD based prediction~\cite{ivanov} is in qualitative 
agreement with the measurement of the 15 matrix elements, in that it 
reproduces the observed hierarchy between the amplitudes which are measured 
to be non-zero and the magnitude of the matrix element \rczz.

The \ttra\ distribution for \rh\ electroproduction has been studied and the
exponential slope parameter $b$ is found to decrease when \qsq\ increases from
photoproduction to the deep-inelastic domain.

The $\gamma^*p \rightarrow \rho p$ cross section has been measured over the 
domain 1 $<$ \qsq\ $<$ 35~GeV$^2$ and follows a \qsq\
dependence of the form $1/(\qsq\ + m_{\rho}^2)^n$, with $n$ = 2.24 $\pm$ 0.09.
This dependence is well described
by a model based on QCD~\cite{cudell}.

The \W\ dependence of the $\gamma^*p \rightarrow \rho p$ cross section has 
been measured for six values of \qsq. 
The measurements suggest that the intercept of the effective trajectory 
governing high \qsq\ \rh\ electroproduction is larger than that describing 
elastic and total hadronic cross sections.

\section*{Acknowledgements}
We are grateful to the HERA machine group whose outstanding efforts
made this experiment possible. We appreciate the immense effort of the
engineers and technicians who constructed and maintained the detector.
We thank the funding agencies for their financial support of the
experiment. We wish to thank the DESY directorate for the support
and hospitality extended to the non-DESY members of the collaboration.
We thank further J.-R.~Cudell, D.Yu.~Ivanov, I.~Royen and T.~Teubner for useful 
discussions and for providing us with their model predictions.

\newpage

{\Large\normalsize}

%

\newpage


\vspace*{6.0cm}
\begin{table}[htbp]
    \centering
    \begin{tabular}{|c|c|r|r|r|}
    \hline
    \hline
      & Element & 2.5 $<$ \qsq\ $<$ 3.5 \gevsq & 3.5 $<$ \qsq\ $<$ 6.0 \gevsq
& 6.0 $<$ \qsq\ $<$ 60 \gevsq \\
    \hline

 1 & \rzqzz    &  0.639 $\pm$ 0.031 \mbig{$^{+0.013}_{-0.010}$} \ 
 &  0.695 $\pm$ 0.031 \mbig{$^{+0.019}_{-0.018}$} \ 
 &  0.748 $\pm$ 0.033 \mbig{$^{+0.037}_{-0.011}$} \   \\

 2 & Re \rzquz &  0.018 $\pm$ 0.020 \mbig{$^{+0.004}_{-0.004}$} \ 
 & -0.019 $\pm$ 0.020 \mbig{$^{+0.009}_{-0.003}$} \ 
 &  0.036 $\pm$ 0.022 \mbig{$^{+0.006}_{-0.012}$} \   \\

 3 & \rzqumu   & -0.020 $\pm$ 0.023 \mbig{$^{+0.002}_{-0.003}$} \ 
 & -0.020 $\pm$ 0.022 \mbig{$^{+0.008}_{-0.001}$} \ 
 &  0.016 $\pm$ 0.023 \mbig{$^{+0.003}_{-0.012}$} \   \\

 4 & \ruzz     & -0.011 $\pm$ 0.081 \mbig{$^{+0.013}_{-0.022}$} \ 
 & -0.085 $\pm$ 0.082 \mbig{$^{+0.021}_{-0.013}$} \ 
 & -0.078 $\pm$ 0.092 \mbig{$^{+0.024}_{-0.002}$} \   \\

 5 & \ruuu     & -0.019 $\pm$ 0.057 \mbig{$^{+0.016}_{-0.003}$} \ 
 &  0.021 $\pm$ 0.057 \mbig{$^{+0.001}_{-0.006}$} \ 
 &  0.010 $\pm$ 0.063 \mbig{$^{+0.001}_{-0.009}$} \   \\

 6 & Re \ruuz  &  0.003 $\pm$ 0.028 \mbig{$^{+0.016}_{-0.003}$} \ 
 & -0.022 $\pm$ 0.028 \mbig{$^{+0.015}_{-0.016}$} \ 
 & -0.042 $\pm$ 0.030 \mbig{$^{+0.005}_{-0.007}$} \   \\ 

 7 & \ruumu    &  0.147 $\pm$ 0.032 \mbig{$^{+0.013}_{-0.007}$} \ 
 &  0.103 $\pm$ 0.031 \mbig{$^{+0.003}_{-0.005}$} \ 
 &  0.081 $\pm$ 0.031 \mbig{$^{+0.010}_{-0.013}$} \   \\

 8 & Im \rduz  &  0.006 $\pm$ 0.028 \mbig{$^{+0.008}_{-0.004}$} \ 
 &  0.049 $\pm$ 0.028 \mbig{$^{+0.015}_{-0.015}$} \ 
 &  0.007 $\pm$ 0.030 \mbig{$^{+0.005}_{-0.011}$} \   \\

 9 & Im \rdumu & -0.156 $\pm$ 0.032 \mbig{$^{+0.014}_{-0.010}$} \ 
 & -0.098 $\pm$ 0.031 \mbig{$^{+0.018}_{-0.007}$} \ 
 & -0.067 $\pm$ 0.032 \mbig{$^{+0.007}_{-0.025}$} \   \\

10 & \rczz     &  0.099 $\pm$ 0.040 \mbig{$^{+0.016}_{-0.002}$} \ 
 &  0.081 $\pm$ 0.041 \mbig{$^{+0.009}_{-0.013}$} \ 
 &  0.107 $\pm$ 0.047 \mbig{$^{+0.015}_{-0.004}$} \   \\

11 & \rcuu     & 0.002 $\pm$ 0.029 \mbig{$^{+0.002}_{-0.007}$} \ 
 & 0.005 $\pm$ 0.029 \mbig{$^{+0.006}_{-0.006}$} \ 
 &  0.014 $\pm$ 0.032 \mbig{$^{+0.006}_{-0.012}$} \   \\

12 & Re \rcuz  &  0.149 $\pm$ 0.013 \mbig{$^{+0.002}_{-0.002}$} \ 
 &  0.142 $\pm$ 0.013 \mbig{$^{+0.003}_{-0.002}$} \ 
 &  0.133 $\pm$ 0.014 \mbig{$^{+0.003}_{-0.004}$} \   \\ 

13 & \rcumu    & -0.019 $\pm$ 0.016 \mbig{$^{+0.005}_{-0.001}$} \ 
 &  0.004 $\pm$ 0.016 \mbig{$^{+0.006}_{-0.006}$} \ 
 &  0.003 $\pm$ 0.016 \mbig{$^{+0.004}_{-0.010}$} \   \\

14 & Im \rsuz  & -0.124 $\pm$ 0.013 \mbig{$^{+0.001}_{-0.001}$} \ 
 & -0.146 $\pm$ 0.013 \mbig{$^{+0.004}_{-0.005}$} \ 
 & -0.146 $\pm$ 0.014 \mbig{$^{+0.004}_{-0.002}$} \   \\

15 & Im \rsumu &  0.022 $\pm$ 0.016 \mbig{$^{+0.001}_{-0.002}$} \ 
 & -0.014 $\pm$ 0.016 \mbig{$^{+0.003}_{-0.004}$} \ 
 &  -0.001 $\pm$ 0.016 \mbig{$^{+0.006}_{-0.003}$} \   \\

 \hline
 \hline
\end{tabular}
\caption{Spin density matrix elements for elastic electroproduction 
  of \rh\ mesons, measured for three values of \qsq\ with the 1996 data sample. 
  The first errors are statistical, the second systematic.}
\label{table:matqsq}
\end{table}
%
%
\begin{table}[htbp]
    \centering
    \begin{tabular}{|c|c|r|r|r|}
    \hline
    \hline
      & Element & 40 $<$ \W\ $<$ 60 \gev & 60 $<$ \W\ $<$ 80 \gev
& 80 $<$ \W\ $<$ 100 \gev \\
    \hline
 1 & \rzqzz    &  0.671 $\pm$ 0.031 \mbig{$^{+0.031}_{-0.025}$} 
 &  0.719 $\pm$ 0.031 \mbig{$^{+0.051}_{-0.033}$}  & 
   0.687 $\pm$ 0.033 \mbig{$^{+0.031}_{-0.016}$} \\
 2 & Re \rzquz & -0.011 $\pm$ 0.020 \mbig{$^{+0.010}_{-0.007}$} 
 &  0.025 $\pm$ 0.020 \mbig{$^{+0.011}_{-0.009}$}  &
   0.052 $\pm$ 0.021 \mbig{$^{+0.006}_{-0.003}$} \\
 3 & \rzqumu   & -0.021 $\pm$ 0.023 \mbig{$^{+0.005}_{-0.006}$} 
 &  0.000 $\pm$ 0.023 \mbig{$^{+0.011}_{-0.010}$}  &
  -0.028 $\pm$ 0.024 \mbig{$^{+0.010}_{-0.004}$} \\
 4 & \ruzz     & -0.048 $\pm$ 0.081 \mbig{$^{+0.021}_{-0.019}$} 
 & -0.151 $\pm$ 0.082 \mbig{$^{+0.020}_{-0.010}$}  &
   0.043 $\pm$ 0.089 \mbig{$^{+0.029}_{-0.026}$} \\
 5 & \ruuu     & -0.013 $\pm$ 0.057 \mbig{$^{+0.008}_{-0.009}$} 
 &  0.080 $\pm$ 0.057 \mbig{$^{+0.003}_{-0.006}$}  &
  -0.060 $\pm$ 0.062 \mbig{$^{+0.021}_{-0.024}$} \\
 6 & Re \ruuz  & -0.002 $\pm$ 0.028 \mbig{$^{+0.006}_{-0.006}$} 
 & -0.018 $\pm$ 0.028 \mbig{$^{+0.013}_{-0.022}$}  &
  -0.023 $\pm$ 0.030 \mbig{$^{+0.022}_{-0.027}$} \\
 7 & \ruumu    &  0.225 $\pm$ 0.031 \mbig{$^{+0.002}_{-0.005}$} 
 &  0.113 $\pm$ 0.031 \mbig{$^{+0.007}_{-0.010}$} &
   0.083 $\pm$ 0.033 \mbig{$^{+0.052}_{-0.044}$} \\
 8 & Im \rduz  & -0.030 $\pm$ 0.028 \mbig{$^{+0.012}_{-0.008}$} 
 &  0.105 $\pm$ 0.028 \mbig{$^{+0.006}_{-0.008}$} &
   0.032 $\pm$ 0.030 \mbig{$^{+0.009}_{-0.006}$} \\
 9 & Im \rdumu & -0.132 $\pm$ 0.032 \mbig{$^{+0.016}_{-0.010}$} 
 & -0.151 $\pm$ 0.031 \mbig{$^{+0.013}_{-0.009}$} &
  -0.068 $\pm$ 0.033 \mbig{$^{+0.020}_{-0.024}$} \\
10 & \rczz     &  0.030 $\pm$ 0.041 \mbig{$^{+0.003}_{-0.003}$} 
 &  0.192 $\pm$ 0.041 \mbig{$^{+0.033}_{-0.013}$} &
   0.114 $\pm$ 0.045 \mbig{$^{+0.007}_{-0.002}$} \\
11 & \rcuu     & -0.009 $\pm$ 0.029 \mbig{$^{+0.004}_{-0.001}$} 
 & -0.015 $\pm$ 0.029 \mbig{$^{+0.004}_{-0.010}$} &
   0.027 $\pm$ 0.031 \mbig{$^{+0.007}_{-0.008}$} \\
12 & Re \rcuz  &  0.177 $\pm$ 0.013 \mbig{$^{+0.015}_{-0.009}$} 
 &  0.118 $\pm$ 0.013 \mbig{$^{+0.011}_{-0.009}$} &
   0.125 $\pm$ 0.014 \mbig{$^{+0.009}_{-0.009}$} \\
13 & \rcumu    & -0.014 $\pm$ 0.016 \mbig{$^{+0.004}_{-0.008}$} 
 & -0.005 $\pm$ 0.016 \mbig{$^{+0.007}_{-0.002}$} &
  -0.018 $\pm$ 0.017 \mbig{$^{+0.007}_{-0.006}$} \\
14 & Im \rsuz  & -0.148 $\pm$ 0.013 \mbig{$^{+0.003}_{-0.004}$} 
 & -0.160 $\pm$ 0.013 \mbig{$^{+0.005}_{-0.006}$} &
  -0.115 $\pm$ 0.014 \mbig{$^{+0.013}_{-0.031}$} \\
15 & Im \rsumu & -0.005 $\pm$ 0.016 \mbig{$^{+0.002}_{-0.002}$} 
 &  0.021 $\pm$ 0.016 \mbig{$^{+0.002}_{-0.002}$} &
  -0.014 $\pm$ 0.017 \mbig{$^{+0.009}_{-0.011}$} \\
 \hline
 \hline
\end{tabular}
\caption{Spin density matrix elements for elastic electroproduction 
  of \rh\ mesons, measured for three values of \w\ with the 1996 data sample. 
  The first errors are statistical, the second systematic.}
\label{table:matrelw}
\end{table}
%
%
%
\begin{table}[htbp]
    \centering
    \begin{tabular}{|c|c|r|r|r|}
    \hline
    \hline
      & Element & 0.0 $<$ $|t|$ $<$ 0.1 \gevsq & 0.1 $<$ $|t|$ $<$ 0.2 \gevsq
& 0.2 $<$ $|t|$ $<$ 0.5 \gevsq \\
    \hline

 1 & \rzqzz    &  0.686 $\pm$ 0.031 \mbig{$^{+0.050}_{-0.039}$} \ 
 &  0.706 $\pm$ 0.031 \mbig{$^{+0.059}_{-0.041}$} \ 
 &  0.634 $\pm$ 0.033 \mbig{$^{+0.042}_{-0.032}$} \  \\

 2 & Re \rzquz &  0.010 $\pm$ 0.020 \mbig{$^{+0.009}_{-0.014}$} \ 
 &  0.021 $\pm$ 0.020 \mbig{$^{+0.009}_{-0.002}$} \ 
 &  -0.001 $\pm$ 0.021 \mbig{$^{+0.013}_{-0.005}$} \   \\

 3 & \rzqumu   & -0.011 $\pm$ 0.023 \mbig{$^{+0.012}_{-0.005}$} \ 
 & -0.011 $\pm$ 0.022 \mbig{$^{+0.013}_{-0.009}$} \ 
 & -0.005 $\pm$ 0.024 \mbig{$^{+0.005}_{-0.012}$} \  \\

 4 & \ruzz     & -0.083 $\pm$ 0.083 \mbig{$^{+0.026}_{-0.031}$} \ 
 & -0.005 $\pm$ 0.082 \mbig{$^{+0.021}_{-0.008}$} \ 
 & -0.058 $\pm$ 0.087 \mbig{$^{+0.015}_{-0.006}$} \  \\

 5 & \ruuu     &  0.016 $\pm$ 0.058 \mbig{$^{+0.018}_{-0.016}$} \ 
 &  0.003 $\pm$ 0.057 \mbig{$^{+0.007}_{-0.007}$} \ 
 & -0.030 $\pm$ 0.061 \mbig{$^{+0.006}_{-0.006}$} \  \\

 6 & Re \ruuz  & -0.032 $\pm$ 0.028 \mbig{$^{+0.022}_{-0.013}$} \ 
 & -0.044 $\pm$ 0.028 \mbig{$^{+0.007}_{-0.018}$} \ 
 &  0.029 $\pm$ 0.030 \mbig{$^{+0.010}_{-0.013}$} \  \\

 7 & \ruumu    &  0.098 $\pm$ 0.031 \mbig{$^{+0.013}_{-0.015}$} \ 
 &  0.134 $\pm$ 0.030 \mbig{$^{+0.006}_{-0.012}$} \ 
 &  0.170 $\pm$ 0.033 \mbig{$^{+0.014}_{-0.009}$} \  \\

 8 & Im \rduz  &  0.020 $\pm$ 0.028 \mbig{$^{+0.014}_{-0.010}$} \ 
 &  0.045 $\pm$ 0.028 \mbig{$^{+0.005}_{-0.008}$} \ 
 &  0.023 $\pm$ 0.031 \mbig{$^{+0.009}_{-0.005}$} \  \\

 9 & Im \rdumu & -0.136 $\pm$ 0.031 \mbig{$^{+0.007}_{-0.003}$} \ 
 & -0.143 $\pm$ 0.031 \mbig{$^{+0.011}_{-0.007}$} \ 
 & -0.078 $\pm$ 0.033 \mbig{$^{+0.025}_{-0.007}$} \   \\ 

10 & \rczz     &  0.090 $\pm$ 0.041 \mbig{$^{+0.055}_{-0.038}$} \ 
 &  0.069 $\pm$ 0.041 \mbig{$^{+0.033}_{-0.012}$} \ 
 &  0.132 $\pm$ 0.044 \mbig{$^{+0.020}_{-0.050}$} \  \\

11 & \rcuu     &  -0.003 $\pm$ 0.029 \mbig{$^{+0.015}_{-0.027}$} \ 
 & 0.015 $\pm$ 0.029 \mbig{$^{+0.007}_{-0.012}$} \ 
 &  0.012 $\pm$ 0.031 \mbig{$^{+0.012}_{-0.013}$} \  \\

12 & Re \rcuz  &  0.155 $\pm$ 0.013 \mbig{$^{+0.005}_{-0.011}$} \ 
 &  0.138 $\pm$ 0.013 \mbig{$^{+0.009}_{-0.010}$} \ 
 &  0.138 $\pm$ 0.014 \mbig{$^{+0.012}_{-0.002}$} \  \\

13 & \rcumu    & -0.021 $\pm$ 0.016 \mbig{$^{+0.014}_{-0.007}$} \ 
 &  0.014 $\pm$ 0.016 \mbig{$^{+0.001}_{-0.007}$} \ 
 & 0.003 $\pm$ 0.017 \mbig{$^{+0.003}_{-0.007}$} \  \\

14 & Im \rsuz  & -0.143 $\pm$ 0.013 \mbig{$^{+0.005}_{-0.006}$} \ 
 & -0.122 $\pm$ 0.013 \mbig{$^{+0.004}_{-0.006}$} \ 
 & -0.152 $\pm$ 0.014 \mbig{$^{+0.004}_{-0.001}$} \  \\

15 & Im \rsumu &  0.004 $\pm$ 0.016 \mbig{$^{+0.002}_{-0.000}$} \ 
 & -0.002 $\pm$ 0.016 \mbig{$^{+0.005}_{-0.003}$} \ 
 &  0.001 $\pm$ 0.017 \mbig{$^{+0.004}_{-0.004}$} \  \\

 \hline
 \hline
\end{tabular}
\caption{Spin density matrix elements for elastic electroproduction 
  of \rh\ mesons, measured for three values of \ttra\ with the 1996 data sample. 
  The first errors are statistical, the second systematic.}
\label{table:matrelt}
\end{table}
%
\begin{table}[htbp]
\begin{center}
\begin{tabular}{|c|l l l|}
\hline
\hline 
\qsq\ (\gevsq) & \multicolumn{3}{|c|}{$R = \sigma_L / \sigma_T$} \\[0.2 cm]
\hline
\hline
1.8  & 1.03 & \mbig{$^{+0.16}_{-0.16}$} & \mbig{$^{+0.10}_{-0.10}$} \\[0.2 cm]
2.7  & 1.75 & \mbig{$^{+0.36}_{-0.29}$} & \mbig{$^{+0.30}_{-0.28}$} \\[0.2 cm]
3.4  & 2.25 & \mbig{$^{+0.42}_{-0.34}$} & \mbig{$^{+0.13}_{-0.10}$} \\[0.2 cm]
4.8  & 2.22 & \mbig{$^{+0.46}_{-0.36}$} & \mbig{$^{+0.14}_{-0.06}$} \\[0.2 cm]
7.2  & 2.67 & \mbig{$^{+0.70}_{-0.50}$} & \mbig{$^{+0.32}_{-0.10}$} \\[0.2 cm]
10.9 & 3.38 & \mbig{$^{+1.30}_{-0.82}$} & \mbig{$^{+0.28}_{-0.46}$} \\[0.2 cm]
19.7 & 2.60 & \mbig{$^{+1.19}_{-0.72}$} & \mbig{$^{+0.27}_{-0.08}$} \\[0.2 cm]
\hline
\hline
\end{tabular}
\end{center}
\caption{Measurement of the ratio $R = \sigma_L / \sigma_T$ for seven values
of \qsq, with $\langle W \rangle = 75$ GeV, 
obtained from the measurement of the matrix element \rzzzz, assuming  
SCHC. The first errors are statistical, the second systematic.}
\label{table:R}
\end{table}
\begin{table}[htbp]
    \centering
    \begin{tabular}{|c|c|c|c|}
    \hline
    \hline
\qsq\ (\gevsq) & \W\ (GeV) & $|t|$ (\gevsq) & $\cos\delta$ \\  
    \hline
2.5 - 3.5 & 30 - 100 & 0.0 - 0.5 & 0.867 $\pm$ 0.051 \mbig{$^{+0.007}_{-0.019}$}  \\
3.5 - 6.0 & 30 - 120 & 0.0 - 0.5 & 0.841 $\pm$ 0.056 \mbig{$^{+0.020}_{-0.003}$}  \\
6.0 - 60. & 30 - 140 & 0.0 - 0.5 & 0.964 $\pm$ 0.071 \mbig{$^{+0.012}_{-0.012}$}  \\
    \hline
2.5 - 60. & 40 - 60  & 0.0 - 0.5 & 0.922 $\pm$ 0.053 \mbig{$^{+0.020}_{-0.019}$}  \\
2.5 - 60. & 60 - 80  & 0.0 - 0.5 & 0.903 $\pm$ 0.064 \mbig{$^{+0.022}_{-0.028}$}  \\
2.5 - 60. & 80 - 100 & 0.0 - 0.5 & 0.690 $\pm$ 0.101 \mbig{$^{+0.035}_{-0.046}$}  \\
    \hline
2.5 - 60. & 30 - 140 & 0.0 - 0.1 & 0.915 $\pm$ 0.039 \mbig{$^{+0.011}_{-0.011}$}  \\
2.5 - 60. & 30 - 140 & 0.1 - 0.2 & 0.904 $\pm$ 0.063 \mbig{$^{+0.041}_{-0.059}$}  \\
2.5 - 60. & 30 - 140 & 0.2 - 0.5 & 0.868 $\pm$ 0.060 \mbig{$^{+0.053}_{-0.006}$}  \\
    \hline
    \hline
    \end{tabular}
\caption{Measurements of the \cosdelta\ parameter as a function of \qsq,
  \W\ and $t$, obtained under NPE and the SCHC approximation from fits to the 
  (\cosths, $\psi$) distributions.
  The first errors are statistical, the second systematic.}
\label{table:cdel}
\end{table}
\begin{table}[ht]
    \centering
    \begin{tabular}{|c|c|c|c|c|}
    \hline
    \hline
 \qsq\ (\gevsq) &  \W\ (GeV) & $|t|$ (\gevsq) & 2 \ruuu + \ruzz & 2 \rcuu + \rczz \\
    \hline
2.5 - 3.0 & 30 - 100 & 0.0 - 0.5 &  0.046 $\pm$ 0.083 \mbig{$^{+0.025}_{-0.009}$} & 0.097 $\pm$ 0.039 \mbig{$^{+0.029}_{-0.005}$}  \\
3.0 - 4.0 & 30 - 100 & 0.0 - 0.5 & -0.140 $\pm$ 0.065 \mbig{$^{+0.011}_{-0.036}$} & 0.115 $\pm$ 0.034 \mbig{$^{+0.011}_{-0.010}$}  \\
4.0 - 6.0 & 30 - 120 & 0.0 - 0.5 & -0.079 $\pm$ 0.072 \mbig{$^{+0.059}_{-0.008}$} & 0.120 $\pm$ 0.036 \mbig{$^{+0.011}_{-0.015}$}  \\
6.0 - 9.0 & 30 - 140 & 0.0 - 0.5 & -0.023 $\pm$ 0.084 \mbig{$^{+0.027}_{-0.029}$} & 0.109 $\pm$ 0.043 \mbig{$^{+0.018}_{-0.005}$}  \\
9.0 - 14. & 30 - 140 & 0.0 - 0.5 &  0.006 $\pm$ 0.119 \mbig{$^{+0.042}_{-0.061}$} & 0.216 $\pm$ 0.054 \mbig{$^{+0.021}_{-0.032}$}  \\
14. - 60. & 30 - 140 & 0.0 - 0.5 & -0.173 $\pm$ 0.156 \mbig{$^{+0.061}_{-0.053}$} & 0.113 $\pm$ 0.077 \mbig{$^{+0.050}_{-0.040}$}  \\
  \hline
2.5 - 60.0 & 40 - 60 & 0.0 - 0.5 & -0.118 $\pm$ 0.066 \mbig{$^{+0.045}_{-0.013}$} & 0.025 $\pm$ 0.033 \mbig{$^{+0.004}_{-0.009}$}  \\
2.5 - 60.0 & 60 - 80 & 0.0 - 0.5 & -0.040 $\pm$ 0.069 \mbig{$^{+0.016}_{-0.025}$} & 0.175 $\pm$ 0.034 \mbig{$^{+0.011}_{-0.012}$}  \\
2.5 - 60.0 & 80 - 100 & 0.0 - 0.5 & -0.106 $\pm$ 0.074 \mbig{$^{+0.024}_{-0.012}$} & 0.183 $\pm$ 0.039 \mbig{$^{+0.018}_{-0.012}$}  \\
 \hline
2.5 - 60.0 & 30 - 140 & 0.0 - 0.1 & -0.060 $\pm$ 0.049 \mbig{$^{+0.027}_{-0.006}$} & 0.092 $\pm$ 0.025 \mbig{$^{+0.028}_{-0.020}$}  \\
2.5 - 60.0 & 30 - 140 & 0.1 - 0.2 &  0.012 $\pm$ 0.068 \mbig{$^{+0.008}_{-0.055}$} & 0.114 $\pm$ 0.033 \mbig{$^{+0.018}_{-0.005}$}  \\
2.5 - 60.0 & 30 - 140 & 0.2 - 0.3 & -0.053 $\pm$ 0.090 \mbig{$^{+0.015}_{-0.041}$} & 0.126 $\pm$ 0.044 \mbig{$^{+0.041}_{-0.023}$}  \\
2.5 - 60.0 & 30 - 140 & 0.3 - 0.5 & -0.182 $\pm$ 0.085 \mbig{$^{+0.074}_{-0.011}$} & 0.196 $\pm$ 0.046 \mbig{$^{+0.010}_{-0.039}$}  \\
    \hline
    \hline
    \end{tabular}
\caption{Measurements of the combinations of matrix elements
  $2 \ruuu + \ruzz$ and $2 \rcuu + \rczz$,
  as a function of \qsq, \W\ and \ttra, obtained from fits to the $\phi$ 
  distributions.
  The first errors are statistical, the second systematic.}
\label{table:rfive}
\end{table}
 \begin{table}[htbp]
\begin{center}
\begin{tabular}{|c|l l l|}
\hline
\hline 
\qsq\ (\gevsq) & \multicolumn{3}{|c|}{$b$ (\gevsqm)} \\[0.2 cm]
\hline
\hline 
1.8  & 8.0 & $\pm$ 0.5  & \mbig{$^{+0.6}_{-0.6}$}  \\[0.2 cm]
3.1  & 7.1 & $\pm$ 0.4  & \mbig{$^{+0.3}_{-0.4}$}  \\[0.2 cm]
4.8  & 5.5 & $\pm$ 0.5  & \mbig{$^{+0.5}_{-0.2}$}  \\[0.2 cm]
7.2  & 6.2 & $\pm$ 0.6  & \mbig{$^{+0.4}_{-0.4}$}  \\[0.2 cm]
10.9 & 5.6 & $\pm$ 0.8  & \mbig{$^{+0.4}_{-0.4}$}  \\[0.2 cm]
19.7 & 4.7 & $\pm$ 1.0  & \mbig{$^{+0.7}_{-0.7}$}  \\[0.2 cm]
\hline
\hline
\end{tabular}
\end{center}
\caption{Measurement of the slope parameter $b$ of the exponential $t$ dependence
for six values of \qsq, with $\langle W \rangle = 75$ GeV.
  The first errors are statistical, the second systematic.}
\label{tab:b}
\end{table}

\begin{table}[htbp]
\begin{center}
\begin{tabular}{|c|l l r|}
\hline
\hline 
\qsq\ (\gevsq) &  \multicolumn{3}{c|}{$\sigma$ 
($\gamma^* p \rightarrow \rho p$) (nb)}\\[0.2 cm]
\hline
\hline
1.1   & 2129 & $\pm$ 369  &  \mbig{$^{+275}_{-275}$} \\[0.2 cm]
1.4   & 1610 & $\pm$ 194  &  \mbig{$^{+207}_{-207}$} \\[0.2 cm]
1.7   & 1186 & $\pm$ 155  &  \mbig{$^{+153}_{-153}$} \\[0.2 cm]
2.3   & 681  & $\pm$ 83   &  \mbig{$^{+88}_{-88}$} \\[0.2 cm]
2.7   & 432  & $\pm$ 39   &  \mbig{$^{+46}_{-31}$} \\[0.2 cm]
3.0   & 399  & $\pm$ 34   &  \mbig{$^{+42}_{-31}$} \\[0.2 cm]
3.3   & 314  & $\pm$ 29   &  \mbig{$^{+41}_{-32}$} \\[0.2 cm]
3.8   & 261  & $\pm$ 24   &  \mbig{$^{+27}_{-24}$} \\[0.2 cm]
4.2   & 206  & $\pm$ 21   &  \mbig{$^{+23}_{-24}$} \\[0.2 cm]
4.7   & 157  & $\pm$ 17   &  \mbig{$^{+14}_{-17}$} \\[0.2 cm]
5.3   & 120  & $\pm$ 14   &  \mbig{$^{+14}_{-17}$} \\[0.2 cm]
6.0   & 106  & $\pm$ 13   &  \mbig{$^{+9}_{-9}$} \\[0.2 cm]
6.7   & 79   & $\pm$ 10   &  \mbig{$^{+11}_{-7}$} \\[0.2 cm]
7.5   & 81   & $\pm$ 10   &  \mbig{$^{+7}_{-11}$} \\[0.2 cm]
8.4   & 50.7 & $\pm$ 7.3  &  \mbig{$^{+4.4}_{-3.7}$} \\[0.2 cm]
9.4   & 47.5 & $\pm$ 6.7  &  \mbig{$^{+4.0}_{-4.9}$} \\[0.2 cm]
10.9  & 27.5 & $\pm$ 4.1  &  \mbig{$^{+2.5}_{-2.5}$} \\[0.2 cm]
13.0  & 19.9 & $\pm$ 3.1  &  \mbig{$^{+1.6}_{-1.6}$} \\[0.2 cm]
15.4  & 17.7 & $\pm$ 3.3  &  \mbig{$^{+1.9}_{-1.7}$} \\[0.2 cm]
18.3  & 11.6 & $\pm$ 2.7  &  \mbig{$^{+1.3}_{-1.3}$} \\[0.2 cm]
22.8  & 6.0  & $\pm$ 1.5  &  \mbig{$^{+0.8}_{-0.6}$} \\[0.2 cm]
35.0  & 1.6  & $\pm$ 0.5  &  \mbig{$^{+0.2}_{-0.2}$} \\[0.2 cm]
\hline
\hline
\end{tabular}
\end{center}
\caption{Measurement of the cross section for the process 
$\gamma^* p \rightarrow \rho p$ as a function of \qsq\ for $W = 75\ \gev$.
The first errors are statistical, the second systematic.}
\label{tab:s_qsq}
\end{table}

 \begin{table}[htbp]
\begin{center}
\begin{tabular}{|c|c|l l r|}
\hline
\hline 
\qsq\ (\gevsq) & $W$ (GeV) & \multicolumn{3}{c|}{$\sigma$ ($\gamma^* p \rightarrow
\rho p)$ (nb)}\\[0.2 cm]
\hline
\hline 
2.0  & 49  & 718   & $\pm$ 85   & \mbig{$^{+92}_{-92}$} \\[0.2 cm]
     & 65  & 991   & $\pm$ 118  & \mbig{$^{+128}_{-128}$} \\[0.2 cm]
     & 86  & 1025  & $\pm$ 117  & \mbig{$^{+133}_{-133}$} \\[0.2 cm]
     & 116 & 1002  & $\pm$ 118  & \mbig{$^{+129}_{-129}$} \\[0.2 cm]
\hline 
3.1  & 40  & 296   & $\pm$ 24   & \mbig{$^{+27}_{-20}$} \\[0.2 cm]
     & 60  & 318   & $\pm$ 28   & \mbig{$^{+27}_{-21}$} \\[0.2 cm]
     & 80  & 410   & $\pm$ 34   & \mbig{$^{+34}_{-28}$} \\[0.2 cm]
\hline
4.8  & 40  & 125   & $\pm$ 13   & \mbig{$^{+11}_{-11}$} \\[0.2 cm]
     & 60  & 137   & $\pm$ 16   & \mbig{$^{+12}_{-10}$} \\[0.2 cm]
     & 80  & 160   & $\pm$ 19   & \mbig{$^{+16}_{-15}$} \\[0.2 cm]
     &100  & 168   & $\pm$ 21   & \mbig{$^{+15}_{-14}$} \\[0.2 cm]
\hline 
7.2  & 50  &  60.3 & $\pm$ 7.9  & \mbig{$^{+5.5}_{-3.9}$}   \\[0.2 cm]
     & 70  &  76.2 & $\pm$ 10.8 & \mbig{$^{+8.9}_{-5.7}$} \\[0.2 cm]
     & 90  &  96.7 & $\pm$ 14.3 & \mbig{$^{+12.2}_{-6.2}$} \\[0.2 cm]
     &110  &  94.7 & $\pm$ 15.0 & \mbig{$^{+8.4}_{-6.1}$} \\[0.2 cm]
     &130  &  76.9 & $\pm$ 18.8 & \mbig{$^{+11.8}_{-6.8}$} \\[0.2 cm]
\hline 
10.9 & 50  &  17.9 & $\pm$ 3.5 & \mbig{$^{+1.7}_{-1.9}$} \\[0.2 cm]
     & 70  &  34.6 & $\pm$ 5.7 & \mbig{$^{+3.7}_{-2.2}$} \\[0.2 cm]
     & 90  &  38.9 & $\pm$ 7.1 & \mbig{$^{+3.0}_{-3.2}$} \\[0.2 cm]
     &110  &  30.1 & $\pm$ 6.8 & \mbig{$^{+2.4}_{-2.0}$} \\[0.2 cm]
     &130  &  42.5 & $\pm$ 9.7 & \mbig{$^{+3.5}_{-2.4}$} \\[0.2 cm]
   \hline
19.7 & 60  &  7.2  & $\pm$ 1.6 & \mbig{$^{+0.6}_{-0.6}$} \\[0.2 cm]
     & 80  &  10.4 & $\pm$ 2.3 & \mbig{$^{+0.7}_{-1.0}$} \\[0.2 cm]
     &100  &  9.4  & $\pm$ 2.4 & \mbig{$^{+1.0}_{-0.7}$} \\[0.2 cm]
     &120  &  14.9 & $\pm$ 3.2 & \mbig{$^{+1.1}_{-1.1}$} \\[0.2 cm]
\hline
\hline
\end{tabular}
\caption{Measurement of the cross section for the process 
$\gamma^* p \rightarrow \rho p$
as a function of $W$, for several values of \qsq. 
The first errors are statistical, the second systematic.}
\label{tab:w}
\end{center}
\end{table}

 \begin{table}[htbp]
\begin{center}
\begin{tabular}{|c|l l l|}
\hline
\hline 
\qsq\ (\gevsq) & \multicolumn{3}{|c|}{$\alpha(0)$} \\[0.2 cm]
\hline
\hline 
2.0   & 1.13 & $\pm$ 0.05  & \mbig{$^{+0.00}_{-0.03}$} \\[0.2 cm]
3.1   & 1.15 & $\pm$ 0.04  & \mbig{$^{+0.00}_{-0.04}$} \\[0.2 cm]
4.8   & 1.12 & $\pm$ 0.04  & \mbig{$^{+0.00}_{-0.04}$} \\[0.2 cm]
7.2   & 1.15 & $\pm$ 0.06  & \mbig{$^{+0.00}_{-0.04}$} \\[0.2 cm]
10.9  & 1.23 & $\pm$ 0.06  & \mbig{$^{+0.00}_{-0.05}$} \\[0.2 cm]
19.7  & 1.27 & $\pm$ 0.11  & \mbig{$^{+0.00}_{-0.05}$} \\[0.2 cm]
\hline
\hline
\end{tabular}
\end{center}
\caption{Measurements of the $\alpha(0)$ parameter (see eqs.
\ref{eq:wdistrdelta} $-$ \ref{eq:alphaprim}) as a function of \qsq.
The first error represents the statistical and non-correlated systematic 
uncertainties on the cross section measurements; the second error 
represents the variation due to the assumption $\alpha^\prime = 0$
(i.e. no shrinkage).}
\label{tab:epsilon}
\end{table}



\begin{thebibliography} {19}
%
\bibitem {H1_rho_94}
  S. Aid et al., H1 Coll., \np {468} {1996} {3}.
%
\bibitem{rho_zeus}
   ZEUS Coll, DESY 98-107, subm. to {\it Eur. Phys. J.}
%
%
\bibitem{CHIO}
   W.D.\ Shambroom et al., CHIO Coll.,
   \prev {D26} {1982} {1}.
%
\bibitem{NMC}
  P.\ Amaudruz et al., NMC Coll., \zp {C54} {1992} {239};\\
  M.\ Arneodo et al., NMC Coll., \np {B429} {1994} {503}.
%
\bibitem{E665}
  M. R. Adams et al., E665 Coll., \zp {C74} {1997} {237}.
%
\bibitem{joos}
  P.\ Joos et al., \np {B113} {1976} {53}.
%
\bibitem {H1_NIM} 
  I. Abt et al.,
  H1 Coll.,
  \nim {386} {1997} {310} and 348.
%
\bibitem {spacal}
  R.D. Appuhn et al., 
  {\it H1 SPACAL Group},    
  \nim {386} {1997} {397}.
%
\bibitem{Koijman_Workshop}
  S.\ Bentvelsen, J.\ Engelen and P.\ Kooijman,
    {\it Reconstruction of (x, \qsq) and extraction of structure functions
    in neutral current scattering at HERA,}
    in: Proc. of the Workshop on Physics at HERA,
    ed. W. Buchm\"uller and G. Ingelman, Hamburg 1992, Vol. 1, p. 23; \\
  K.C.\ Hoeger,
    {\it Measurement of $x$, $y$, \qsq\ in Neutral Current Events,}
    ibid., p. 43.
%
\bibitem{Jacquet_Blondel}
  F. Jacquet, A. Blondel, DESY 79-048 (1979) 377.
%
\bibitem{DIFFVM}
 {\it DIFFVM program,}
 see: B.\ List,
    {\it Diploma Thesis}, Techn. Univ. Berlin, 1993, unpubl.
%
\bibitem{Barbara}
  B. Clerbaux, {\it PhD Thesis}, Univ. Libre de Bruxelles, 1998, unpubl.,
  DESY-THESIS-1999-001.
%
%
\bibitem{HERACLES}
%
A.\ Kwiatkowski, H.-J.\ M\"ohring and H.\ Spiesberger,
 Computer Physics Commun. {\bf 69} (1992) 155; \\
A.\ Kwiatkowski, H.-J.\ M\"ohring and H.\ Spiesberger,
 in: Proc. of the Workshop on Physics at HERA, ed. W. Buchm\"uller and G.
 Ingelman, Hamburg 1992, Vol. 3, p. 1294; \\
H.\ Spiesberger, {\it HERACLES version 4.4}, unpubl. program manual (1993).
%
\bibitem{PDG}
  C.\ Caso et al., 
  {\it Particle Data Group,}
  \epj {C3} {1998} {1}.
%
\bibitem{ZEUS_omega}
  M.\ Derrick et al.,
  ZEUS Coll.,
  \zp {C73} {1996} {73}. 
%
\bibitem{ZEUS_phi}
  M.\ Derrick et al.,
  ZEUS Coll.,
  \pl {B380} {1996} {220}. 
%
\bibitem{H1_phi}
  C. Adloff et al., 
  H1 Coll., 
  \zp {C75} {1997} {607}.
%
\bibitem {H1_rho_95}
  H1 Coll., 
  {\it Elastic Electroproduction of $\rho$ and $\phi$ Mesons 
        for 1 $<$ \qsq\ $<$ 5 \gevsq\ at HERA},
  contributed paper to the Int. Europhys. Conf. on HEP, Jerusalem, Israel,
  1997.
%
\bibitem{Goulianos}
  K. Goulianos,
  \prep {101} {1983} {169}.
%
\bibitem{jetset}
 T.\ Sj\"ostrand, Computer Physics Commun. {\bf 82} (1994) 74.  
%
\bibitem{pdiss_paper}
  X. Janssen, {\it M\'emoire de Licence}, Univ. Libre de Bruxelles, 
  1998, unpubl.
%
\bibitem {drell}
  S.D.\ Drell, 
  \prl {5} {1960} {278}.
%
\bibitem {ross_sto}
  R.\ Ross and V.\ Stodolsky, 
  \prev {149} {1966} {1173}.
%
\bibitem {jackson}
  J.D.\ Jackson, 
  \nc {34} {1964} {1644}. 
%
\bibitem {soding}
  P.\ S\"{o}ding, 
  \pl {B19} {1966} {702}.
%
\bibitem {h1_phot}
  S. Aid et al., H1 Coll., \np {463} {1996} {3}.
%
\bibitem {zeus_phot}
  M. Derrick et al., ZEUS Coll., \zp {C69} {1995} {39}.
%
\bibitem {zeus_phottt}
  J. Breitweg et al., ZEUS Coll., \epj {C2} {1998} {247}.
%
\bibitem {bauer}
  T.H.\ Bauer et al., 
  \rmp {50} {1978} {261}.
%
\bibitem{shilling-wolff}
  K. Schilling and G. Wolf,
  \np {B61} {1973} {381}. 
%
\bibitem{ivanov}
  D.Yu. Ivanov and R. Kirschner,
  \prev {D59} {1998} {114026}.
%
\bibitem{delpapa}
  C.\ del Papa et al., \prev {D19} {1979} {1303}.
%
\bibitem{sss}
  D. Schildknecht, G.A. Schuler and B. Surrow,
  {\it Vector-Meson Electroproduction from Generalized Vector Dominance},
  preprint CERN-TH-98-294 (1998), hep-ph/9810370.
%
\bibitem{mrt}
 A.D. Martin, M.G. Ryskin and T. Teubner, \prev {D55}{1997}{4329}.
%
\bibitem{cudell}
 I. Royen and  J.-R. Cudell, 
  {\it Fermi Motion and Quark Off-shellness in Elastic Vector-Meson
  Production}, 
  preprint UGL-PNT-98-2-JRC (1998), hep-ph/9807294.
%
\bibitem{pdflib}
  H. Plothow-Besch, {\it PDFLIB: Nucleon, Pion and Photon Parton Density
  Functions and $\alpha_{s}$ Calculations, User's Manual - version 7.09},
  W5051 PDFLIB, 02/07/1997, CERN-PPE.
%
\bibitem{mrs}
A. Martin, R. Roberts and  W. Stirling, \pl {B387} {1996} {419}.
%
\bibitem{grv}
M. Gl\"uck, E. Reya and A.Vogt, \zp {C67} {1995} {433}.
%
\bibitem {zeus_phott}
  M. Derrick et al., ZEUS Coll., \zp {C73} {1997} {253}.  
%
\bibitem {hand}
  L. N. Hand, \prev {129} {1963} { 1834}.
%
%
\bibitem{dola}    
    A.\ Donnachie and P.V.\ Landshoff, \pl {B296} {1992} {227}.
%
\bibitem{cudellfit}
  J.-R. Cudell, K. Kang and S. Kim,
  \pl {B395} {1997} {311}.
%
\bibitem{fks}
 L. Frankfurt, W. Koepf and M. Strikman, \prev {D54}{1996}{3194}.
%
%
%
%
%


\end{thebibliography}
\end{document}